\documentclass[12pt]{article}

\usepackage{amsmath}
\usepackage{graphicx}
\usepackage{natbib}
\usepackage{subfig}
\usepackage{amssymb}
\usepackage{amsfonts}
\usepackage{mathtools}
\usepackage{multirow}
\usepackage{amsthm}
\usepackage{smile}
\usepackage{titlesec}
\usepackage{color}
\usepackage{algorithm}
\usepackage{enumerate}
\usepackage[colorlinks,
linkcolor=red,
anchorcolor=blue,
citecolor=blue
]{hyperref}
\usepackage{cleveref}
\usepackage{url}

\newcommand{\blind}{0}

\newcommand{\HomoModel}{MonoSBM}
\newcommand{\HeteroModel}{MultiSBM}
\def\T{\mathrm{\scriptstyle T}}
\def\rF{\mathrm{\scriptstyle F}}
\crefname{equation}{Eq.}{Eqs.}
\crefname{assumption}{assumption}{assumptions}
\crefname{cond}{condition}{conditions}

\usepackage{stmaryrd}
\newcommand{\nset}[1]{\llbracket #1 \rrbracket} % << Additional user-defined commands

\usepackage{xcolor}

\usepackage{pdfpages}  % << arXiv

\begin{document}

\def\spacingset#1{\renewcommand{\baselinestretch}%
{#1}\small\normalsize} \spacingset{1}

\if0\blind{
  \title{\bf Simultaneous estimation of connectivity and dimensionality in samples of networks}
  \author{
  Wenlong Jiang\textsuperscript{1},
  Chris McKennan\textsuperscript{1},
  Jesús Arroyo\textsuperscript{2}, and
  Joshua Cape\textsuperscript{3}\thanks{
    This research was supported in part by the University of Pittsburgh Center for Research Computing through the resources provided. Specifically, this work used the H2P cluster which is supported by the National Science Foundation under grant OAC 2117681. J. Arroyo and J. Cape were supported in part by the National Science Foundation under grants DMS 2413553 and DMS 2413552, respectively. J. Cape was also supported in part by the University of Wisconsin--Madison, Office of the Vice Chancellor for Research and Graduate Education with funding from the Wisconsin Alumni Research Foundation. The authors thank Kevin Z. Lin for assistance with the primate brain dataset.}\\
    \textsuperscript{1}Department of Statistics, University of Pittsburgh\\
    \textsuperscript{2}Department of Statistics, Texas A\&M University\\
    \textsuperscript{3}Department of Statistics, University of Wisconsin--Madison}
  \maketitle
} \fi

\if1\blind
{
  \bigskip
  \bigskip
  \bigskip
  \begin{center}
    {\LARGE\bf Simultaneous estimation of connectivity and dimensionality in samples of networks}
\end{center}
  \medskip
} \fi

\bigskip
\begin{abstract}
    An overarching objective in contemporary statistical network analysis is extracting salient information from datasets consisting of multiple networks. To date, considerable attention has been devoted to node and network clustering, while comparatively less attention has been devoted to downstream connectivity estimation and parsimonious embedding dimension selection. Given a sample of potentially heterogeneous networks, this paper proposes a method to simultaneously estimate a latent matrix of connectivity probabilities and its embedding dimensionality or rank after first pre-estimating the number of communities and the node community memberships. The method is formulated as a convex optimization problem and solved using an alternating direction method of multipliers algorithm. We establish estimation error bounds under the Frobenius norm and nuclear norm for settings in which observable networks have blockmodel structure, even when node memberships are imperfectly recovered. When perfect membership recovery is possible and dimensionality is much smaller than the number of communities, the proposed method outperforms conventional averaging-based methods for estimating connectivity and dimensionality. Numerical studies empirically demonstrate the accuracy of our method across various scenarios. Additionally, analysis of a primate brain dataset demonstrates that posited connectivity is not necessarily full rank in practice, illustrating the need for flexible methodology.
\end{abstract}

\noindent%
\vfill

%\newpage
\spacingset{1.75} % DON'T change the spacing!
\section{Introduction}
\label{sec:intro}

Networks describe interactions among entities, such as friendships among people in social networks \citep{yang2011like,leung2014interactive,ameri2023strangers}, gene-gene associations in biological networks \citep{tong2004global,xia2015testing,al-aamri2019analyzing}, and region-region connectivity in brain networks \citep{sporns2012simple,zuo2012network,vecchio2017connectome}.
Network analysis has emerged as an active research area in the past several decades due in part to the abundance of high quality graph-structured and network-valued data.
Often, individual networks are viewed as instances of random graphs, where edges (interactions) reflect observable or inferred connectivity patterns among nodes (entities). 
In practice, networks are structured, say, with nodes of a similar type sharing similar traits and connectivity patterns, in contrast to nodes of different types or with different traits. 
These concepts can be codified using the formalism of latent node communities or blocks.

This paper focuses on the stochastic blockmodel (SBM) \citep{holland1983stochastic} as the exemplar random graph model with community structure.
Stochastic blockmodel random graphs partition $n$ nodes into $K$ communities, typically with $K \ll n$, on the basis of latent block (community) memberships.
Conditional on the node memberships, edges between nodes are independently drawn from different Bernoulli distributions, where the Bernoulli edge connectivity probabilities are specified in a symmetric $K \times K$ matrix. 
As such, three key estimation tasks emerge from the study of SBMs as generative models for network data: to estimate the community memberships of nodes, to estimate the number of communities, and to estimate the matrix of connectivity probabilities.

The problem of estimating node memberships, also called \emph{community detection}, has been well-studied in SBMs and is surveyed in \cite{abbe2017community}.
Popular techniques include modularity maximization \citep{newman2004finding,bickel2009nonparametric}, likelihood-based approaches \citep{celisse2012consistency,choi2012stochastic,zhao2012consistency,amini2013pseudo}, and spectral clustering \citep{von2007tutorial,rohe2011spectral,lei2015consistency,paul2020spectral,lei2022bias}. 
Traditionally, community detection requires a pre-estimated or pre-specified (known) number of communities.
The problem of estimating the number of communities $K$ has itself received significant attention in recent years, via the use of spectral methods \citep{le2022estimating}, network cross-validation methods \citep{chen2018network,li2020network}, likelihood-based methods \citep{wang2017likelihood}, model selection methods \citep{yang2021simultaneous}, and stepwise testing methods \citep{jin2023optimal}.

Comparatively fewer works focus on estimating the connectivity probability matrix and its embedding dimensionality or rank in SBMs.
This is due in part to the fact that connectivity probabilities are viewed as nuisance parameters when exclusively seeking to cluster nodes.
Further, stylized versions of SBMs, such as the homogeneous planted partition model with connectivity probability matrix $\Bb$ satisfying $B_{k k} = p$ and $B_{k l} = q$ for $k \neq l$, are full rank and have historically received outsized attention in the literature. 
In such settings, the most commonly used edge probability estimator, namely averaging over edges within or between communities \citep{choi2012stochastic,li2020network}, has competitive mean squared error \citep{tang2022asymptotically} and is almost always full rank itself.

In practice, though, the underlying or inferred probability matrix need not be full rank, particularly when $K$ is large \citep{cai2022consensus}.
Namely, structural dependencies can exist between communities which can give rise to systematic relationships among connectivity probabilities, in which case the matrix rank (or an approximation thereof) may intrinsically determine connectivity estimation accuracy better than the number of communities. 
Consequently, methods assuming the full-rankness of $\Bb$ may not yield parsimonious dimensionality reduction for network embeddings, accurate estimates of connectivity probabilities, or appropriate conclusions for real data.
These observations suggest the potential usefulness of considering alternative strategies for estimating general $\Bb$ matrices as investigated here.

For single network analysis, various network estimation methods have been proposed based on thresholding, smoothing, and mixing techniques \citep{airoldi2013stochastic,chatterjee2015matrix,zhang2017estimating,li2023network}.
For SBMs specifically, \cite{tang2022asymptotically} studies competing methods for estimating the matrix $\Bb$ and investigates their asymptotic efficiency properties.
The results therein require that the rank of $\Bb$, interpretable as its embedding dimensionality, be given or at least consistently estimable, and the network itself must be not too sparse.
For multiple, potentially quite sparse networks with shared latent structure, the challenge of estimating embedding dimensionality or rank is even more pronounced while remaining equally important for downstream tasks such as hypothesis testing and link prediction.
These observations motivate the present paper and its focus on datasets comprising multiple, potentially quite sparse networks.

\subsection{Contributions}

This paper proposes a method to simultaneously estimate the connectivity probability matrix and its dimensionality in samples of stochastic blockmodel random graphs (SBMs).
The method is formulated as a convex optimization problem and solved using an alternating direction method of multiplier (ADMM) algorithm \citep{eckstein1992splitting,boyd2010distributed}.
To the best of our knowledge, this approach is new in the setting of statistical network analysis.
We provide theoretical support for the proposed method by establishing error bounds under the Frobenius norm and the nuclear norm for estimating the matrix of connectivity probabilities.
Our bounds quantify the influence of network sparsity and node misclustering rate on estimation accuracy without imposing stringent assumptions on either.
Numerical studies empirically demonstrate the accuracy of our method in various settings, even when individual graphs in the sample are themselves very sparse and hence contribute only limited usable information.
Further, when perfect membership recovery is possible and dimensionality is much smaller than the number of communities, the proposed method outperforms conventional estimation based on entrywise averaging.
Finally, a real data example shows that the proposed method is useful in practice when the posited connectivity probability matrix is plausibly rank degenerate.
Our contributions can be integrated with existing approaches for community detection \citep{paul2020spectral,jing2021community,lei2022bias,xu2023covariate} and estimating $K$ \citep{ lei2016goodness,wang2017likelihood,chen2018network,yang2021simultaneous,jin2023optimal} that do not require matrices of connectivity probabilities to be full rank.

\subsection{Notation}

Given a positive integer $n$, let $\nset{n} \coloneqq \{1, \dots, n\}$.
For any $p$-dimensional vector $\bu \coloneqq (u_{1}, \dots, u_{p})^{\T} \in \RR^{p}$ and $1 \le q < \infty$, let $\|\bu\|_{q} \coloneqq (\sum_{j=1}^{p}\left|u_{j}\right|^{q})^{1/q}$ denote the $\ell_{q}$ norm.
Let $\|\bu\|_{0} \coloneqq \sum_{j=1}^{p} \II(u_{j} \ne 0)$ denote the number of nonzero entries of $\bu$, where $\II(\cdot)$ denotes the zero-one indicator function.
For any two vectors $\bu, \bv \in \RR^{p}$, denote the Euclidean inner product by $\langle \bu, \bv \rangle \coloneqq \bu^{\T} \bv$.
Given an $m \times n$ matrix $\Ab$, let $\Ab_{i \bullet}$ denote the $i$-th row of $\Ab$ and $\Ab_{\bullet j}$ denote the $j$-th column of $\Ab$.
Given index sets $\cI \subseteq \nset{m}, \cJ \subseteq \nset{n}$, let $\Ab_{\cI\cJ}$ denote the submatrix of $\Ab$ containing the corresponding rows and columns.
We write $\|\Ab\|_{\rF} \coloneqq \sqrt{\sum_{i,j} A_{ij}^{2}}$ to denote the Frobenius norm of $\Ab$.
We write $\|\Ab\|_{\max} \coloneqq \max_{i,j}|A_{ij}|$ to denote the maximum absolute entry of $\Ab$.
We write $\|\Ab\|_{*} \coloneqq \sum_{k=1}^{\min\{m,n\}}\sigma_{k}$ to denote the nuclear norm of $\Ab$, where $\sigma_{k} \equiv \sigma_{k}(\Ab)$ denotes the $k$-th largest singular value of $\Ab$.
We write $|\Ab|$ to denote the matrix of element-wise absolute values of $\Ab$.
We write $\operatorname{rk}(\Ab) \coloneqq \max\{j: \sigma_{j}(\Ab) > 0\}$ to denote the rank of $\Ab$.
We write $\mathbf{\Lambda} \in \RR^{n \times n}$ to denote the diagonal matrix whose diagonal elements are the eigenvalues $\lambda_{1}, \dots, \lambda_{n}$ of a pre-specified matrix.
In contrast, we use $\lambda$ (without a subscript) to denote a tuning parameter.
We call $\Zb \in \{0, 1\}^{n \times K}$ a membership matrix when each row has exactly one $1$ and $(K-1)$ zeroes.
Here, the community membership of node $i$ is denoted by $g_{i} \in \nset{K}$, hence $\Zb_{i g_{i}} = 1$.
For each $k \in \nset{K}$, let $n_{k}$ denote the number of nodes belonging to the $k$-th community.
The zero matrix and identity matrix are denoted by $\mathbf{0}$ and $\Ib$, respectively, when their sizes are clear from context.
The vectorization of $\Ab$ is obtained by sequentially stacking its columns and is written as $\mathrm{vec}(\Ab)$.
The (right) Kronecker product of $\Ab \in \RR^{m \times n}$ and $\Bb \in \RR^{p \times q}$ is written in shorthand as $\Ab \otimes \Bb = (A_{ij} \Bb) \in \RR^{pm \times qn}$.
Given two non-negative sequences of real numbers $\{a_{n}\}_{n \ge 1}$ and $\{b_{n}\}_{n \ge 1}$, we write $a_{n} \lesssim b_{n}$ if there exist constants $C, n_{0} > 0$ such that $a_{n} \le C b_{n}$ for all $n \ge n_{0}$. 
Analogously, we write $a_{n} \gtrsim b_{n}$ when $b_{n} \lesssim a_{n}$.

\section{Estimation using samples of networks}
\label{sec:setting}

In this paper, we consider samples or collections of node-aligned networks, $\{\Ab^{(1)}, \dots, \Ab^{(L)}\}$, where $\Ab^{(\ell)}$ denotes the adjacency matrix for the $\ell$-th layer (network) in the collection.
In settings such as neuroimaging, samples of networks (e.g., derived from brain scans) are often collected on a fixed, shared set of nodes (e.g., brain regions) and are anticipated to exhibit similar connectivity patterns, for example, based on subjects' characteristics such as disease status.
With this in mind, we specify the setting of multiple independent and identically distributed stochastic blockmodel graphs \citep{holland1983stochastic} as follows, writing `mono' to emphasize that all adjacency matrices (layers) are generated via a single, common matrix of connectivity probabilities.

\begin{definition}[\textmd{Monolayer Stochastic Blockmodel (\HomoModel{})}]
    \label{def1}
    Let $K$ be the number of communities, $n$ be the number of nodes, and $L$ be the number of layers.
    Across all layers, let $\Zb \in \{0,1\}^{n \times K}$ be the fixed membership matrix with $Z_{ik}=1$ if and only if node $i$ belongs to community $k$.
    The matrix of connectivity probabilities is denoted by $\Bb^{*} \in [0,1]^{K \times K}$ with $\operatorname{rk}(\Bb^{*}) = d \le K$.
    For all $i \le j$, $A_{ij}$ is independently drawn from $\operatorname{Bernoulli}(\rho B^{*}_{g_{i}g_{j}})$, yielding the symmetric adjacency matrix $\Ab \in \{0,1\}^{n \times n}$.
    We call the collection $\Ab^{(\ell)} \overset{\text{i.i.d.}}{\sim} \operatorname{SBM}(\Bb^{*},\Zb)$ for $\ell \in \nset{L}$ a \emph{monolayer stochastic blockmodel}.
    Here, $\EE(\Ab^{(\ell)}) = \rho \Zb \Bb^{*} \Zb^{\T}$, where the sparsity term $\rho \equiv \rho_{n} \in (0,1]$ is known.
\end{definition}

\Cref{def1} specifies four population-level quantities of interest: $K$, $\Zb$, $d$, and $\Bb^{*}$.
As mentioned in \Cref{sec:intro}, numerous methods exist for estimating $K$ and $\Zb$, whereas this paper seeks to simultaneously estimate $d$ and $\Bb^{*}$.
Here, $n$ and $L$ are conventionally given from context, while assuming that $\rho$ is given avoids the ambiguity of individual finite-$n$ graphs belonging to multiple asymptotic sparsity regimes wherein $n \rho_{n} \rightarrow \infty$.

Community memberships or block labels are identifiable only up to global relabelling, hence it is necessary to make estimation statements about $\Zb$ modulo orthogonal transformations, specifically column permutations.
This leads us to write
\begin{equation}
    \label{model}
    \overline{\Ab}
    \coloneqq
    \frac{1}{L} \sum_{\ell=1}^{L} \Ab^{(\ell)}
    =
    \EE(\Ab) + \mathbf{\Delta}_{1},
    \quad
    \rho \hat{\Zb} \hat{\bm{\Pi}} \Bb^{*} \hat{\bm{\Pi}}^{\T} \hat{\Zb}^{\T}
    =
    \EE(\Ab) + \mathbf{\Delta}_{2}
    ,
\end{equation}
where the permutation matrix $\hat{\bm{\Pi}} \in \RR^{K \times K}$ satisfies $\hat{\bm{\Pi}} = \underset{\bm{\Pi}}{\arg\min}\|\hat{\Zb}\bm{\Pi} - \Zb\|_{0}$ given an estimated membership matrix $\hat{\Zb}$.
Here, each element of the former error matrix, $\mathbf{\Delta}_{1}$, converges to zero in probability as $L \rightarrow \infty$ by the law of large numbers.
\Cref{model} suggests $\overline{\Ab}$ as a natural, interpretable estimator of $\EE(\Ab)$.
Depending on the choice of clustering method used (e.g., see \cite{lei2015consistency,lei2022bias}), each element in the latter error matrix, $\mathbf{\Delta}_{2}$, may concentrate near zero with high probability even for very sparse individual graphs.
If $\Zb$ can be perfectly or exactly recovered with high probability, then $\hat{\Zb}\hat{\bm{\Pi}} = \Zb$, hence $\mathbf{\Delta}_{2} = \mathbf{0}$ holds with high probability.

Given $K$ and $\hat{\Zb}$, we wish to estimate the dimensionality and the connectivity per $\hat{\bm{\Pi}} \Bb^{*} \hat{\bm{\Pi}}^{\T}$, denoted simply by $\Bb^{*}$.
One approach for estimating $\Bb^{*}_{\rho} \equiv \rho\Bb^{*}$ is to solve
\begin{equation}
  \label{eq:e1}
  \underset{
  \Bb_{\rho}:~\Bb_{\rho}
  =
  \Bb_{\rho}^{\T}
  }{\mathrm{minimize}}~
  \|\overline{\Ab} -\hat{\Zb}\Bb_{\rho}\hat{\Zb}^{\T}\|_{\rF}^{2}
\end{equation}
and then to set $\hat{\Bb} \coloneqq \hat{\Bb}_{\rho}/\rho$.
The solution to \Cref{eq:e1} is equivalent to blockwise averaging which yields full rank estimates in general.
To simultaneously estimate the rank, we shall introduce a penalty term into \Cref{eq:e1} to obtain the modified problem
\begin{equation}
    \label{eq:e2}
    \underset{
    \Bb_{\rho}:~\Bb_{\rho}
    =
    \Bb_{\rho}^{\T}
    }{\mathrm{minimize}}~\|\overline{\Ab} - \hat{\Zb}\Bb_{\rho}\hat{\Zb}^{\T}\|_{\rF}^{2}
    +
    \lambda \|\Bb_{\rho}\|_{*}
    ,
\end{equation}
where $\lambda > 0$ is a tuning parameter and the nuclear norm $\|\cdot\|_{*}$ encourages low-rank solutions.
\Cref{eq:e2} and the scaled, rewritten version in \Cref{eq:e3} are the focus of this paper.
To the best of our knowledge, this formulation together with \Cref{eq:e1} is novel for analyzing structured networks in the context of statistical network analysis.

Under \HomoModel{}, each layer shares the same probability matrix $\Bb^{*}$.
This is not always realistic in practice, though, such as for the primate brain data in \cref{sec:real_data}, originally collected in \cite{bakken2016comprehensive} and investigated in \cite{lei2022bias}, where layers correspond to time points. 
\cite{lei2022bias} showed that adjacency matrices across layers have different connectivity patterns, suggesting that the underlying probability matrices differ.
This gives rise to the concept of the multilayer stochastic blockmodel \citep{han2015consistent,paul2016consistent,lei2022bias} which permits heterogeneous connectivity probability matrices in each layer but with the same node memberships across layers.

\begin{definition}[\textmd{Multilayer Stochastic Blockmodel (\HeteroModel{})}]
    \label{def2}
    Let $K$ be the number of communities, $n$ be the number of nodes, and $L$ be the number of layers.
    The membership matrix $\Zb \in \{0,1\}^{n \times K}$ is fixed across all layers with $Z_{ik}=1$ if and only if node $i$ belongs to community $k$.
    Let $\rho \equiv \rho_{n} \in (0,1]$ denote the known sparsity term.
    For $1 \le \ell \le L$, suppose $\Ab^{(\ell)}$ are independent adjacency matrices from among $\tilde{L} \le L$ different SBMs with connectivity probability matrices $\Bb^{*(\tilde{\ell})}$ and $\operatorname{rk}(\Bb^{*(\tilde{\ell})}) = d_{\tilde{\ell}}$, for $\tilde{\ell} \in \nset{\tilde{L}}$, which we call a \emph{multilayer stochastic blockmodel}. 
    Here, $\EE(\Ab^{(\ell)}) = \rho \Zb \Bb^{*(\tilde{\ell})} \Zb^{\T}$ if $\ell \in \GG_{\tilde{\ell}}$, where $\GG_{\tilde{\ell}}$ is the index set of $\Ab^{(\ell)}$ associated to $\operatorname{SBM}(\Bb^{*(\tilde{\ell})}, \Zb)$.
\end{definition}

Unlike \HomoModel{}, \HeteroModel{} necessitates estimating the different matrices $\Bb^{*(\tilde{\ell})}$.
Beforehand, to first estimate $\Zb$ in \HeteroModel{}, we employ the bias-adjusted spectral clustering method proposed by \cite{lei2022bias}, i.e., we cluster the rows of the leading eigenvectors of the matrix $\sum_{\ell = 1}^{L} [(\Ab^{(\ell)})^{2} - \operatorname{diag}(\Ab^{(\ell)} \mathbf{1})]$.
We then adopt a two-stage method to estimate $\Bb^{*(\tilde{\ell})}$ in \HeteroModel{} when $\tilde{L}$ is given.
Namely, we first estimate the probability matrix for each layer and cluster these matrices into $\tilde{L}$ groups based on their vectorized upper triangular elements, yielding $\hat{\GG}_{\tilde{\ell}}$.
The layers in each $\hat{\GG}_{\tilde{\ell}}$ belong to \HomoModel{} with $\Bb^{*} = \Bb^{*(\tilde{\ell})}$, so finally we solve \Cref{eq:e2} to estimate each $\Bb^{*(\tilde{\ell})}$.

\section{ADMM algorithm}
\label{sec:algorithm}

Here, we derive a method to solve the optimization problem in \Cref{eq:e2}.
The problem is convex, so we employ the alternating direction method of multiplier (ADMM) algorithm \citep{eckstein1992splitting,boyd2010distributed}.
In fact, we present an ADMM algorithm to solve a more general problem.
Given a symmetric matrix $\Yb \in \RR^{n \times n}$ and a feature matrix $\Xb \in \RR^{n \times K}$, we seek a low-rank matrix $\Wb \in \RR^{K \times K}$ so that $\Yb \approx \Xb\Wb\Xb^{\T}$, namely
\begin{equation}
    \label{eq:ADMM1}
    \begin{aligned}
        &\underset{\Wb}{\mathrm{minimize}}~
        \|\Yb - \Xb\Wb\Xb^{\T}\|_{\rF}^{2}
        +
        \lambda\|\Wb\|_{*},
        &\mathrm{subject\ to}\ \Wb = \Wb^{\T}
        .
    \end{aligned}
\end{equation}
\Cref{eq:ADMM1} can be rewritten as
\begin{equation}
    \label{eq:ADMM2}
    \begin{aligned}
        &\underset{\mathbf{W,V}}{\mathrm{minimize}}~
        \|\Yb - \Xb\Wb\Xb^{\T}\|_{\rF}^{2}
        +
        \lambda\|\Vb\|_{*},
        &\mathrm{subject\ to}\ \Vb = \Wb
        . 
    \end{aligned}
\end{equation}
Consequently, the scaled augmented Lagrangian for \Cref{eq:ADMM2} is given by
\begin{equation*}
    \label{eq:srl1b}
    \cL(\Wb, \Vb, \mathbf{\Theta})
    = \|\Yb - \Xb\Wb\Xb^{\T}\|_{\rF}^{2}
    + \lambda\|\Vb\|_{*}
    + \frac{\rho_{1}}{2} \|\Vb - \Wb
    + \mathbf{\Theta}\|_{\rF}^{2}
    ,
\end{equation*}
where $\mathbf{\Theta}$ is the dual variable and $\Wb$, $\Vb$ are the primal variables. 
Here, $\Wb$ and $\Vb$ can be iteratively updated.
Solving the general problem in \Cref{eq:ADMM2} is summarized in \Cref{alg}, and the corresponding derivation is provided in the supplement.

\begin{algorithm}[ht]
\caption{ADMM algorithm for solving~\Cref{eq:ADMM2}}
\begin{enumerate}
  \item Initialize the parameters:
  \begin{enumerate}[(a)]
    \item Primal variables, $\hat{\Wb}^{0}$ and $\hat{\Vb}^{0}$, each as the zero matrix.
    \item Dual variable, $\hat{\mathbf{\Theta}}^{0}$, as the zero matrix.
    \item Tuning parameter $\lambda > 0$, Lagrangian parameter $\rho_{1} > 0$, and tolerance $\epsilon>0$.
  \end{enumerate}
  \item Iterate until the stopping criterion $\|\hat{\Wb}^{t} - \hat{\Wb}^{t-1}\|_{\rF}^{2} / K^{2} \le \epsilon$ is met, where $\hat{\Wb}^{t}$ is the value of $\hat{\Wb}$ obtained at the $t$-th iteration:
  \begin{enumerate}[(a)]
    \item Update $\hat{\Wb}^{t}$ and $\hat{\Vb}^{t}$:
    \begin{enumerate}[(i)]
      \item Here, $\hat{\Wb}^{t}$ is the matrix form of $\hat{\bw}^{t}$, namely $\hat{\bw}^{t} = \mathrm{vec}(\hat{\Wb}^{t})$, where $\hat{\bw}^{t} \leftarrow (2\Cb^{\T}\Cb + \rho_{1}\Ib)^{-1}[2\Cb^{\T} \yb + \rho_{1}(\hat{\bm{\theta}}^{t-1} + \hat{\bv}^{t-1})]$, with $\Cb = \Xb \otimes \Xb$, $\by = \mathrm{vec}(\Yb)$, $\hat{\bm{\theta}}^{t-1} = \mathrm{vec}(\hat{\mathbf{\Theta}}^{t-1})$, and $\hat{\bv}^{t-1} = \mathrm{vec}(\hat{\Vb}^{t-1})$.
      \item Assign $\hat{\Vb}^{t} \leftarrow \sum_{j=1}^{p}\max(\sigma_{j} -\lambda/\rho_{1},0)\bm{\alpha}_{j}\bm{\beta}_{j}^{\T}$, where $\sum_{j=1}^{p}\sigma_{j}\bm{\alpha}_{j}\bm{\beta}_{j}^{\T}$ is the singular value decomposition of $\hat{\Wb}^{t} - \hat{\mathbf{\Theta}}^{t-1}$.
    \end{enumerate}
    \item Update $\hat{\mathbf{\Theta}}^{t}$: 
  \end{enumerate}
  \begin{enumerate}[(i)]
    \item Assign $\hat{\mathbf{\Theta}}^{t} \leftarrow \hat{\mathbf{\Theta}}^{t-1} + \hat{\Vb}^{t} - \hat{\Wb}^{t}$.
  \end{enumerate}
\end{enumerate}
\label{alg}
\end{algorithm}

\begin{remark}[Estimating matrix entries]
    Using \Cref{alg} to estimate $\Bb^{*}_{\rho}$ does not guarantee that $\hat{B}_{\rho,kl} \in [0,1]$. 
    Nevertheless, in our numerical studies, $\hat{B}_{\rho,kl} \in [0,1]$ holds for reasonable values of $\lambda$ selected by cross-validation procedures as described below.
    If in practice $\hat{B}_{\rho,kl} < 0$ or $\hat{B}_{\rho,kl} > 1$, then one can simply set $\hat{B}_{\rho,kl} = 0$ or $\hat{B}_{\rho,kl} = 1$.
\end{remark}

\subsection{Tuning parameter selection}
\label{sec:algorithm_tuning}

The convergence of \Cref{alg} can be sped up with the help of a warm start.
To do so, begin with the smallest $\lambda$ value among candidate initial values, and then use the output from the current trial as the initial value for the next choice of $\lambda$.
One may consider a range of values starting from a small positive number, as \Cref{thm1} indicates a reasonable $\lambda$ value should be away from zero.
An upper bound for $\lambda$ is to consider the smallest value which ensures $\hat{\Vb}^{t}$ in \Cref{alg} is the zero matrix.

When multiple adjacency matrices are available (corresponding to different layers), we propose to choose the tuning parameter $\lambda$ via the cross-validation procedure in \Cref{cv2}.
If instead only a single observation of $\Ab$ is available, then we consider an alternative approach detailed in the supplement.

\begin{algorithm}[ht]%htbp
    \caption{$M$-fold cross-validation for multiple networks}
    Input: Adjacency matrices $\Ab^{(\ell)}$ for $1 \le \ell \le L$, estimated membership matrix $\hat{\Zb}$, tuning parameter value(s) $\lambda$, number of folds $M$.
    \begin{enumerate}
      \item Sample splitting:
      Randomly split $\Ab^{(\ell)}$, $1 \le \ell \le L$, into $M$ equal-sized subsets. Let $\cI^{(m)}$ denote the matrix indices in the $m$-th fold.
      \item For each $1 \le m \le M$ and each value $\lambda$: 
      \begin{enumerate}[(a)]
        \item Set $\overline{\Ab}^{(m)} \coloneqq \frac{1}{|\cI^{(m)}|}\sum_{\ell \in \cI^{(m)}}\Ab^{(\ell)}$ and $\overline{\Ab}^{(-m)} \coloneqq \frac{1}{|\cI^{(-m)}|}\sum_{\ell \in \cI^{(-m)}}\Ab^{(\ell)}$.
        \item Apply \Cref{alg} to $\overline{\Ab}^{(-m)}$ and $\hat{\Zb}$ to obtain $\Bb_{\rho}^{\lambda}$.
        \item Compute the validation loss $\operatorname{Loss}(\lambda)^{(m)} \coloneqq \|\overline{\Ab}^{(m)} - \hat{\Zb}\Bb_{\rho}^{\lambda}\hat{\Zb}^{\T}\|_{\rF}^{2}$. 
      \end{enumerate}
      \item Define $\operatorname{Loss}(\lambda) \coloneqq \sum_{m=1}^{M}\operatorname{Loss}(\lambda)^{(m)}$, and return $\hat{\lambda} = \underset{\lambda}{\arg\min}\ \operatorname{Loss}(\lambda)$.
    \end{enumerate}
    \label{cv2}
\end{algorithm}

\section{Theoretical guarantees}
\label{sec:theory}

This section investigates the theoretical properties of our estimation method under the \HomoModel{} setting by leveraging the techniques in \cite{klopp2014noisy,hamdi2022lowrank}.
In particular, we establish expectation and high probability error bounds for the difference between $\hat{\Bb}$ and $\Bb^{*}$, where $\hat{\Bb}$ is obtained by solving the optimization problem
\begin{equation}
    \label{eq:e3}
    \hat{\Bb}
    =
    \underset{\mathbf{B}:~\Bb=\Bb^{\T}}{\mathrm{minimize}}~
    \frac{1}{n} \|\overline{\Ab} -\rho\hat{\Zb}\Bb\hat{\Zb}^{\T}\|_{\rF}^{2} 
    +
    \lambda \|\rho\Bb\|_{*}
    .
\end{equation}
The first term in \Cref{eq:e3} differs from \Cref{eq:e2} by the scaling $n^{-1}$, thus leading to different selections of $\lambda$ in the two formulations, but does not affect the resulting error bounds.

In preparation for \Cref{thm1}, we make the following assumption.

\begin{assumption}[Commensurate ground-truth and estimated community sizes]
    \label{assumption1}
    The true community sizes $\{n_{k}\}_{k=1}^{K}$ and the estimated community sizes $\{\hat{n}_{k}\}_{k=1}^{K}$ are both commensurate, namely, there exist constants $c_{1}, c_{2} \ge 1$ such that $n_{k} \in [n/(c_{1}K), c_{1}n/K]$ and $\hat{n}_{k} \in [n/(c_{2}K),c_{2}n/K]$ for all $1 \le k \le K$. Here, $K$ can be fixed or grow with $n$.
\end{assumption}

The assumption of commensurate ground-truth community sizes is common in the literature \citep{fan2022alma,lei2022bias}, and assuming commensurate estimated community sizes is reasonable due to existing clustering guarantees, as discussed in \Cref{sec:discussion}.
Nevertheless, \Cref{assumption1} can indeed be relaxed, if desired, but at the expense of more tedious derivations and complicated resulting bounds.

In further preparation for the main result, we write the set of correctly clustered nodes as $\Omega \coloneqq \{i: \hat{\Zb}_{i\bullet} = \Zb_{i\bullet}\}$, and hence, let $|\Omega^{c}|$ denote the number of misclustered nodes.
Write $\bm{\Delta}_{3} \coloneqq \bm{\Delta}_{1} - \bm{\Delta}_{2} = \overline{\Ab} - \rho\hat{\Zb}\Bb^{*}\hat{\Zb}^{\T}$ for $\bm{\Delta}_{1}$ and $\bm{\Delta}_{2}$ in \Cref{model}.
Finally, we define the benchmark tuning parameter value
\begin{align}
    \Lambda_{\operatorname{val}}
    &\coloneqq
    16\sqrt{2}c_{1}\frac{1}{K}\sqrt{\frac{\rho}{L}}
    \left(\sqrt{K} + \sqrt{\log n}\right)\nonumber\\ 
    &\qquad+ 
        \left(\sqrt{\frac{c_{1}}{K}} + \sqrt{2}\sqrt{\frac{\left|\Omega^{c}\right|}{n}}\right)^{2}
        \left(2\sqrt{2}\rho\sqrt{\frac{c_{1}n}{K}} \sqrt{|\Omega^{c}|}\|\Bb^{*}\|_{\mathrm{op}} + \rho|\Omega^{c}|\|\Bb^{*}\|_{\mathrm{max}}\right)\nonumber\\ 
    &\qquad+ 
        \left(2\frac{|\Omega^{c}|}{n} + 2\sqrt{2}\sqrt{\frac{c_{1}}{K}}\sqrt{\frac{|\Omega^{c}|}{n}}\right) 
        12\sqrt{2}\sqrt{\frac{n\rho \log n}{L}}
        \label{lambda_val}
        ,
\end{align}
which appears in \Cref{thm1}, the general form of our main theoretical result.

\begin{theorem}[Estimation of the connectivity probability matrix]
    \label{thm1}
    Consider a sample of observable networks per \Cref{def1}.
    Suppose \Cref{assumption1} holds and $\operatorname{rk}(\Bb^{*}) = d$.
    Let $\hat{\Bb}$ be the solution to \Cref{eq:e3} for tuning parameter $\lambda > 0$ satisfying
    \begin{equation}
        \label{lambda}
        3\Lambda_{\operatorname{val}} 
        \le
        \lambda
        \le
        3 C \Lambda_{\operatorname{val}},
    \end{equation}
    where $C \ge 1$ is any user-specified constant.
    If $n\sqrt{L\rho} \geq C^{\prime} K(\sqrt{K} + \sqrt{\log n})$ and $L n \rho \geq C^{\prime\prime} \log n$ for some sufficiently large positive constants $C^{\prime}$ and $C^{\prime\prime}$, then
    \begin{equation}
        \label{general_bound}
        \begin{aligned}
            &\|\hat{\Bb} - \Bb^{*}\|_{*}\\
            &\qquad \le
            \sqrt{72d} \|\hat{\Bb} - \Bb^{*}\|_{\rF} \\
            &\qquad \lesssim
            K^{2}d\Bigg[16\sqrt{2}\frac{c_{1}}{K}\sqrt{\frac{1}{Ln^{2}\rho}}\left(\sqrt{K} + \sqrt{\log n}\right)\\
            &\qquad\qquad + 
            \left(\sqrt{\frac{c_{1}}{K}}+\sqrt{2}\sqrt{\frac{|\Omega^{c}|}{n}}\right)^{2}\left(2\sqrt{2}\sqrt{\frac{c_{1}}{K}}\|\Bb^{*}\|_{\mathrm{op}}\sqrt{\frac{|\Omega^{c}|}{n}}+\|\Bb^{*}\|_{\max}\frac{|\Omega^{c}|}{n}\right)\\
            &\qquad\qquad + 
            \left(2\sqrt{2}\sqrt{\frac{c_{1}}{K}}\sqrt{\frac{|\Omega^{c}|}{n}}+2\frac{|\Omega^{c}|}{n}\right)12\sqrt{2}\sqrt{\frac{\log n}{Ln\rho}}
            \Bigg]
        \end{aligned}
    \end{equation}
    with probability at least $1 - O(n^{-8})$.
\end{theorem}

In \Cref{thm1}, for the special case $L=1$, sharper control of $n^{-1}\|\hat{\Zb}^{\T} \bm{\Delta}_{3} \hat{\Zb}\|_{\mathrm{op}}$ in the proof leads to an improved bound.
Namely, by replacing the term $12\sqrt{2}\sqrt{n\rho \log n / L}$ in \Cref{lambda_val} with $C^{\prime\prime\prime}\sqrt{n\rho}$, where $C^{\prime\prime\prime} > 0$ denotes a sufficiently large constant, then an improvement of \Cref{general_bound} is obtained by replacing the term $12\sqrt{2}\sqrt{\log n / (Ln\rho)}$ with $C^{\prime\prime\prime} / \sqrt{n \rho}$.

\begin{remark}[Deterministic versus stochastic estimated communities]
    For simplicity, \Cref{assumption1} and hence \Cref{thm1} considers the estimated community sizes to be deterministic.
    If a specific clustering method with quantifiable performance is employed, then the assumption can be restated as holding with high probability.
    Namely, define the event
    \begin{equation*}
        E
        \coloneqq
        \left\{
        \hat{n}_{k} \in [n/(c_{2}K), c_{2}n/K] 
        \text{ for some }
        c_{2} \ge 1
        \text{ for all }
        1 \le k \le K
        \right\}
        .
    \end{equation*}
    Then, the upper bound in \Cref{thm1} holds with probability at least $1 - \PP(E^{c}) - O(n^{-8})$.
\end{remark}

The proof of \Cref{thm1} relies on the following key inequality implied by \Cref{assumption1}.

\begin{lemma}[Frobenius norm lower bound]
    \label{balance_ineq}
    Under \Cref{assumption1}, for any $\Bb \in \RR^{K \times K}$,
    \begin{equation*}
        \frac{1}{n} \|\hat{\Zb} \Bb \hat{\Zb}^{\T}\|_{\rF}^{2}
        \ge
        \frac{n}{(c_{2}K)^2}\|\Bb\|_{\rF}^{2}
        .
    \end{equation*}
\end{lemma}

In \Cref{thm1}, the choice of tuning parameter $\lambda$ stems from the following lemma.

\begin{lemma}[Operator norm bound for tuning parameter selection]
    \label{op_bound}
    Invoke \Cref{assumption1}.
    If $n\sqrt{L\rho} \geq C^{\prime} K(\sqrt{K} + \sqrt{\log n})$ and $Ln\rho \geq C^{\prime\prime} \log n$ hold for some sufficiently large positive constants $C^{\prime}$ and $C^{\prime\prime}$, then
    \begin{equation*}
        \begin{aligned}
        \frac{1}{n}\|\hat{\Zb}^{\T} \bm{\Delta}_{3} \hat{\Zb}\|_{\mathrm{op}}
        &\le
        \Lambda_{\operatorname{val}}
        \end{aligned}
    \end{equation*}
    with probability at least $1 - O(n^{-8})$.
    If $L = 1$, then the bound can be improved by replacing the term $12\sqrt{2}\sqrt{n\rho \log n / L}$ in the definition of $\Lambda_{\operatorname{val}}$ with $C^{\prime\prime\prime} \sqrt{n\rho}$ for some constant $C^{\prime\prime\prime} > 0$, per the earlier comment.
\end{lemma}

The following corollary presents special cases of \Cref{thm1} and highlights the performance of the estimation method when node memberships are perfectly recovered.

\begin{corollary}[Special case of \Cref{thm1}]
    \label{cor1}
    Assume the hypotheses in \Cref{thm1}.
    \begin{enumerate}
        \item If $d, K = O(1)$, then
        \begin{equation}
            \|\hat{\Bb} - \Bb^{*}\|_{\rF}
            \lesssim
            \sqrt{\frac{1}{n}}
            \sqrt{\frac{\log n}{Ln\rho}}
            +
            \sqrt{\frac{|\Omega^{c}|}{n}}
            \left(
            \|\Bb^{*}\|_{\mathrm{op}}
            +
            \sqrt{\frac{|\Omega^{c}|}{n}}
            +
            \sqrt{\frac{\log n}{Ln\rho}}
            \right)
        \end{equation}
        with probability at least $1 - O(n^{-8})$.
        
    \item For general $d$ and $K$, if $|\Omega^{c}| = 0$, then
    \begin{equation}
        \label{eq:frobenius_bound_opti_corollary}
        \|\hat{\Bb} - \Bb^{*}\|_{\rF}
        \lesssim
        K \sqrt{d} \left(\sqrt{K} + \sqrt{\log n}\right) 
        \frac{1}{\sqrt{n}}
        \frac{1}{\sqrt{L n \rho}}
    \end{equation}
    with probability at least $1 - O(n^{-8})$.
    Moreover, in expectation,
    \begin{equation}
        \EE\left[\|\hat{\Bb} - \Bb^{*}\|_{\rF}\right] \lesssim K^{3/2}\sqrt{d}\frac{1}{\sqrt{n}}\frac{1}{\sqrt{Ln\rho}}
        .
    \end{equation}
    \end{enumerate}
\end{corollary}

\begin{remark}[Comparing estimators]
    For comparison, consider the average-based estimator $\hat{\Bb}^{\mathrm{Avg}}$ when $\hat{\Zb} = \Zb$, i.e., $|\Omega^{c}| = 0$.
    To understand the concentration properties of $\|\hat{\Bb}^{\mathrm{Avg}} - \Bb^{*}\|_{\rF}$, it holds by direct computation that the expected squared Frobenius difference satisfies
    \begin{equation*}
        \frac{K^{4}}{L n ^{2} \rho}
        ~\lesssim~
        \mathbb{E}\left[
        \|\hat{\Bb}^{\mathrm{Avg}}
        -
        \Bb^{*} \|_{\rF}^{2}\right]
        ~\lesssim~
        \frac{K^{4}}{L n^{2} \rho}
        .
    \end{equation*}
    Hence, by Jensen's inequality,
    \begin{equation*}
        \label{eq:expected_frobenius_bound_avg}
        \mathbb{E}\left[
        \|\hat{\Bb}^{\mathrm{Avg}}
        -
        \Bb^{*} \|_{\rF}\right]
        \lesssim
        K^{2} \frac{1}{\sqrt{n}} \frac{1}{\sqrt{L n \rho}}
        .
    \end{equation*}
    Consequently, $\hat{\Bb}$ per \Cref{cor1} is preferable to $\hat{\Bb}^{\mathrm{Avg}}$ when $K$ is large and $d \ll K$.
    This is precisely the improvement one can hope to achieve using the proof techniques in \cite{klopp2014noisy,hamdi2022lowrank}.
\end{remark}

\begin{remark}[Properties of norms]
    \label{rem:nuclear_frobenius_upperbound_improvement}
    We pause here to comment on the significance of the perturbation analysis conducted to obtain \Cref{thm1} and \Cref{cor1}.
    First, note that $\|\Bb^{*}\|_{*}$ and $\|\Bb^{*}\|_{\rF}$ denote the $\ell_{1}$ norm and $\ell_{2}$ norm of the vector of singular values of $\Bb^{*}$, respectively.
    By elementary properties of norms,
    \begin{equation*}
        \|\Bb^{*}\|_{\rF}
        \le
        \|\Bb^{*}\|_{*}
        \le
        \sqrt{K}\|\Bb^{*}\|_{\rF},
        \qquad
        \Bb^{*} \in \mathbb{R}^{K \times K}
        .
    \end{equation*}
    Since $\operatorname{rk}(\Bb^{*}) = d \le K$, the stated results yield improved upper bounds on $\|\cdot\|_{*}$ in terms of $\|\cdot\|_{\rF}$ by replacing $\sqrt{K}$ with $C\sqrt{\operatorname{rk}(\Bb^{*})}$ for some dimension-free universal constant $C > 0$.
\end{remark}

\begin{remark}[Avoiding suboptimal dependence on $K$]
    The na\"ive bound $\|\Bb^{*}\|_{\mathrm{op}} \le K$ can be unnecessarily coarse.
    Note that if $\Bb^{*} = \bv \bv^{\T}$, then $\|\Bb^{*}\|_{\operatorname{op}} = \|\bv\|_{2}^{2}$.
    Thus, if for example $\bv=(p,p^{2},\dots,p^{K})^{\T}$ for some $p \in (0,1)$, then $\|\bv\|_{2}^{2} = \sum_{k=1}^{K}p^{2k} = \frac{p^{2}(1-p^{2K})}{1-p^{2}} \le \frac{p^{2}}{1-p^{2}}$ irrespective of $K$.
    Similarly, $\|\Bb^{*}\|_{\mathrm{op}}$ does not depend on $K$ when $\Bb$ is a diagonal matrix.
\end{remark}

\begin{remark}[Network sparsity]
    Network sparsity is traditionally introduced and included to (at least partially) quantify problem difficulty (by reflecting the amount of available network information) and justify the applicability of analysis procedures to real-world non-dense graphs.
    The typical sparsity condition for single-graph statistical analysis and clustering is of the form $n \rho \geq C \log n$, whereas the hypotheses in \Cref{thm1} and \Cref{cor1} show that the availability of multiple graphs (layers), $L$, compensates for having even sparser individual graphs; see also \cite{lei2022bias}.
    As expected, observe that the estimation error bounds degrade as the sparsity term $\rho$ (in the denominator) gets smaller. 
    Further, having more sparsity can in turn lead to a larger misclustering rate, $n^{-1} |\Omega^{c}|$, but this aspect and its subsequent influence on estimation error bounds is implicit since the stated results are not confined to a specific choice of clustering algorithm or to specific clustering performance. 
    \Cref{sec:discussion} provides details about the misclustering rate of several existing clustering approaches with explicit dependence on network sparsity.
\end{remark}

\section{Numerical studies}
\label{sec:numerical}

This section illustrates the performance of the proposed method under \HomoModel{} and \HeteroModel{}.
We use \Cref{cv2} to select $\lambda$ from among a range of candidate values.
We run each simulation setup $100$ times and compare the performance of our method with the performance of average-based estimation.
To simplify the notation, $\hat{\Bb}^{\mathrm{Our}}$ and $\hat{\Bb}^{\mathrm{Avg}}$ denote the estimators obtained from \Cref{alg} and the averaging method, respectively.

We first specify the entrywise averaging estimation procedure.
Per our notation, for each $k \in \nset{K}$, let $\hat{\Zb}_{\bullet k} \in \RR^{n}$ denote the $k$-th column of $\hat{\Zb}$, and let $\hat{n}_{k}$ denote the estimated number of nodes in the $k$-th community.
Averaging yields the entrywise estimates
\begin{equation*}
    \label{Avg}
    \hat{\Bb}_{k\ell}^{\mathrm{Avg}}
    =
    \frac{1}{\rho \hat{n}_{k} \hat{n}_{\ell}} \hat{\Zb}_{\bullet k}^{\T} \overline{\Ab} \hat{\Zb}_{\bullet\ell},
    \quad
    \text{ for each }
    (k, \ell) \in \nset{K} \times \nset{K}
    .
\end{equation*}
We do not consider using the low-rank truncation (spectral embedding) of $\overline{\Ab}$, since in the very sparse regime it is difficult to determine how many leading eigenvectors to choose, while in the not-too-sparse regime the spectral embedding does not significantly improve the estimation accuracy.
We give a simple simulation to illustrate this observation.

\begin{figure}[ht]%htbp
    \centering
    \subfloat{\includegraphics[width=0.45\linewidth]{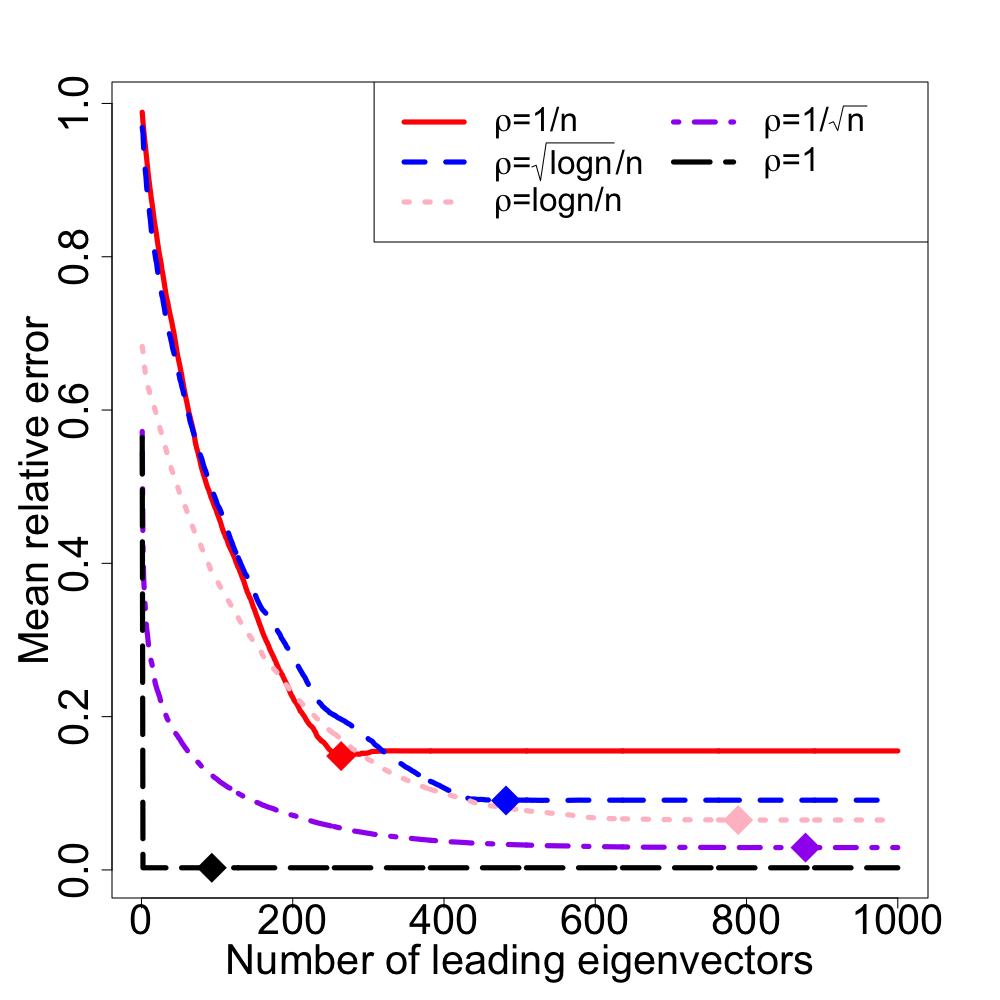}}
    \subfloat{\includegraphics[width=0.45\linewidth]{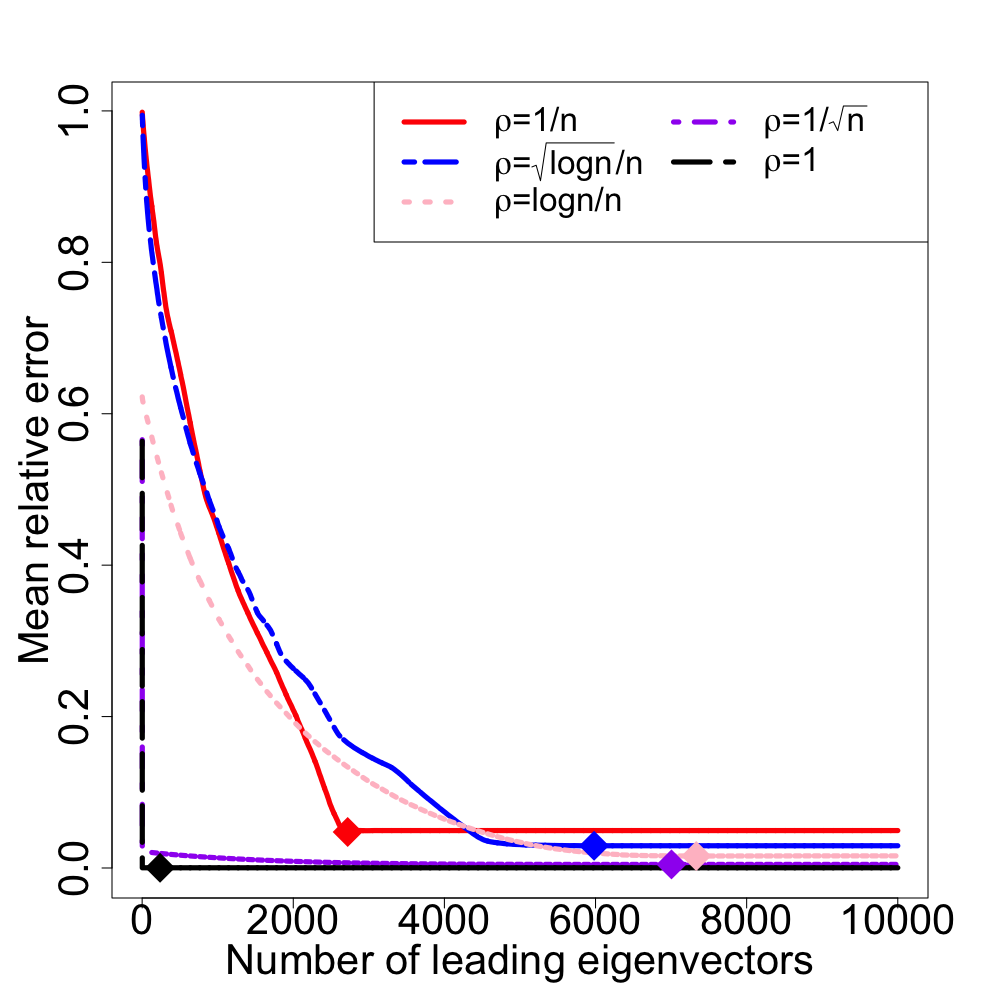}}
    \caption{Simulation example with relative errors averaged over 100 replications, for $n=10^{3}$ (left) and $n=10^{4}$ (right). The diamonds indicate the number of eigenvectors giving the smallest relative error per trajectory. See \Cref{sec:numerical}.}
    \label{Atrunc}
\end{figure}

Consider a single simulated SBM adjacency matrix $\Ab$ with $n \in \{10^{3},10^{4}\}$, $K=2$, $d=2$, $\rho \in \{1/n,\sqrt{\log n}/n, \log n/n, 1/\sqrt{n}, 1\}$, and obtain $\Ab^{\hat{d}}$, the low-rank spectral truncation of $\Ab$ with $\hat{d}$ leading eigenvectors and eigenvalues.
We average the elements of $\Ab^{\hat{d}}$ to estimate $\Bb^{*}$ and denote the estimator as $\hat{\Bb}^{\hat{d}}$.
The per-iterate relative error $\mathrm{RE}_{\hat{d}} \coloneqq \|\hat{\Bb}^{\hat{d}} - \Bb^{*}\|_{\rF} / \|\Bb^{*}\|_{\rF}$ measures the accuracy of $\hat{\Bb}^{\hat{d}}$.
The average of $\mathrm{RE}_{\hat{d}}$ from $100$ Monte Carlo replicates is summarized in \Cref{Atrunc} which shows that the preferable number of eigenvectors to choose is much larger than $K$ and $d$ except when $\rho_{n} = 1$.

\subsection{MonoSBM with perfectly estimated node memberships}
\label{MonoTrueZ}
Here, $K$ and $\Zb$ are considered given, as a proxy for the fact that many previously cited methods can consistently estimate them under appropriate signal-to-noise conditions.
We explore how the estimation accuracy of $\EE(\Ab)$ can affect the estimation accuracy of $\Bb^{*}$ obtained from \Cref{alg} when $d = K$, $d < K$, and $d \ll K$.
The results for $d \ll K$ are summarized below, and the other settings are provided in the supplement.
   
\textbf{Dimensionality $d \ll K$}:
We show that our method is applicable when $d$ is unknown and can significantly improve the estimation error when $\hat{\Zb} = \Zb$ with $d \ll K$.
We consider SBMs with $K = 10$, $d = 1$, $L = 100$, $\rho = 0.1$, $n \in \{10^{3}, 10^{4}\}$, $n_{i} = 0.15n, i \in \{1,2\}$, $n_{j} = 0.1n$ for $j \in \{3,4,5\}$, $n_{k} = 0.08n$ for $k \in \{6,\dots,10\}$, and probability matrix
\begin{equation}
  \label{B3}
  \Bb^{*}
  =
  \bu\bu^{\T},
  \quad
  \bu
  =
  (0.9, 0.9^{2}, \dots, 0.9^{10})^{\T}.
\end{equation}
We calculate the estimation error under the Frobenius norm and compute the average estimated dimensionality across 100 replications, accompanied by the standard error.
We anticipate that a good estimate of $\Bb^{*}$ will significantly differ from the estimate obtained from the averaging method, since when $d \ll K$, the noise from the additional eigenvectors of $\hat{\Bb}^{\mathrm{Avg}}$ dramatically affects the accuracy.
\Cref{t3} confirms this.
The averaging method continues to produce full-rank estimates even though the ground-truth matrix is rank degenerate.

\begin{table}[htbp]
\centering
\resizebox{\columnwidth}{!}{%
\begin{tabular}{ccccccc}
\hline
&
\begin{tabular}[c]{@{}c@{}}
    Our Method \\
    $n = 10^3$
\end{tabular}
& &
\begin{tabular}[c]{@{}c@{}}
    Avg. Method \\
    $n=10^{3}$
\end{tabular}
& &
\begin{tabular}[c]{@{}c@{}}
    Our Method \\
    $n=10^4$
\end{tabular}
&
\begin{tabular}[c]{@{}c@{}}
    Avg. Method \\
    $n=10^{4}$
\end{tabular} \\ \hline
Estimation error & $0.0081(2 \times 10^{-4})$ &  & $0.0188(2 \times 10^{-4})$ &  & $0.0008(2 \times 10^{-5})$ & $0.0019(2 \times 10^{-5})$ \\
Dimensionality estimation & 1.15(0.04) &  & 10(0) &  & 1.15(0.04) & 10(0) \\ \hline
\end{tabular}%
}
\caption{Estimation accuracy comparisons in \HomoModel{} when $\hat{\Zb} = \Zb$. See \Cref{MonoTrueZ}.}
\label{t3}
\end{table}

\subsection{MonoSBM with imperfectly estimated node memberships}
\label{sec:estimated_Z}

Here, we explore how the estimation accuracy of $\Zb$ affects the estimation accuracy and rank estimation of $\Bb^{*}$ for our method and the averaging method.
First, we estimate $\Zb$ via spectral clustering with Gaussian mixture modeling (GMM).
Specifically, let the eigendecomposition of $\overline{\Ab}$ be $\overline{\Ab} = \sum_{i=1}^{n}\lambda_{i}\bu_{i}\bu_{i}^{\T}$ with eigenvalues $|{\lambda_{1}}| \ge \dots \ge |{\lambda_{n}}|$ and eigenvectors $\bu_{1},\dots,\bu_{n}$.
Let $\Ub = (\ub_{1} \mid \dots \mid \ub_{K}) \in \RR^{n \times K}$ be the matrix of leading eigenvectors and $\mathbf{\Lambda} = \mathrm{diag}(\lambda_{1},\dots,\lambda_{K}) \in \RR^{K \times K}$ be the diagonal matrix with the leading eigenvalues.
We use GMM-based clustering applied to $\Ub \mathbf{\Lambda}$ to estimate the membership vector $\bg = (g_{1},\dots,g_{n})^{\T}$.
In order to compare $\hat{\Bb}$ and $\Bb^{*}$, we permute $\hat{\bg}$ based on the Hamming distance so that
\begin{equation*}
    \tilde{\bg}
    \coloneqq
    (\hat{\pi}(\hat{g}_{1}),\dots,\hat{\pi}(\hat{g}_{n}))^{\T}
    \in
    \underset{\pi}{\arg\min}~\frac{1}{n}\sum_{i=1}^{n}\II(g_{i} \ne \pi(\hat{g}_{i}))
    ,
\end{equation*} 
where $\pi : \nset{K} \to \nset{K}$ is a permutation map.
We then take $\hat{\Zb}$ to be the membership matrix of $\tilde{\bg}$.
We consider $d \ll K$ here, while the supplement presents the results for $d = K$, $d < K$, as well as comparison of our method with the spectral embedding method and the low-rank approximation method from \cite{tang2022asymptotically}.

\textbf{Dimensionality $d \ll K$}:
Here, we consider the same model setting as in \Cref{MonoTrueZ} with $\Bb^{*}$ in \Cref{B3}, to facilitate comparison with the previous scenario $\hat{\Zb} = \Zb$.
\Cref{t6} only shows results for $n=10^{4}$ since increasing $n$ improves the performance of all methods.

\begin{table}[htbp]
  \centering
  \resizebox{!}{1.25cm}{%
  \begin{tabular}{cccc}
  \hline
  & Our Method &  & Avg. Method \\ \hline
  Estimation error & 0.5596(0.0447) &  & 0.5599(0.0447)\\
  Dimensionality estimation & 1.93(0.09) &  & 10(0)\\ \hline
  \end{tabular}%
  }
  \caption{Estimation accuracy comparisons in \HomoModel{} when $\hat{\Zb} \ne \Zb$ with the smallest (average) ARI $=0.8142(0.9406)$. See \Cref{sec:estimated_Z}.} 
  \label{t6}
\end{table}

\begin{figure}[ht]%htbp
    \centering
    \subfloat{\includegraphics[width=0.4\linewidth]{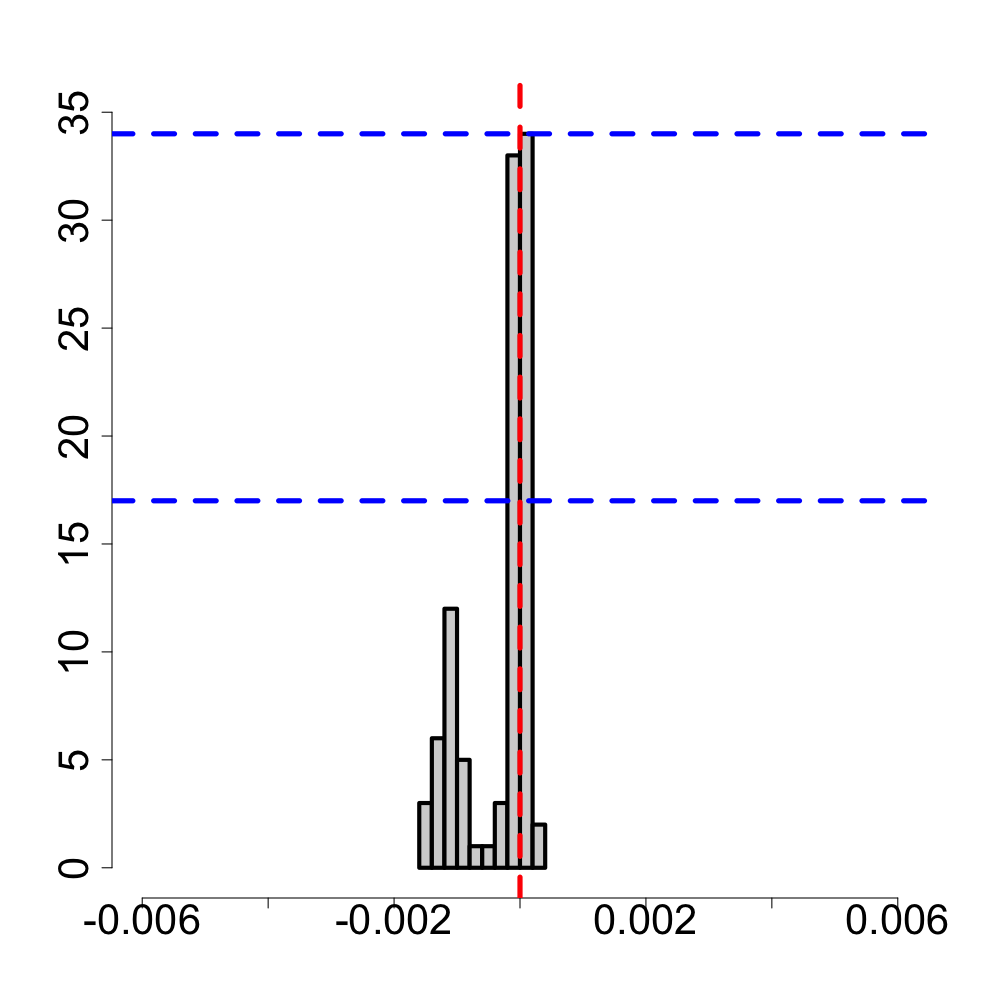}}
    \subfloat{\includegraphics[width=0.4\linewidth]{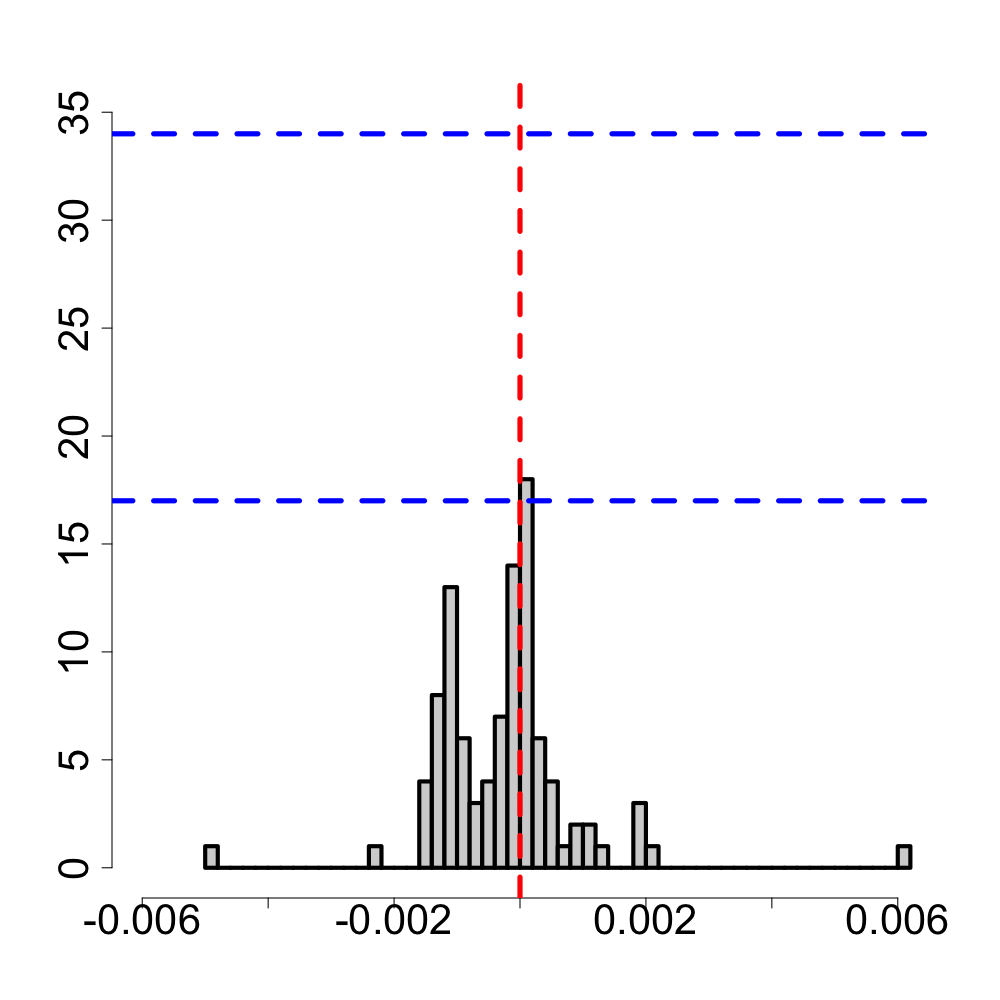}}
    \caption{The left panel shows $\mathbf{\Delta}^{(i)} = \|\hat{\Bb}^{\mathrm{Our}(i)} - \Bb^{*}\|_{\rF} - \|\hat{\Bb}^{\mathrm{Avg}(i)} - \Bb^{*}\|_{\rF}$, and the right panel shows $\mathbf{\Delta}^{(i)} = \|\hat{\Bb}^{\mathrm{AvgLR}(i)} - \Bb^{*}\|_{\rF} - \|\hat{\Bb}^{\mathrm{Avg}(i)} - \Bb^{*}\|_{\rF}$, where $i \in \nset{100}$. The dashed vertical line shows $\mathbf{\Delta}^{(i)} = 0$. The upper (bottom) horizontal line gives the maximum (half maximum) frequency across the two histograms. See \Cref{sec:estimated_Z}.}
    \label{K10estZ}
\end{figure}

Per \Cref{t6}, when $d \ll K$ the estimation accuracy can be similar if $\hat{\Zb} \neq \Zb$, but the average estimated dimensionality of our method is still close to the true rank $d = 1$.
\Cref{K10estZ} shows that the histogram plotting differences between our method and the averaging method is left-skewed, while the histogram plotting differences between the low-rank approximation method with true rank $d=1$ and the averaging method exhibits greater variation.
Even though the true rank is known, the low-rank approximation method with true $d$ may produce a worse estimate of $\Bb^{*}$ compared to the averaging method, but our method is routinely better than the averaging method. 
Thus, choosing a reasonable rank value for the low-rank approximation method appears to be important.
Additional simulation results in the supplement indicate that for the low-rank approximation method, the resulting average estimated rank (using cross-validation) is larger than $\hat{d}$ from our method and has larger standard error.

To summarize, under \HomoModel{}, we find that regardless of whether $\Zb$ can be perfectly recovered, our method more accurately estimates the dimensionality compared to traditional approaches.
If $\hat{\Zb} = \Zb$ and $d \ll K$, then our method can noticeably improve upon the estimation accuracy of the averaging method. 
Therefore, to get a more accurate estimate of the connectivity probability matrix $\Bb^{*}$, estimating $\Zb$ plays a crucial role.

\subsection{MonoSBM with re-estimated node memberships}
\label{sec:re-estimate}

In \HomoModel{}, one of the most popular methods to estimate $\Zb$ is by spectral clustering using the $K$-means algorithm, where the true embedding dimension is $d$.
If $d \ll K$ but $\hat{d}$ is chosen to be much larger than $d$, then $K$-means clustering might not yield stable results, namely running $K$-means multiple times for different initializations can give noticeably different estimated node partitions.
In addition, $K$-means with a much larger embedding dimension might have a large misclustering error rate, leading to large estimation error for $\Bb^{*}$.
Based on the above simulation results, our method shows competitive rank estimation even when $\hat{\Zb} \ne \Zb$, which can help us to choose a better embedding dimension for applying the $K$-means algorithm.
We illustrate the idea using the following simulation setting. 

\textbf{Dimensionality $d \ll K$:}
Consider SBMs with $n = 10^{3}$, $K = 10$, $d = 2$, $\rho = 0.15$, $L = 50$, having the same community sizes as in \Cref{MonoTrueZ}, and with
\begin{equation*}
  \label{B4}
  \Bb^{*}
  =
  \left(\bu_{1}\bu_{1}^{\T}
  +
  \bu_{2}\bu_{2}^{\T}\right)/2,
  \quad
  \bu_{i}
  \overset{\text{i.i.d.}}{\sim}
  \mathrm{Uniform}(0.2,0.9)^{K}
  \in
  \RR^{K},
  \quad
  i
  \in
  \{1, 2\}
  .
\end{equation*}

We perform the following steps to re-estimate $\Zb$ and $\Bb^{*}$:
\begin{enumerate}
    \item Run the $K$-means algorithm with embedding dimension $K$ to obtain $\hat{\Zb}_{K}$.
    \item Given $\hat{\Zb}_{K}$, apply our method to obtain $\hat{d}$.
    \item If $\hat{d} \ne K$, then run $K$-means with embedding dimension $\hat{d}$ to obtain $\hat{\Zb}_{\hat{d}}$.
    \item Given $\hat{\Zb}_{\hat{d}}$ and $\hat{d}$, use the low-rank approximation method to compute $\hat{\Bb}$.
\end{enumerate}

In Step~$4$ above, we use the low-rank approximation method since it is computationally fast and we have a good estimate $\hat{d}$.
We could also use our method to obtain $\hat{\Bb}$, but the estimation accuracy will presumably be similar based on previous simulation results.
Here, \Cref{reestimateZ} shows that $\hat{d}$ is still close to $d$.
After re-estimating $\Zb$ and $\Bb^{*}$, the misclustering error rate and estimation error both improve.

\begin{table}[htbp]
    \centering
    \resizebox{0.8\textwidth}{!}{%
    \begin{tabular}{ccc}
        \hline
        & Without Re-estimation & With Re-estimation \\
        \hline
        Misclustering error rate
        & 0.1393(0.0051)
        & 0.0225(0.0035) \\
        Estimation error
        & 0.6936(0.0292)
        & 0.0648(0.0192) \\
        \hline
    \end{tabular}%
    }
    \caption{Performance with and without node membership re-estimation. The average (standard error) of $\hat{d}$ is 2.82(0.11). See \Cref{sec:re-estimate}.}
    \label{reestimateZ}
\end{table}

\subsection{MultiSBM with bias-adjusted spectral clustering for estimating node memberships}
\label{sec:multiSBM_Zhat_BASC}

Here, we consider the \HeteroModel{} setting with $\Bb^{*(i)}$, $i \in \nset{4}$, to illustrate how $\Bb^{*(i)}$ can be estimated from the multilayer networks.
For each $\Bb^{*(i)}$, we generate 50 layers, which results in $L = 200$.
We use bias-adjusted spectral clustering \citep{lei2022bias} to estimate $\Zb$.
As in the previous section, we consider $K = 3$, $d \in \{1,2,3\}$, $n = 10^{3}$, $(n_{1}, n_{2}, n_{3}) = (0.25n, 0.25n, 0.5n)$, $\rho \in \{\sqrt{\log n} / n,\log n / n\}$, and we construct $\Bb^{*(i)}$ so that the adjusted Rand index (ARI) is at least $0.8$ in each layer when $\rho = \sqrt{\log n} / n$.
In more detail, for $\Bb^{*(i)}$, $i \in \{1,2,3\}$, we modify $\Bb^{*}$ in \cite{lei2022bias}, which gives
{\footnotesize
  \begin{equation*}
  \begin{aligned}
  \Ub &= \begin{bmatrix}
      1/2 & 1/2 & -\sqrt{2}/2 \\
      1/2 & 1/2 & \sqrt{2}/2 \\
      \sqrt{2}/2 & -\sqrt{2}/2 & 0 
      \end{bmatrix},\quad &\Bb^{*(1)} = \Ub\begin{bmatrix}
      1.2 & 0 & 0 \\
      0 & 0.6 & 0 \\
      0 & 0 & -0.7
      \end{bmatrix} \Ub^{\T} \\ 
  \Bb^{*(2)} &= \Ub\begin{bmatrix}
      1.2 & 0 & 0 \\
      0 & 0.6 & 0 \\
      0 & 0 & 0.7
      \end{bmatrix} \Ub^{\T}, \quad &\Bb^{*(3)} = \Ub\begin{bmatrix}
      1.7 & 0 & 0 \\
      0 & 0 & 0 \\
      0 & 0 & -0.6
      \end{bmatrix} \Ub^{\T},
  \end{aligned}
  \end{equation*}
}
while $\Bb^{*(4)} = \bu\bu^{\T}$, with $\bu = (0.8,0.8^{2},0.8^{3})^{\T},$ so that $d_{1} = d_{2} = 3$, $d_{3} = 2$, and $d_{4} = 1$. 
After estimating $\Zb$, we use the averaging method to estimate $\Bb^{*}$ in each layer and then apply GMM-based clustering on the upper triangular elements of $\hat{\Bb}$ to cluster each layer into $\tilde{L}$ groups.
For simplicity, we assume $\tilde{L}=4$ is given.
If $\tilde{L}$ were unknown, then we could construct a matrix $\Gb \in \RR^{L \times n(n+1)/2}$, where $\Gb_{\ell\bullet}$ has the upper triangular elements of $\hat{\Ab}^{(\ell)}$, and use the scree plot of the singular values of $\Gb$ to estimate $\tilde{L}$.
Finally, for each group $i$, we estimate $\Bb^{*(i)}$ as for the \HomoModel{}.  

The estimation errors in \Cref{t7} indicate this two-stage method can be a useful approach to estimate connectivity probability matrices in \HeteroModel{}.
\Cref{t8} shows that our method outperforms alternative approaches for estimating dimensionality and gives competitive connectivity matrix estimates.
If $\tilde{L}$ is very large, then our method is arguably more convenient, since the rank is automatically estimated by tuning $\lambda$, and we do not need to estimate the rank of $\Bb^{*(\tilde{\ell})}$ for $\ell \in \nset{\tilde{L}}$ in advance. 

\begin{table}[htbp]
    \centering
    \resizebox{\columnwidth}{!}{%
    \begin{tabular}{ccccccccccc}
    \hline
    & & Our Method &  & Avg. Method &  & AvgLR Method  &  & Spectral Embedding with $K$ &  & Spectral Embedding with $d$ \\ \hline
    \multirow{4}{*}{$\rho=\frac{\sqrt{\log n}}{n}$} & $\hat{\Bb}^{(1)}$ & $0.1374(6.5 \times 10^{-3})$ &  & $0.1374(6.5 \times 10^{-3})$ &  & $0.1374(6.5\times 10^{-3})$ &  & $0.1443(6.4 \times 10^{-3})$ &  & $0.1443(6.4 \times 10^{-3})$ \\
    & $\hat{\Bb}^{(2)}$ & $0.1413(6.9 \times 10^{-3})$ &  & $0.1413(6.9 \times 10^{-3})$ &  & $0.1413(6.9 \times 10^{-3})$ &  & $0.1483(6.7 \times 10^{-3})$ &  & $0.1483(6.7 \times 10^{-3})$ \\
    & $\hat{\Bb}^{(3)}$ & $0.1199(5.6 \times 10^{-3})$ &  & $0.1201(5.6 \times 10^{-3})$ &  & $0.1200(5.6 \times 10^{-3})$ &  & $0.1449(5.3 \times 10^{-3})$ &  & $0.1450(5.3 \times 10^{-3})$ \\
    & $\hat{\Bb}^{(4)}$ & $0.0244(1.1 \times 10^{-3})$ &  & $0.0274(1.0 \times 10^{-3})$ &  & $0.0248(1.1 \times 10^{-3})$ &  & $0.0225(1.0 \times 10^{-3})$ &  & $0.0224(1.0 \times 10^{-3})$ \\
    \multirow{4}{*}{$\rho=\frac{\log n}{n}$} & $\hat{\Bb}^{(1)}$ & $0.0104(4.0 \times 10^{-4})$ &  & $0.0104(4.0 \times 10^{-4})$ &  & $0.0104(4.0 \times 10^{-4})$ &  & $0.0105(4.0 \times 10^{-4})$ &  & $0.0105(4.0 \times 10^{-4})$ \\
    & $\hat{\Bb}^{(2)}$ & $0.0131(5.0 \times 10^{-4})$ &  & $0.0131(5.0 \times 10^{-4})$ &  & $0.0131(5.0 \times 10^{-4})$ &  & $0.0132(5.0 \times 10^{-4})$ &  & $0.0132(5.0 \times 10^{-4})$ \\
    & $\hat{\Bb}^{(3)}$ & $0.0113(4.0 \times 10^{-4})$ &  & $0.0120(4.0 \times 10^{-4})$ &  & $0.0114(4.0 \times 10^{-4})$ &  & $0.0120(4.0 \times 10^{-4})$ &  & $0.0120(4.0 \times 10^{-4})$ \\
    & $\hat{\Bb}^{(4)}$ & $0.0110(5.0 \times 10^{-4})$ &  & $0.0131(5.0 \times 10^{-4})$ &  & $0.0113(6.0 \times 10^{-4})$ &  & $0.0092(4.0 \times 10^{-4})$ &  & $0.0091(4.0 \times 10^{-4})$ \\ \hline     
    \end{tabular}%
    }
    \caption{Average (standard error) estimation errors under the Frobenius norm. For $\rho = \sqrt{\log n} / n$, the smallest (average) ARI with $n = 10^{3}$ is 0.8198(0.9406). See \Cref{sec:multiSBM_Zhat_BASC}.}
    \label{t7}
\end{table}

\begin{table}[t]%htbp
    \centering
    \resizebox{!}{3cm}{%
    \begin{tabular}{ccccccc}
    \hline
    & & Our Method & & Avg. Method & & AvgLR Method \\ \hline
    \multirow{4}{*}{$\rho=\frac{\sqrt{\log n}}{n}$} & $\hat{\Bb}^{(1)}$ & 3(0) &  & 3(0) &  & 3(0) \\
    & $\hat{\Bb}^{(2)}$ & 3(0) &  & 3(0) &  & 3(0) \\
    & $\hat{\Bb}^{(3)}$ & 2.12(0.03) &  & 3(0) &  & 2.25(0.04) \\
    & $\hat{\Bb}^{(4)}$ & 1.18(0.05) &  & 3(0) &  & 1.31(0.07) \\
    \multirow{4}{*}{$\rho=\frac{\log n}{n}$} & $\hat{\Bb}^{(1)}$ & 3(0) &  & 3(0) &  & 3(0) \\
    & $\hat{\Bb}^{(2)}$ & 3(0) &  & 3(0) &  & 3(0) \\
    & $\hat{\Bb}^{(3)}$ & 2.12(0.03) &  & 3(0) &  & 2.21(0.04) \\
    & $\hat{\Bb}^{(4)}$ & 1.31(0.06) &  & 3(0) &  & 1.43(0.07) \\ \hline
    \end{tabular}%
    }
    \caption{Average (standard error) estimated dimensionality. See \Cref{sec:multiSBM_Zhat_BASC}.}
    \label{t8}
\end{table}

\section{Application to primate brain gene co-expression}
\label{sec:real_data}

We apply our method to a microarray dataset from \cite{bakken2016comprehensive} obtained from the medial prefrontal cortex of developing rhesus monkeys originally containing $9173$ genes with $10$ layers.
The dataset consists of gene co-expression representing 10 different developmental time points, spanning from 40 days (E40) in the embryo to 48 months (48M) after birth.
Each time point corresponds to post-mortem tissue samples collected from multiple rhesus monkeys.
Previous studies demonstrated meaningful findings when applying stochastic blockmodel methodology to this dataset, with a focus on community detection rather than investigating the connectivity within and between communities \citep{lei2020consistent,lei2022bias}.
Here, our goal is instead to determine whether or not connectivity is plausibly full rank, with the aim of enabling more comprehensive analysis.

We use the original dataset together with the preprocessing \textsf{R} code provided in \cite{lei2022bias}.
\cite{lei2022bias} preprocessed the original dataset into $10$ adjacency matrices by hard-thresholding the Pearson correlation matrix for each layer and then removed all the genes corresponding to nodes whose total degree across the $L = 10$ layers is less than $90$. 
After preprocessing, each layer contains $n = 7836$ genes. 
\cite{lei2020consistent,lei2022bias} found that the connectivity patterns change dramatically across the $10$ layers, so we posit that this gene co-expression network belongs to \HeteroModel{}.

We first use bias-adjusted spectral clustering and the choice $K = 8$, as found in \cite{lei2022bias}, to estimate the membership matrix $\Zb$.
The resulting estimated membership matrix here, $\hat{\Zb}$, is therefore the same as in \cite{lei2022bias}; see \cite[Table 1]{lei2022bias} for details about the scientific interpretation regarding each community of genes.

\begin{figure}[t]%htbp
\begin{minipage}{0.18\textwidth}
\centering
\includegraphics[width=\linewidth]{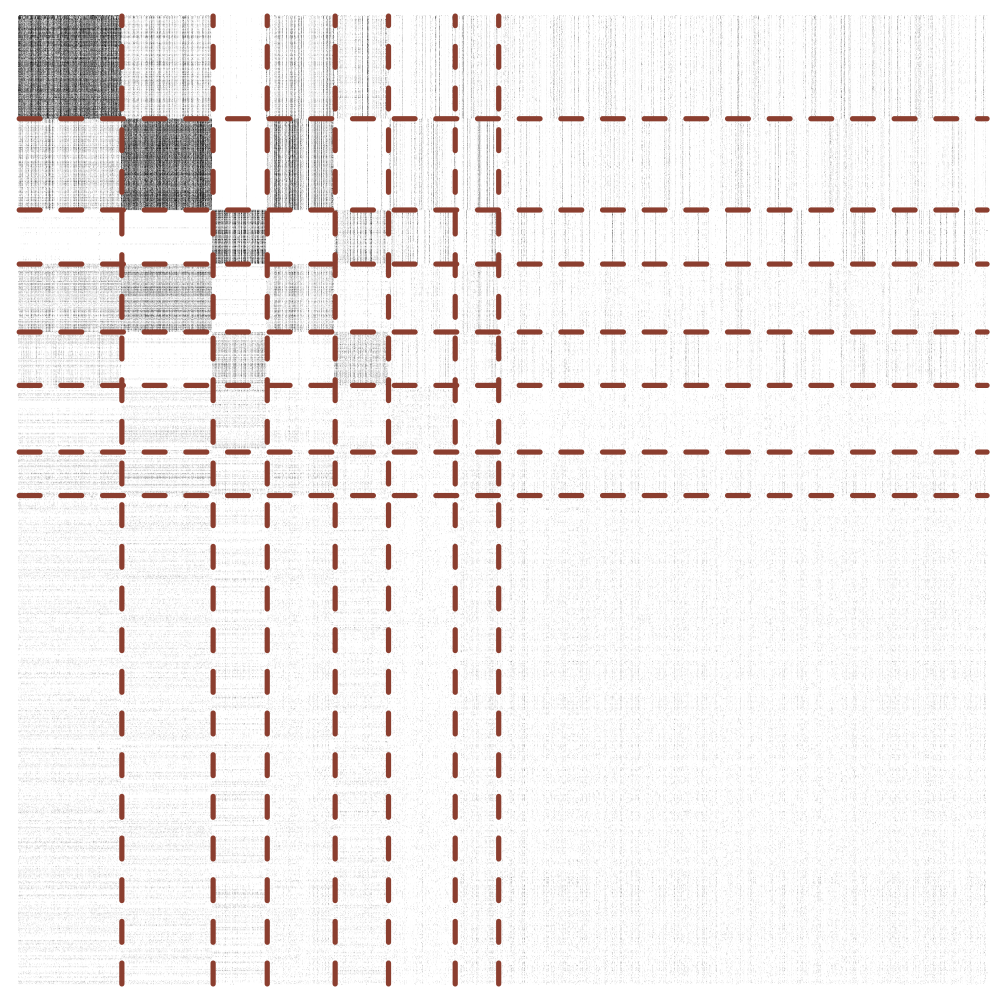}
\end{minipage}\hfill
\begin{minipage}{0.18\textwidth}
\centering
\includegraphics[width=\linewidth]{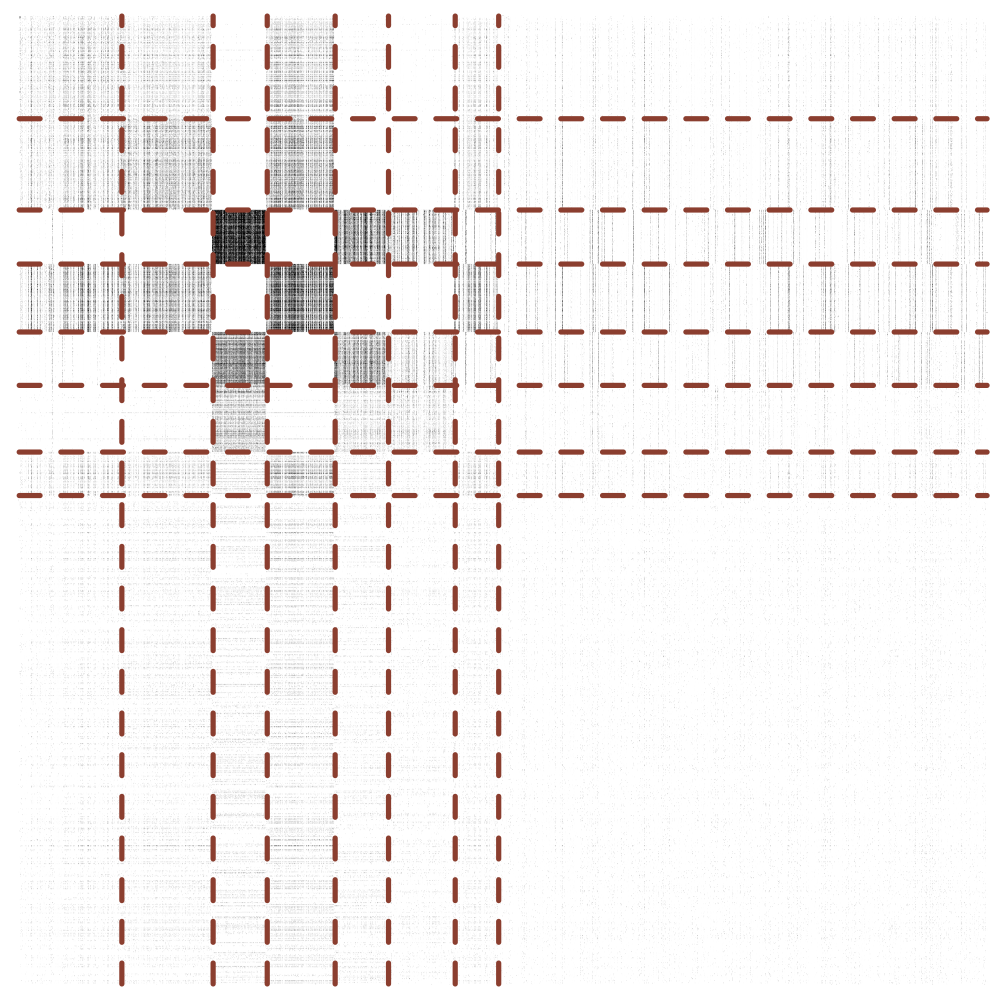}
\end{minipage}\hfill
\begin{minipage}{0.18\textwidth}
\centering
\includegraphics[width=\linewidth]{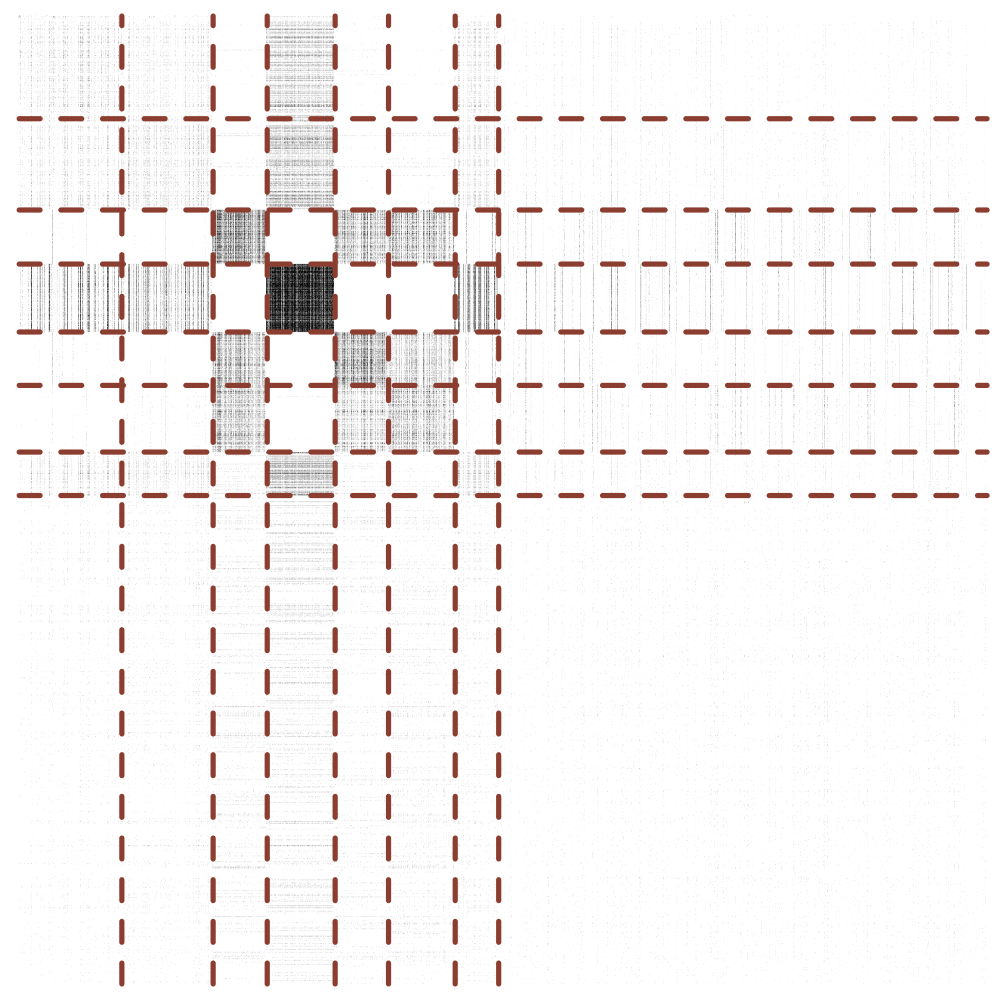}
\end{minipage}\hfill
\begin{minipage}{0.18\textwidth}
\centering
\includegraphics[width=\linewidth]{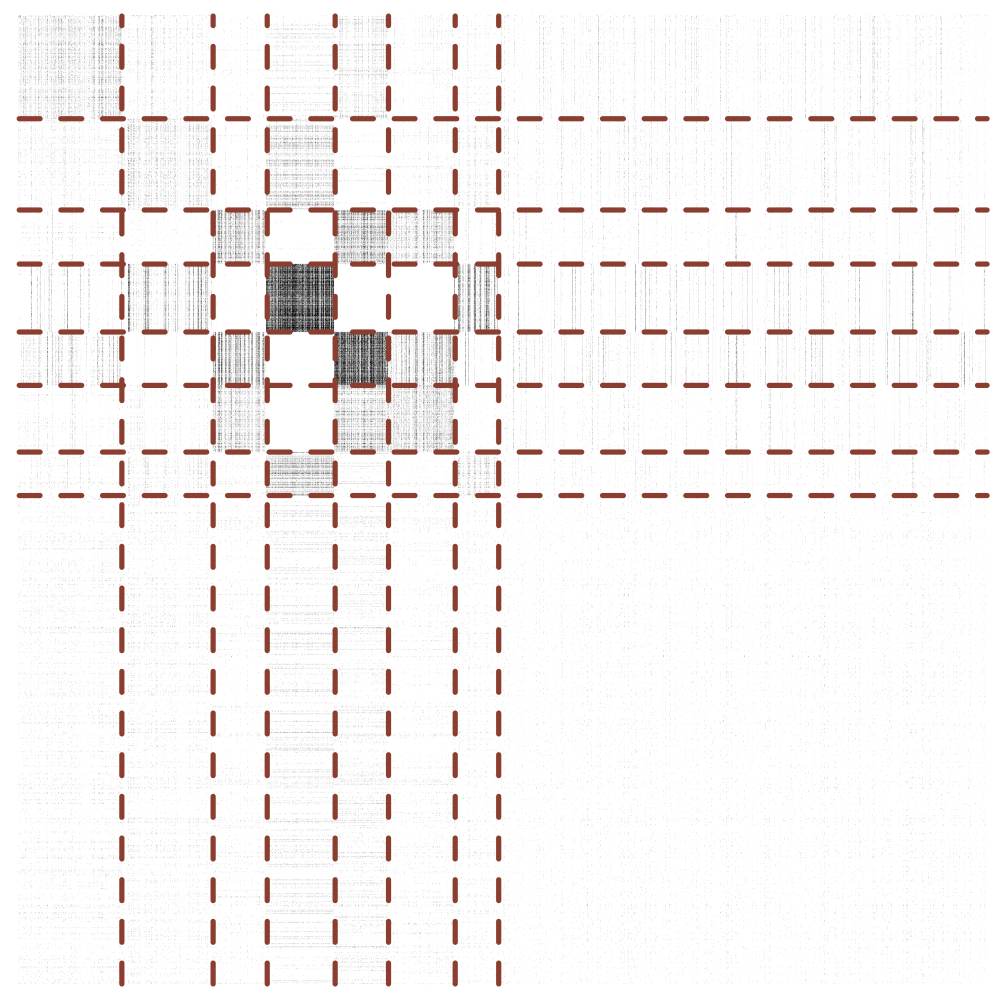}
\end{minipage}\hfill
\begin{minipage}{0.18\textwidth}
\centering
\includegraphics[width=\linewidth]{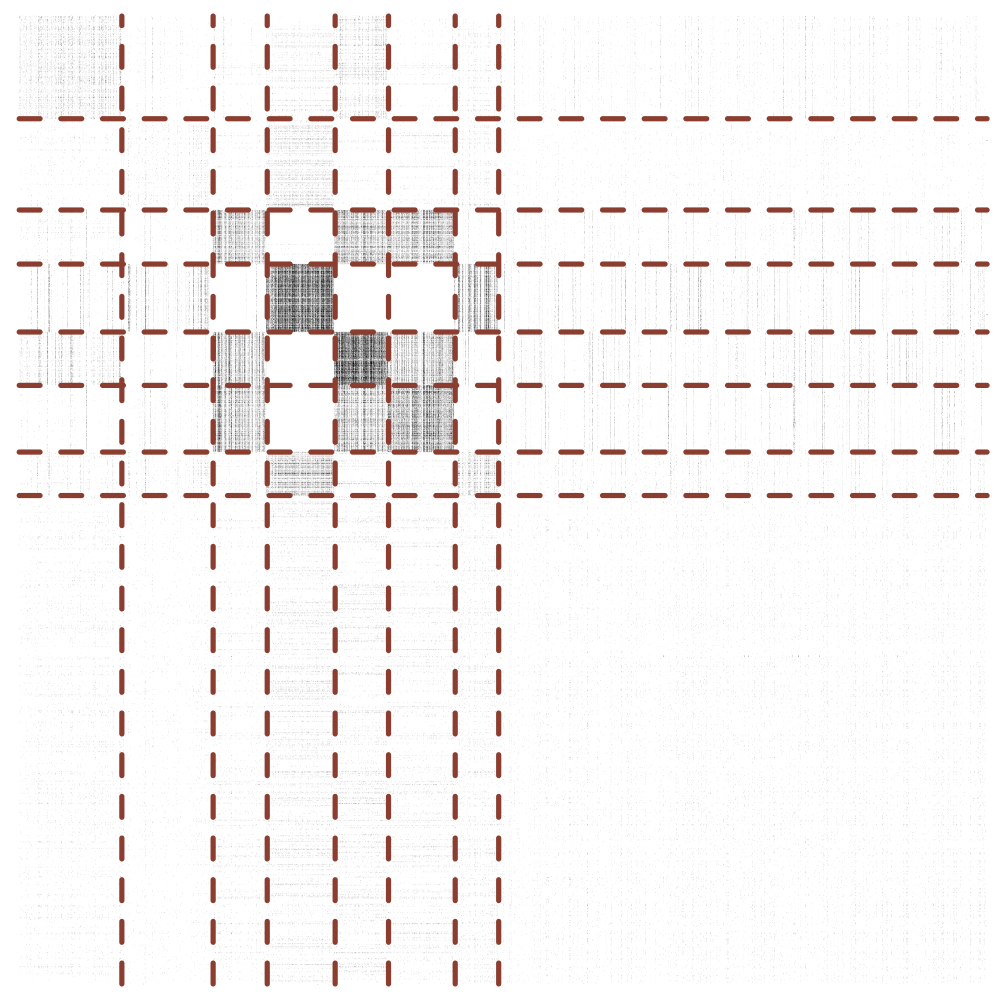}
\end{minipage}\par
\vspace*{0.5cm}
\begin{minipage}{0.18\textwidth}
\centering
\includegraphics[width=\linewidth]{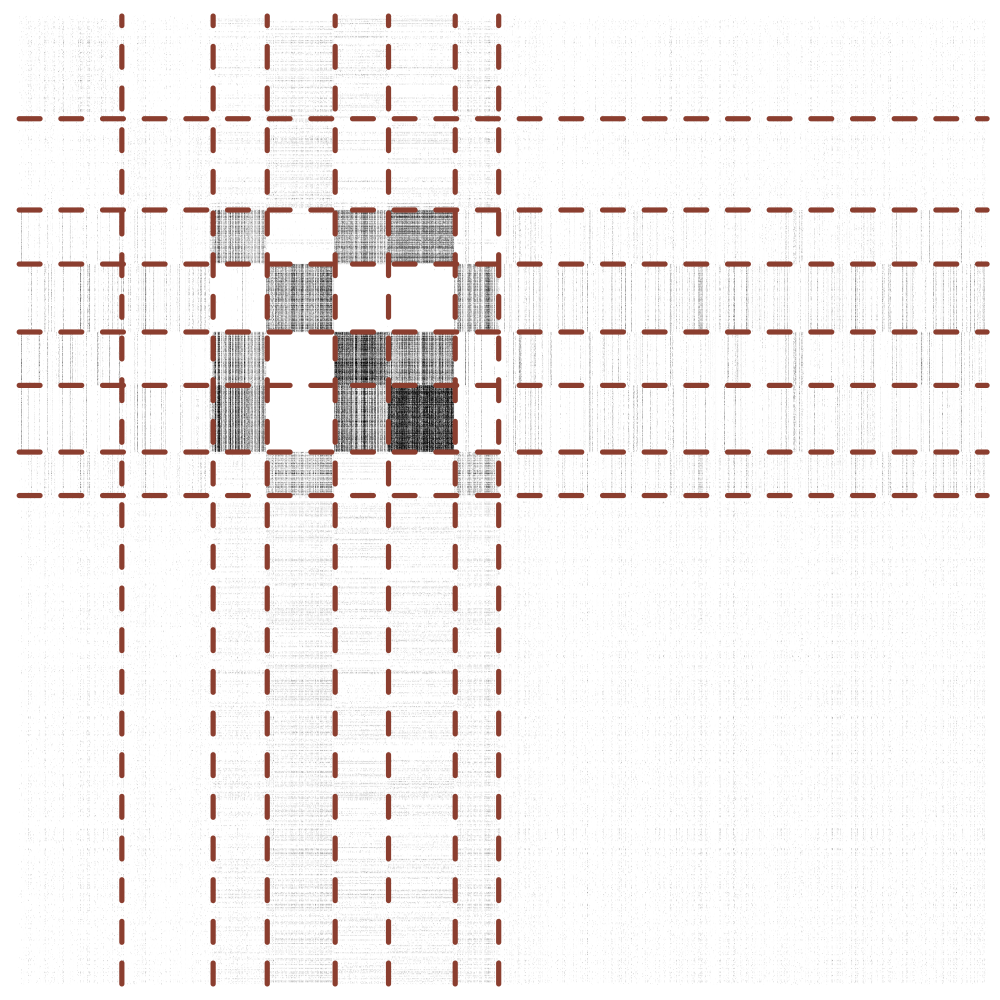}
\end{minipage}\hfill
\begin{minipage}{0.18\textwidth}
\centering
\includegraphics[width=\linewidth]{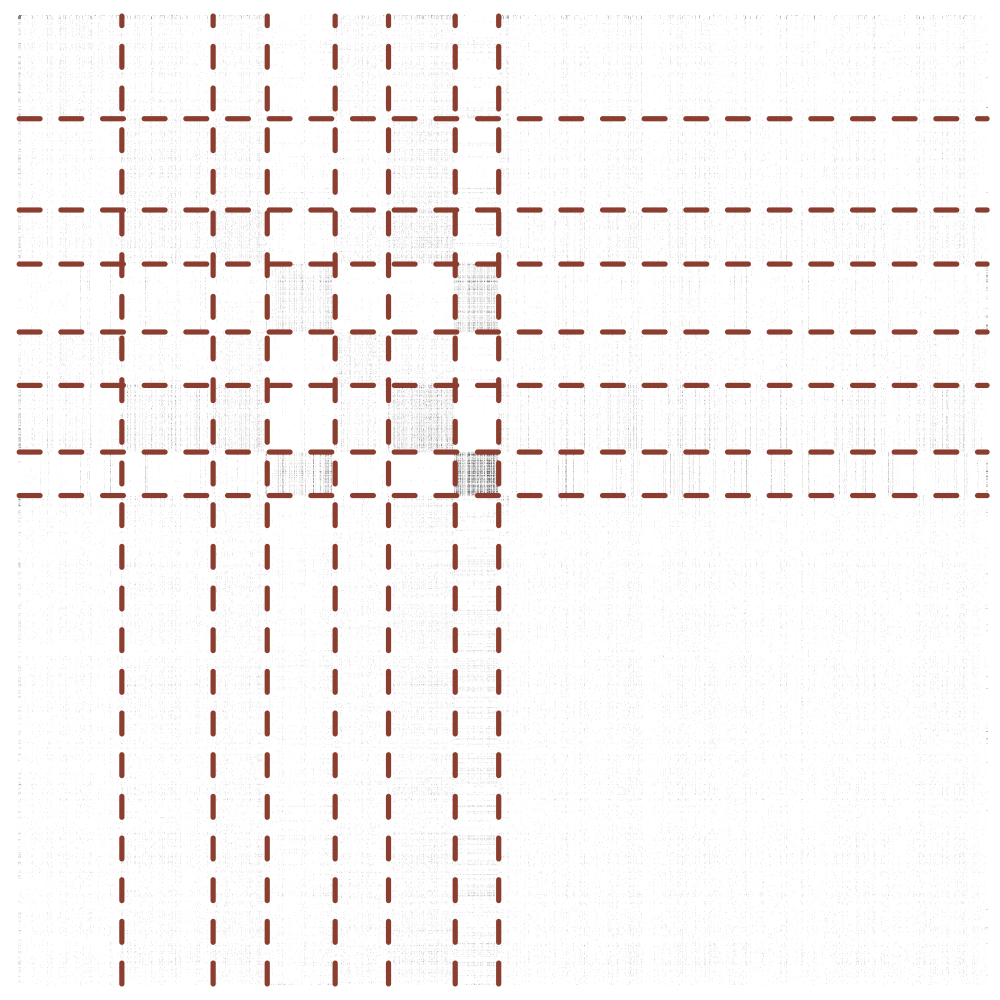}
\end{minipage}\hfill
\begin{minipage}{0.18\textwidth}
\centering
\includegraphics[width=\linewidth]{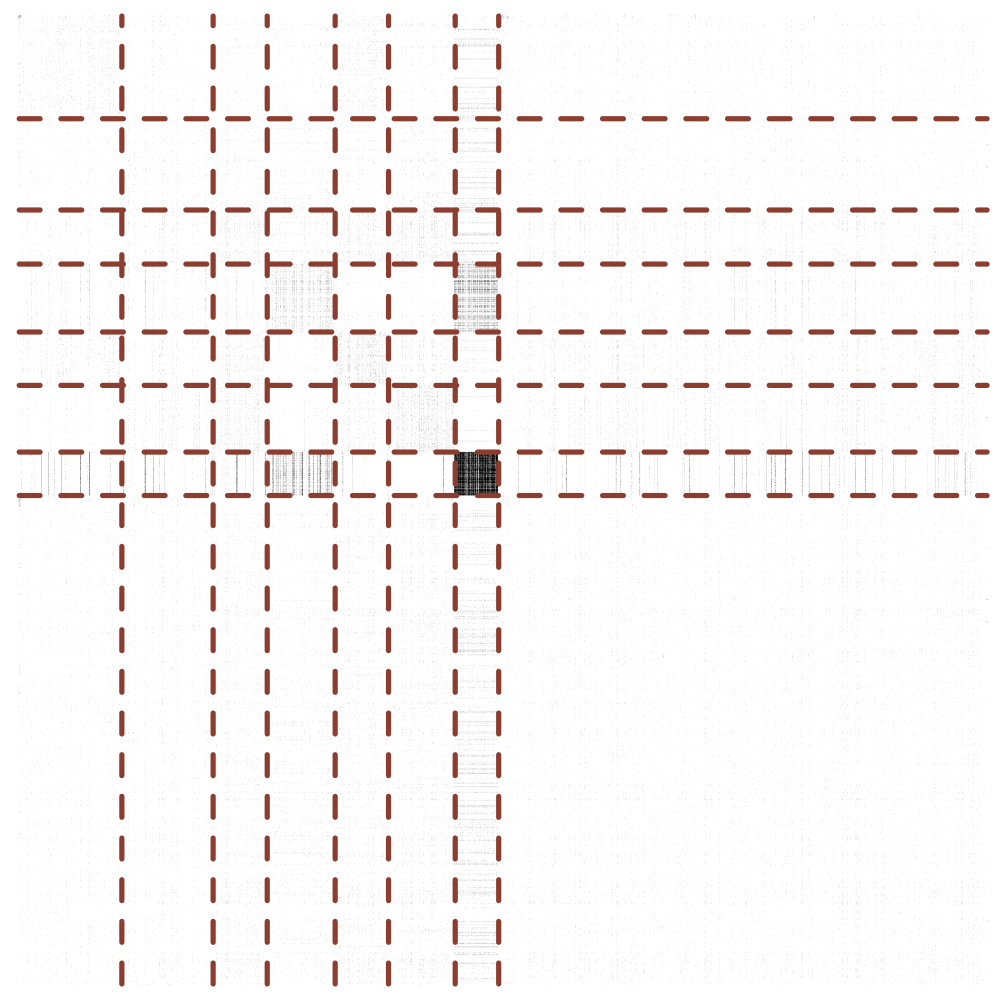}
\end{minipage}\hfill
\begin{minipage}{0.18\textwidth}
\centering
\includegraphics[width=\linewidth]{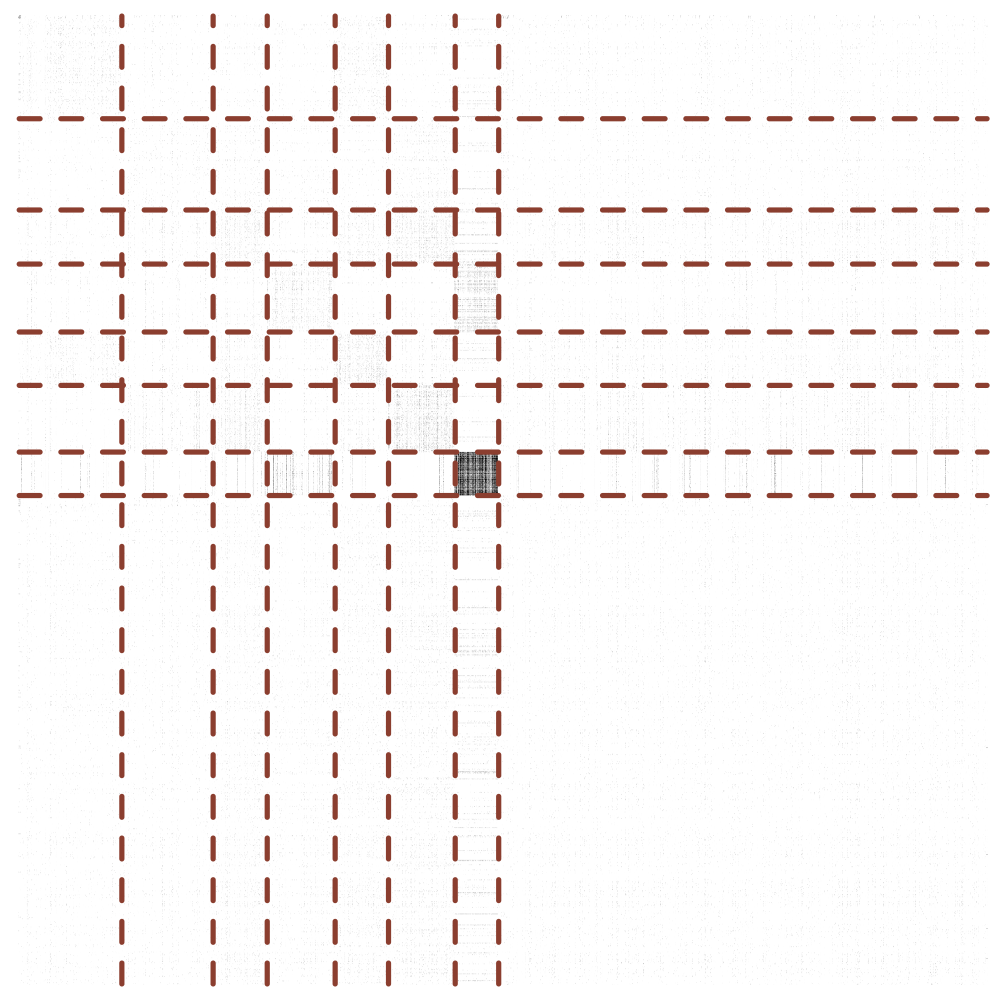}
\end{minipage}\hfill
\begin{minipage}{0.18\textwidth}
\centering
\includegraphics[width=\linewidth]{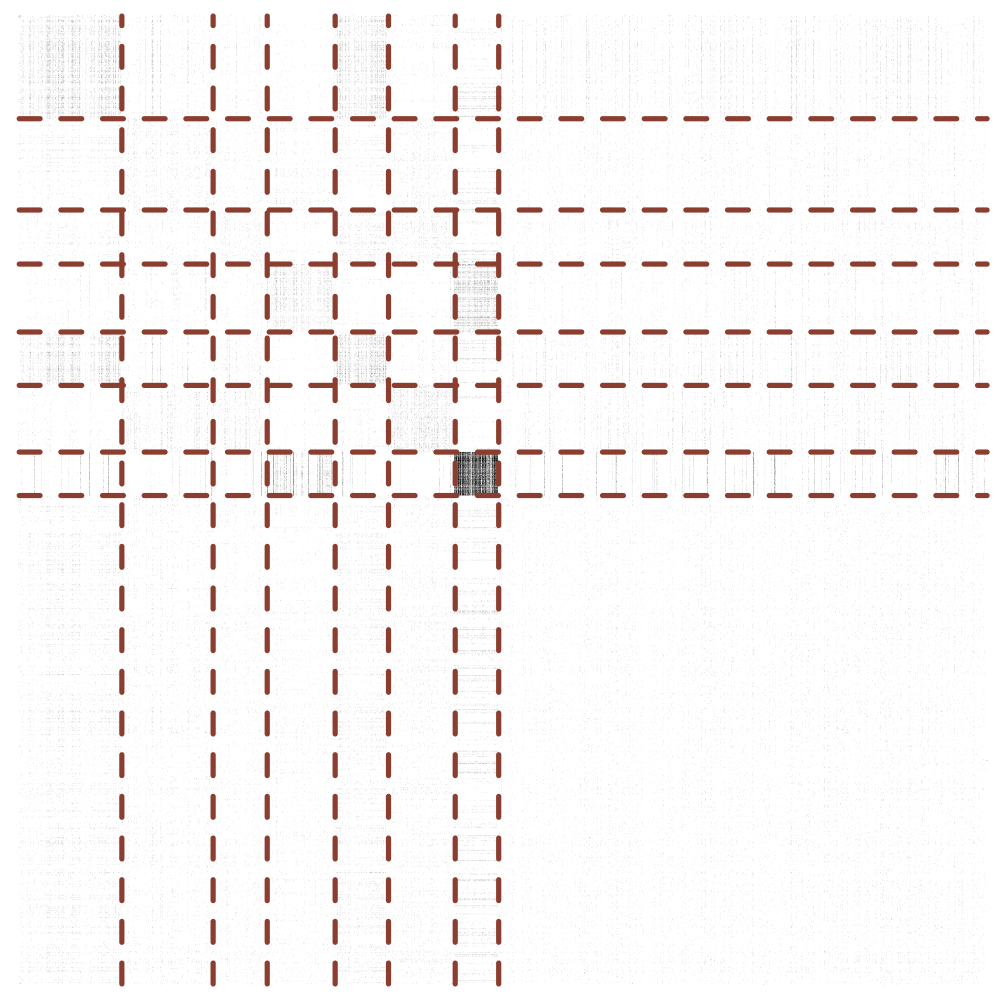}
\end{minipage}\par
\caption{Heatmaps of adjacency matrices for the $10$ layers. Grey pixels indicate edges. The dashed vertical and horizontal lines distinguish the $K = 8$ communities. In the first row, time points are from E40 to E90 (left to right). In the second row, time points are from E120 to 48M (left to right). See \Cref{sec:real_data}.}
\label{pnasA}
\end{figure}

\Cref{pnasA} shows heatmaps depicting the adjacency matrices for the $10$ layers, where the genes are ordered according to the estimated communities.
We first estimate the connectivity probability matrix in each layer and then cluster the $10$ probability matrices into $\tilde{L} = 3$ groups.
We choose $\tilde{L} = 3$ based on the scree plot (provided in the supplement) and \Cref{pnasA} which suggests that the adjacency matrices at time point E40, from time points E50 to E120, and from time points 0M to 48M have similar patterns, respectively.
In other words, we have $3$ groups, where the first group contains a single time point E40, the second group contains time points E50--E120, and the third group contains time points 0M--48M.

\begin{figure}[t]%htbp
    \centering\subfloat{\includegraphics[width=0.25\linewidth]{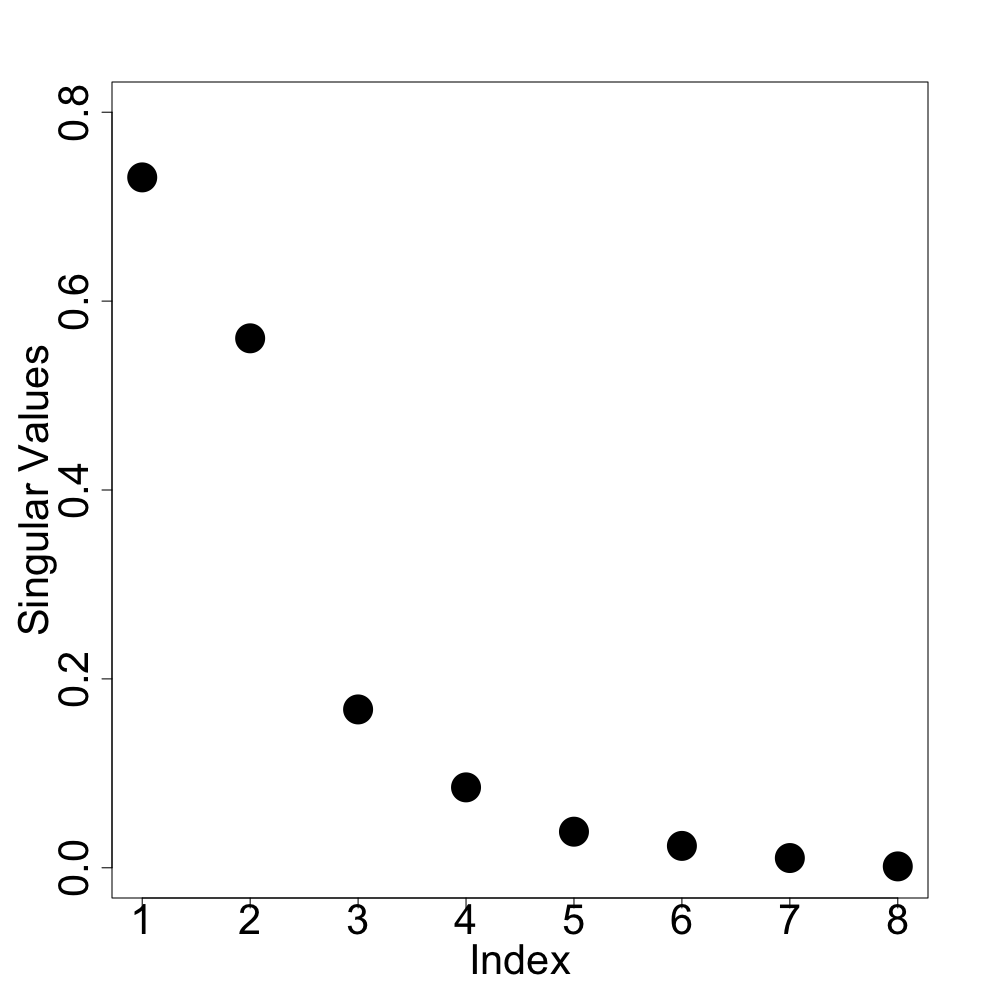}}\nolinebreak\subfloat{\includegraphics[width=0.25\linewidth]{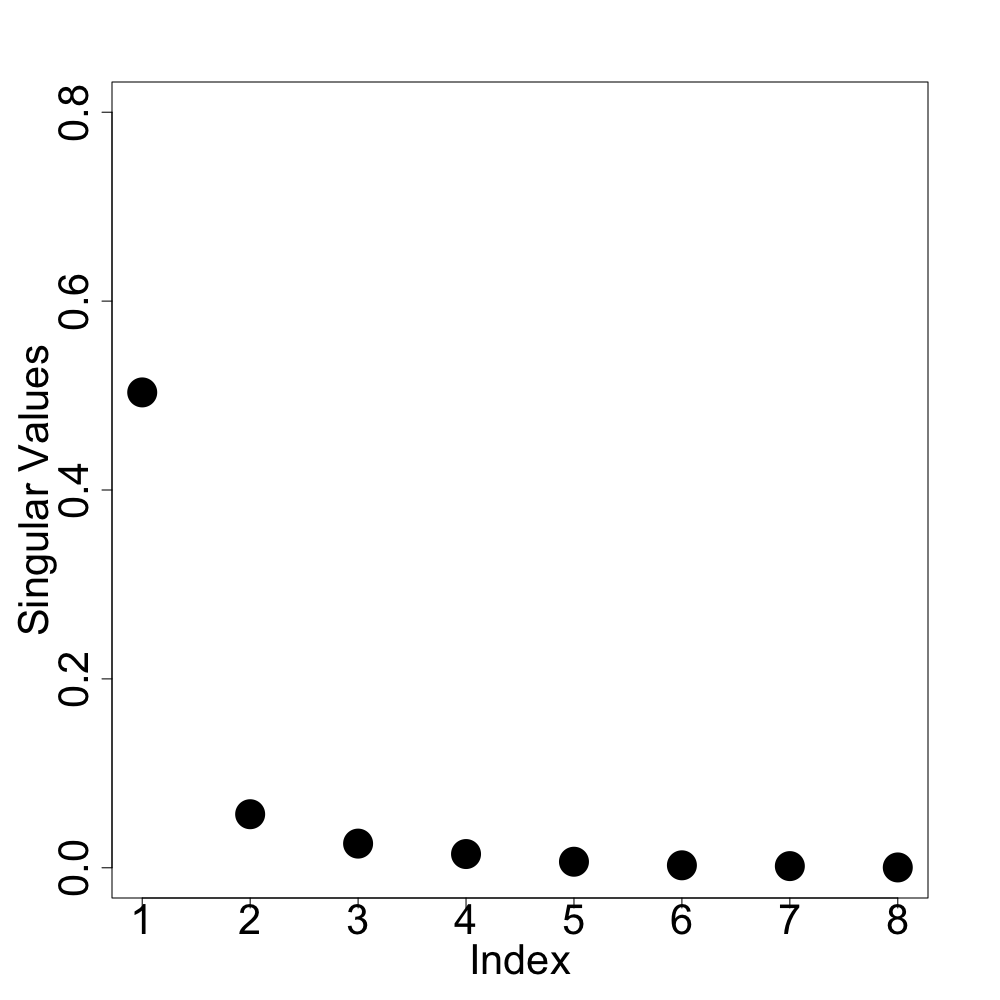}}
    \caption{The scree plots for $\Bb^{\mathrm{Avg}(2)}$ (left) and $\Bb^{\mathrm{Avg}(3)}$ (right). See \Cref{sec:real_data}.}
    \label{pnasSVD}
\end{figure}

We focus on the second and third groups, since these each have multiple adjacency matrices, and we find that the scree plots in \Cref{pnasSVD} reveal potential low-rank patterns in the estimated probability matrices.
So, we apply our method to these two groups with a wide range of $\lambda$ values and use cross-validation to select $\lambda$.
Since the number of adjacency matrices in each group is small and each adjacency matrix is very sparse, we choose to use the repeated cross-validation method instead of the $M$-fold cross-validation method.
In more detail, for the second group containing $5$ layers, we use $3$ layers as the training data and $2$ layers as the validation data, and run our method on a total of $10$ combinations to select $\lambda$.
For the third group containing $4$ layers, we use $2$ layers as the training data and the other $2$ layers as the validation data.

Our method yields $\hat{d} = 8$ for the second group and $\hat{d} = 7$ for the third group.
It might be surprising to find that even though there are large gaps in the scree plots for both groups, our method only shows that the probability matrix $\hat{\Bb}$ is low rank in the third group and the rank $\hat{d} = 7$ is not much smaller than $K = 8$.
We provide some commentary here.

\begin{figure}[htb]
    \begin{minipage}{0.4\textwidth}
    \centering
    \includegraphics[width=0.8\linewidth]{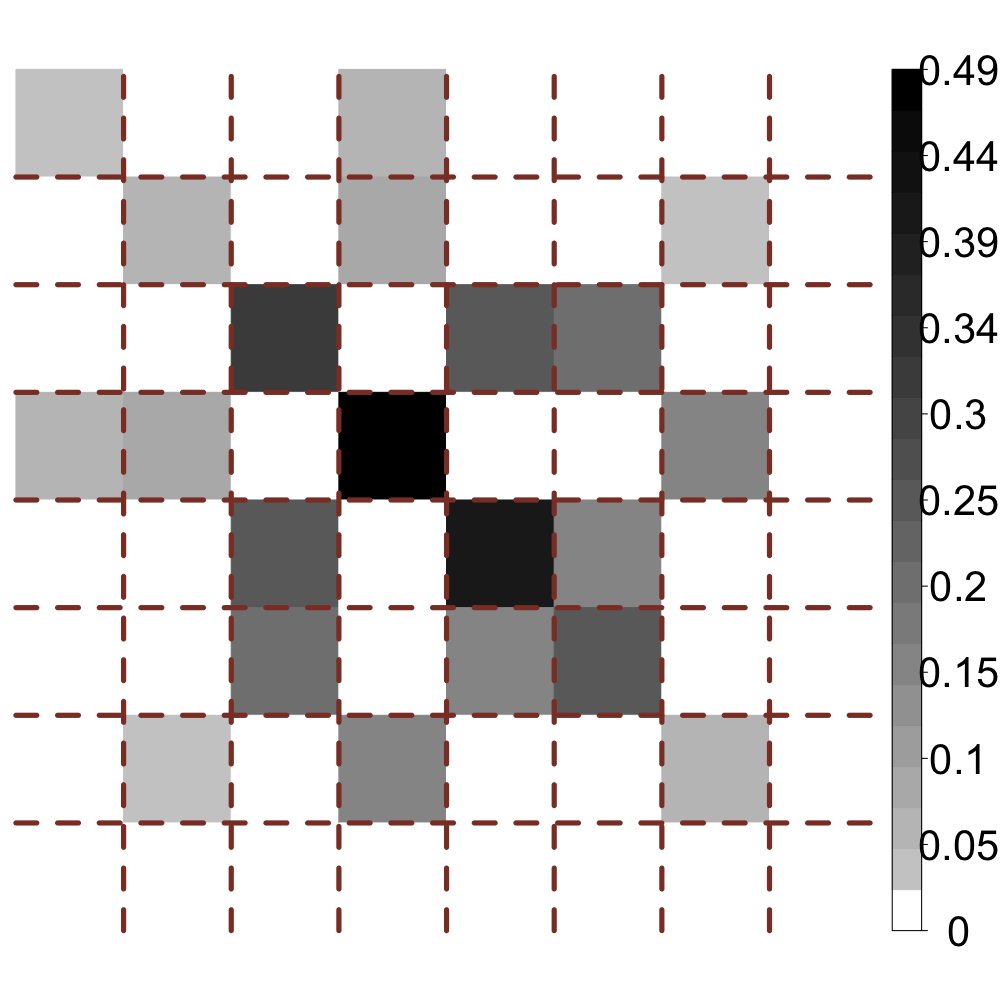}
    \end{minipage}\hfill
    \begin{minipage}{0.4\textwidth}
    \centering
    \includegraphics[width=0.8\linewidth]{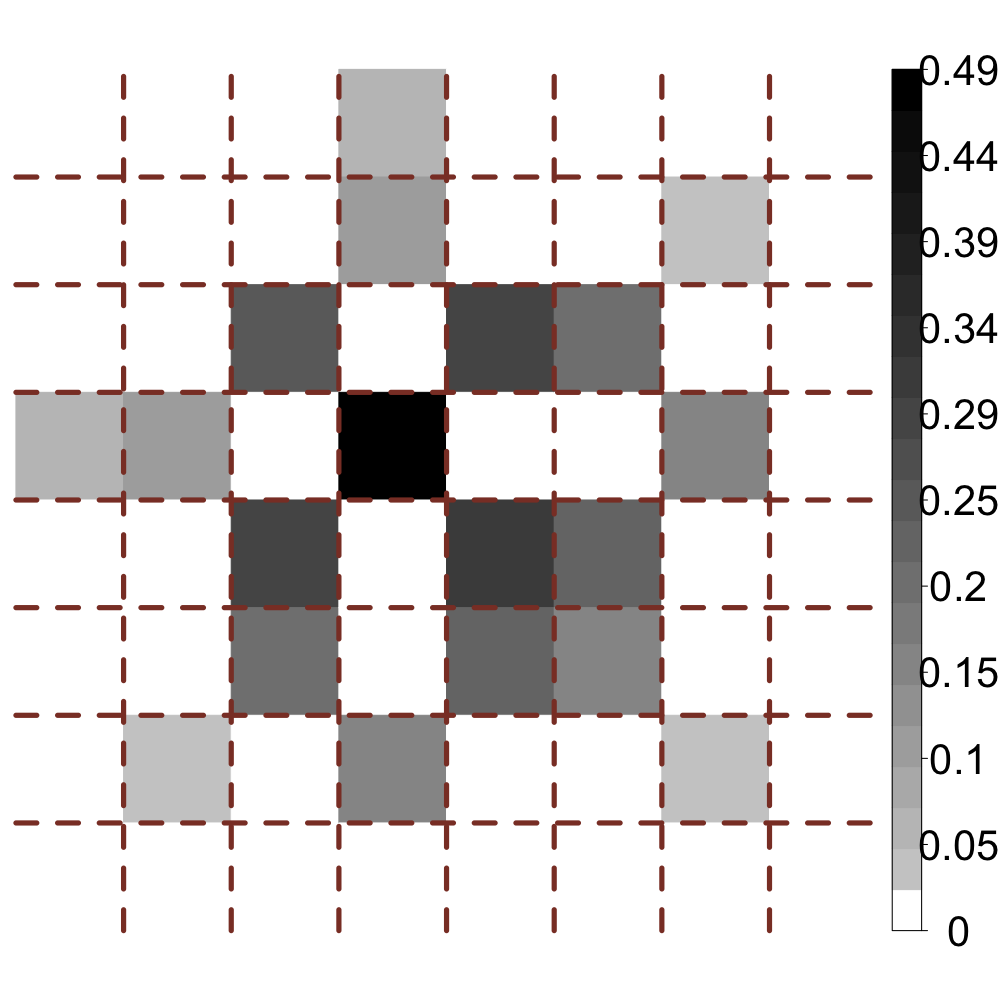}
    \end{minipage}\par
    \vspace*{-0.2cm}
    \begin{minipage}{0.4\textwidth}
    \centering
    \includegraphics[width=0.8\linewidth]{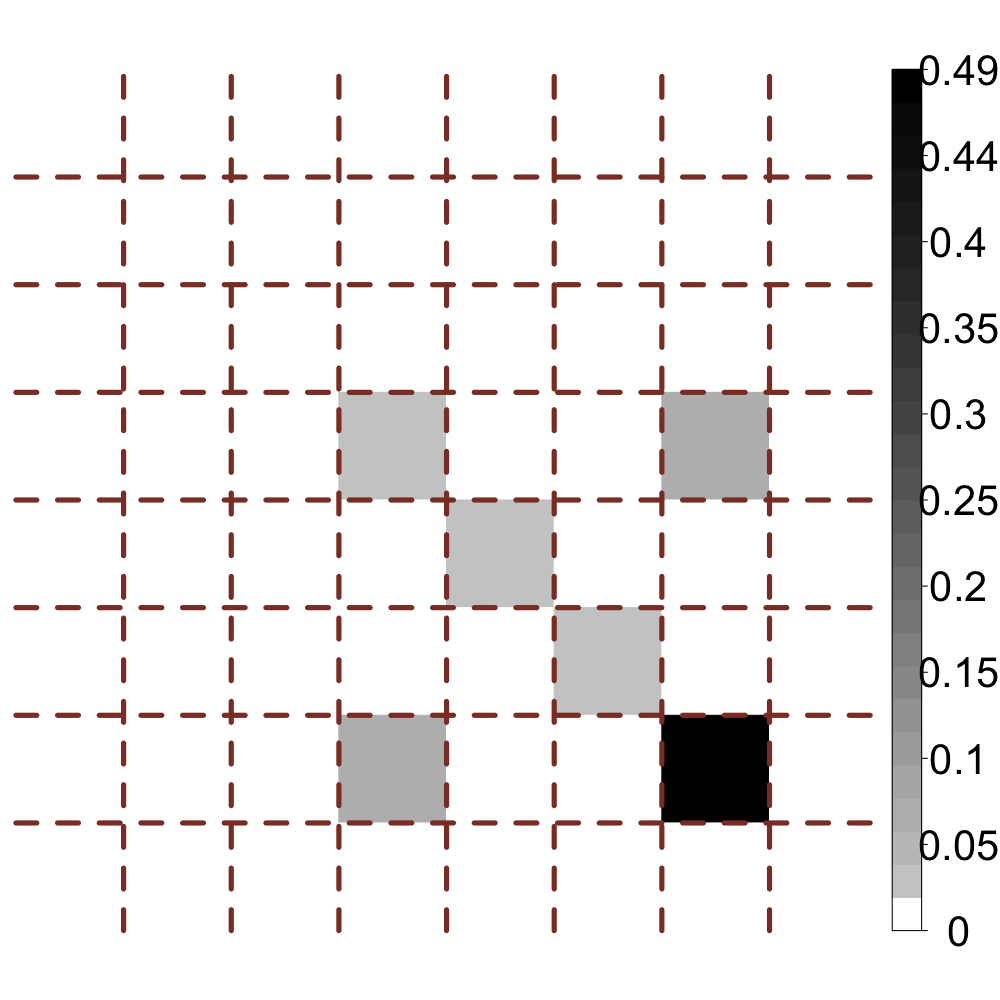}
    \end{minipage}\hfill
    \begin{minipage}{0.4\textwidth}
    \centering
    \includegraphics[width=0.8\linewidth]{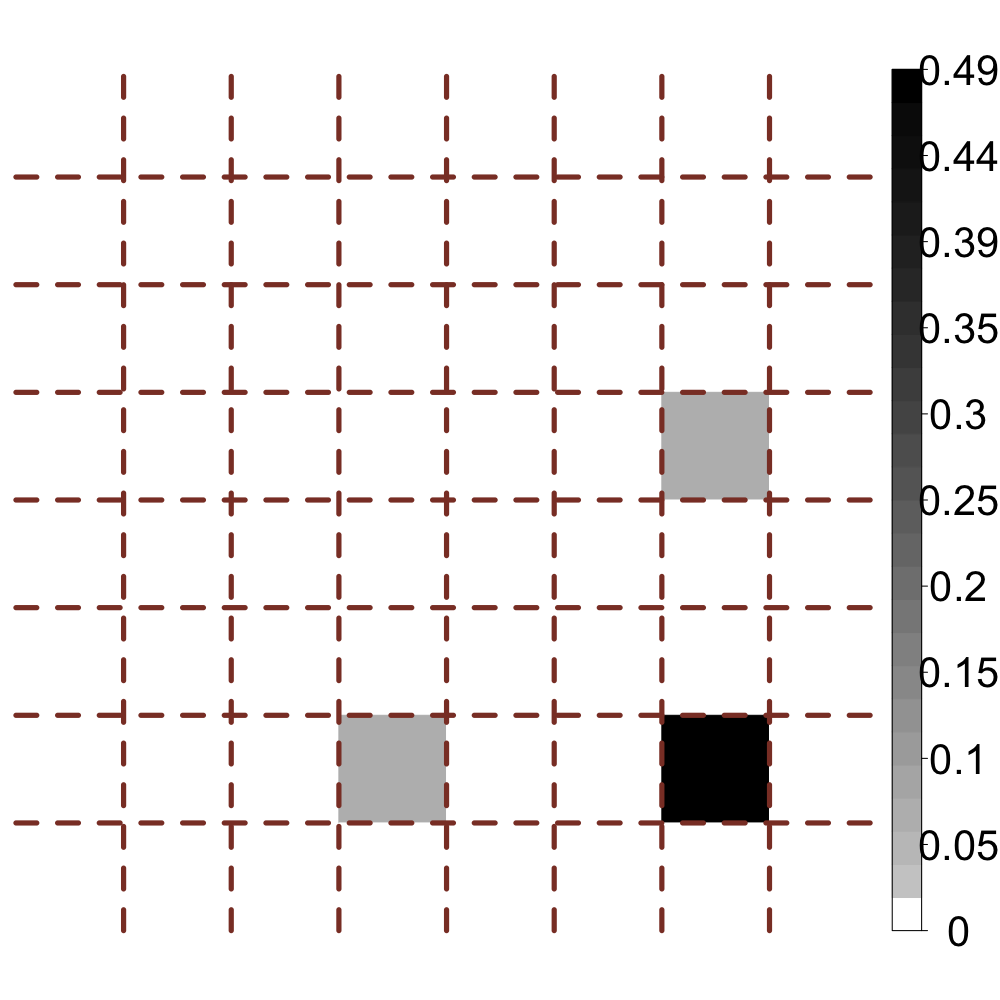}
    \end{minipage}\par
    \caption{Heatmaps of $\hat{\Bb}$. Darker colors indicate larger probabilities. Dashed lines distinguish $K = 8$ communities. First row: second group, with $\hat{d} = 8$ (left) and $\hat{d} = 2$ (right). Second row: third group, with $\hat{d} = 8$ (left) and $\hat{d} = 2$ (right). See \Cref{sec:real_data}.}
    \label{Bmat}
\end{figure}

First, \Cref{sec:estimated_Z} shows that an estimated probability matrix with the true rank value is not necessarily best to recover the true connectivity due to having noisy observations.
Second, the adjacency matrices are very sparse in this gene co-expression dataset, and we only have five and four adjacency matrices in the second and third group, respectively, which makes it difficult to select $\lambda$. 
However, by checking the selected $\lambda$ value for each combination of the training data and the validation data, we find that for both groups, $\hat{d}$ using $\lambda$ selected by some combinations is in fact much smaller than $K = 8$, and the corresponding $\hat{\Bb}$ noticeably differs from $\hat{\Bb}^{\mathrm{Avg}}$.
\Cref{Bmat} shows heatmaps of $\hat{\Bb}$ with full rank $\hat{d} = 8$ and low rank $\hat{d} = 2$ for the second and the third groups for the $\lambda$ value selected by one combination of the training data and the validation data.
We find that for the second group, the strength of the connectivity in community one and community two decreases for $\hat{d} = 2$, and for the third group, the strength of the connectivity in community four, community five, and community six decreases for $\hat{d} = 2$.
Moreover, even though the low-rank estimate $\hat{\Bb}$ indicates weaker connectivity strength in some communities, it retains the same strong connectivity as in the full-rank $\hat{\Bb}$ (e.g., community four in the second group and community seven in the third group).
Thus, \Cref{Bmat} further suggests that we should not simply estimate $\Bb^{*}$ to have full rank per the real data.
If we can estimate the membership matrix suitably well and have access to more layers, then estimating $\Bb^{*}$ with more accurate rank may provide more meaningful understanding of the connectivity within and between communities.

\section{Discussion}
\label{sec:discussion}

Our estimation error bounds under the Frobenius norm and the nuclear norm can be specialized to specific clustering algorithms by leveraging existing misclustering rate results in the literature.
For example, \cite{lei2022bias} obtains a performance guarantee for bias-adjusted spectral clustering of the form
\begin{equation*}
    n^{-1}|\Omega^{c}|
    =
    O_{P}\left(K({n_{\min}n})^{-2} + K\log(L+n)(n_{\min}\sqrt{L}n\rho)^{-2}\right)
    .
\end{equation*}
Elsewhere, \cite{xu2023covariate} develops a covariate-assisted method which in the absence of covariates achieves
\begin{equation*}
    n^{-1}|\Omega^{c}|
    =
    O_{P}\left(n_{\max}n^{4}n_{\min}^{-6}\log n (L\rho)^{-1}\right)
    .
\end{equation*}
For $L = 1$, \cite{zhang2022randomized} considers spectral clustering with random projection and random sampling; there, in the special case $B^{*}_{ii} = \rho a$ and $B^{*}_{ij} = \rho b$ with balanced community sizes $n/K$, it is shown that
$
n^{-1}|\Omega^{c}|
=
O_{P}(K^{3}(n\rho)^{-1})
$,
as well as
\begin{equation*}
    \|\hat{\Bb} - \Bb^{*}\|_{\max}
    =
    O_{P}\left(
    (n\rho)^{-1}K^{3/2}\log n
    \left\{
    1+(1-K^3/\log n)^{-1}
    \right\}
    \right)
    ,
\end{equation*}
provided $n\rho \ge c_{0} \log n$ for some suitable positive constant $c_{0}$.

Next, we briefly elaborate on applying our method to \HeteroModel{} per \Cref{def2}.
We cluster $L$ layers into $\tilde{L}$ groups so that graphs in each group belong to \HomoModel{}.
Often, $\tilde{L}$ is given in multilayer network analysis due to domain-specific and problem-specific contextual knowledge \citep{pensky2024clustering,jing2021community,fan2022alma,noroozi2024sparse}.
If $\tilde{L}$ is unknown, then we can use an elbow-based scree plot method applied to the matrix $\Gb \in \RR^{L \times n(n+1)/2}$ ($\tilde{\Gb} \in \RR^{L \times K(K+1)/2}$), where $\Gb_{\ell\bullet}$ ($\tilde{\Gb}_{\ell\bullet}$) consists of the upper triangular elements of $\hat{\Ab}^{(\ell)}$ ($\hat{\Bb}^{(\ell)}$). 
For more general multilayer networks, we can either employ the elbow method on different matrices or utilize tensor decomposition techniques.
For \HeteroModel{}, we use spectral clustering on $\Gb$ ($\tilde{\Gb}$) to cluster $L$ layers, which works well in our numerical studies and real data analysis.
For other types of multilayer networks, we can use alternative methods which enjoy guarantees for the between-layer clustering error rate, $R_{\mathrm{BL}}$, defined as
\begin{equation*}
    R_{\mathrm{BL}}
    \coloneqq
    \underset{\pi}{\min}~
    L^{-1}
    \sum_{\tilde{\ell}=1}^{\tilde{L}} |\GG_{\tilde{\ell}} \setminus \hat{\GG}_{\pi(\tilde{\ell})}|
    ,
\end{equation*}
where $\pi : \nset{\tilde{L}} \to \nset{\tilde{L}}$ is a permutation.
In particular, the TWIST algorithm of \cite{jing2021community} exhibits
$
R_{\mathrm{BL}}
=
O_{P}
\left(
\log n/(Ln\rho)
\right)
$
while the ALMA algorithm of \cite{fan2022alma} exhibits
\begin{equation*}
    R_{\mathrm{BL}}
    =
    O_{P}\left(\log^{4}(n+L)/(n^{2}\rho) + \log^{4}(n+L)/(n\{\min(n,L)\rho\})^{2}\right)
    ,
\end{equation*}
both of which are iterative methods developed for multilayer SBMs.
In general, we note that caution is needed in method selection, as $\Bb^{*}$ might be low rank, and some methods mentioned above only study $R_{\mathrm{BL}}$ in full rank settings.

In the multilayer network literature, there is growing interest in the use of tensor techniques such as the Tucker decomposition. 
\cite{han2022exact} develops a higher-order spectral clustering and higher-order Lloyd algorithm to achieve exact clustering in a tensor block model.
More closely related to our problem (where we do not assume all $\Bb^{*}$ to be full rank), aside from the aforementioned method in \cite{xu2023covariate} for \HeteroModel{}, is the TWIST method in \cite{jing2021community} which has the error rate for the $\tilde{\ell}$-th layer group
$n^{-1}|\Omega_{\tilde{\ell}}^{c}|
=
O_{P}
(\tilde{L}{K_{\tilde{\ell}}}^{2} \log n (L n \rho)^{-1})$,
where $K_{\tilde{\ell}}$ is the number of node communities in the $\tilde{\ell}$-th layer group.

In summary, the results in this paper complement and can be integrated with the aforementioned existing bounds, providing a point of departure for future work on multiple network analysis. 
Collectively, the theoretical and empirical investigations presented here suggest the potential usefulness of the proposed approach, either in a stand-alone fashion or more realistically as one of several methods to be applied in tandem when analyzing networks.
While it is perhaps tempting to work with average-based estimators, given their conceptual and practical simplicity, doing so can be sub-optimal or inadvertently lead to overlooking structural properties of data-generating mechanisms.
As such, this paper provides one principled alternative for examining connectivity and dimensionality in samples of structured networks.

\section*{Supplementary materials}

\begin{description}
    \item[Appendix:] Document containing additional discussion, the derivation of \Cref{alg}, the $M$-fold cross-validation method for a single network, additional simulation results, and all proofs.

    \item [R code and data:] Code to reproduce the simulations and real data analysis, subject to data sharing permission. See also \url{https://github.com/wej24/ADMM_SBM}.
\end{description}

\bibliography{reference}



\section*{Supplementary material}

# Supplement for "Simultaneous estimation of connectivity and dimensionality in samples of networks"

Wenlong Jiang[1], Chris McKennan[1], Jesús Arroyo[2], and Joshua Cape[3]

[1]Department of Statistics, University of Pittsburgh
[2]Department of Statistics, Texas A&M University
[3]Department of Statistics, University of Wisconsin–Madison

August 17, 2025

**Supplementary Material**

This supplementary material document contains the derivation of Algorithm 1 in the main text, the $M$-fold cross-validation method for a single observed network, additional simulations, additional real data details, and proofs of the stated theoretical results.

# S1   Derivation of Algorithm 1 in the main text

Recall that in Algorithm 1 in the main text, we seek to minimize the function

$$\mathcal{L}(\mathbf{W}, \mathbf{V}, \mathbf{\Theta}) = \|\mathbf{Y} - \mathbf{X}\mathbf{W}\mathbf{X}^{\mathrm{T}}\|_{\mathrm{F}}^2 + \lambda \|\mathbf{V}\|_* + \frac{\rho_1}{2} \|\mathbf{V} - \mathbf{W} + \mathbf{\Theta}\|_{\mathrm{F}}^2.$$

Here, $\mathbf{W}$ and $\mathbf{V}$ can be iteratively computed, namely, first update $\mathbf{W}$ assuming $\mathbf{V}$ is fixed and then update $\mathbf{V}$ assuming $\mathbf{W}$ is fixed.

**Update for $\mathbf{W}$:** Solve the optimization problem

$$\underset{\mathbf{W}}{\text{minimize}} \ \|\mathbf{Y} - \mathbf{X}\mathbf{W}\mathbf{X}^{\mathrm{T}}\|_{\mathrm{F}}^2 + \frac{\rho_1}{2} \|\mathbf{V} - \mathbf{W} + \mathbf{\Theta}\|_{\mathrm{F}}^2,$$

which is equivalent to solving

$$\underset{\boldsymbol{w}}{\text{minimize}} \ \|\boldsymbol{y} - \mathbf{C}\boldsymbol{w}\|_2^2 + \frac{\rho_1}{2} \|\boldsymbol{v} - \boldsymbol{w} + \boldsymbol{\theta}\|_2^2,$$

where $\boldsymbol{y} = \text{vec}(\mathbf{Y})$, $\boldsymbol{w} = \text{vec}(\mathbf{W})$, $\boldsymbol{v} = \text{vec}(\mathbf{V})$, $\boldsymbol{\theta} = \text{vec}(\boldsymbol{\Theta})$, and $\mathbf{C} = \mathbf{X} \otimes \mathbf{X}$. We obtain $\widehat{\boldsymbol{w}} = \left(2\mathbf{C}^{\text{T}}\mathbf{C} + \rho_1 \mathbf{I}\right)^{-1} \left[2\mathbf{C}^{\text{T}}\boldsymbol{y} + \rho_1(\boldsymbol{\theta} + \boldsymbol{v})\right]$, where $\widehat{\boldsymbol{w}} = \text{vec}(\widehat{\mathbf{W}})$ and $\mathbf{I} \in \mathbb{R}^{K \times K}$ is the identity matrix.

**Update for V:** Solve the optimization problem

$$\underset{\mathbf{V}}{\text{minimize}} \ \frac{\lambda}{\rho_1} \|\mathbf{V}\|_* + \frac{1}{2} \|\mathbf{V} - \mathbf{W} + \boldsymbol{\Theta}\|_{\text{F}}^2.$$

This yields $\widehat{\mathbf{V}} = \sum_{j=1}^{K} \max(\sigma_j - \lambda/\rho_1, 0) \boldsymbol{\alpha}_j \boldsymbol{\beta}_j^{\text{T}}$, where $\sum_{j=1}^{K} \sigma_j \boldsymbol{\alpha}_j \boldsymbol{\beta}_j^{\text{T}}$ denotes the full singular value decomposition of $\mathbf{W} - \boldsymbol{\Theta}$.

## S2    M-fold cross-validation for a single network

Chen and Lei (2018) proposes an $M$-fold network cross-validation procedure to select the number of communities $K$. The authors randomly split the nodes into $M$ equal-sized subsets, estimate the model parameters using the rectangular submatrix $\mathbf{A}_{\mathcal{I}^{(-m)}\bullet}$ by removing the rows of $\mathbf{A}$ in the $m$-th subset, and calculate the predictive loss using the square submatrix $\mathbf{A}_{\mathcal{I}^{(m)}\mathcal{I}^{(m)}}$ for nodes in the $m$-th subset. See (Chen and Lei, 2018, Algorithm 1) for details. Here, we adapt their algorithm to choose $\lambda$ in our problem. Algorithm 3 summarizes the entire procedure for choosing $\lambda$ based on the single observation $\mathbf{A}$.

## S3    Additional numerical studies

### S3.1    MonoSBM for a single network

We conduct two additional simulations with $L = 1$ to further explore how $\mathbf{A}$ can affect the estimation accuracy of $d$ and $\mathbf{B}^*$. To remove the noisy effect of $\widehat{\mathbf{Z}}$, we consider $\widehat{\mathbf{Z}} = \mathbf{Z}$.

Let $\mathbf{B}^*$ and $\mathbf{Z}$ be as in Section 5.1 of the main text, along with $K = 2$ and $\rho = \log n / n$. We run each setting 100 times and use Algorithm 3 to select the tuning parameter $\lambda$. Table 6 shows that when $\widehat{\mathbf{Z}} = \mathbf{Z}$, even for a moderately sparse SBMs, Algorithm 1 in the

**Algorithm 3** $M$-fold cross-validation for a single network
---
Input: Adjacency matrix $\mathbf{A}$, estimated node membership matrix $\widehat{\mathbf{Z}}$, set of candidate tuning parameter values $\lambda$, number of folds $M$.

1. Block-wise node splitting: Randomly split the nodes for each community into $M$ equal-sized subsets so that $\widehat{n}_i^{(m)}/\widehat{n}_j^{(m)} = \widehat{n}_i/\widehat{n}_j$, $i \neq j$, where $\widehat{n}_i^{(m)}$ is the number of nodes for the $i$-th community in the $m$-th fold. Let $\mathcal{I}^{(m)}$ denote the index set of the nodes in the $m$-th fold. Write $\mathbf{A}_{\mathcal{I}^{(-m)}\mathcal{I}^{(-m)}}$ and $\widehat{\mathbf{Z}}_{\mathcal{I}^{(-m)}\bullet}$ as the submatrices of $\mathbf{A}$ and $\widehat{\mathbf{Z}}$ with nodes not in the $m$-th fold.

2. For each $1 \leq m \leq M$ and each value $\lambda$, do the following:

   (a) Estimate $\mathbf{B}_\rho^*$, denoted as $\mathbf{B}_\rho^\lambda$, using Algorithm 1 on $\mathbf{A}_{\mathcal{I}^{(-m)}\mathcal{I}^{(-m)}}$ and $\widehat{\mathbf{Z}}_{\mathcal{I}^{(-m)}\bullet}$. Compute $\widehat{\mathbf{A}} = \widehat{\mathbf{Z}}\mathbf{B}_\rho^\lambda\widehat{\mathbf{Z}}^{\mathrm{T}}$.

   (b) Compute the validation loss
   $$\mathrm{Loss}(\lambda)^{(m)} := \|\widehat{\mathbf{A}} - \mathbf{A}\|_{\mathrm{F}}^2 - \|\widehat{\mathbf{A}}_{\mathcal{I}^{(-m)}\mathcal{I}^{(-m)}} - \mathbf{A}_{\mathcal{I}^{(-m)}\mathcal{I}^{(-m)}}\|_{\mathrm{F}}^2.$$

3. Set $\mathrm{Loss}(\lambda) := \sum_{m=1}^M \mathrm{Loss}(\lambda)^{(m)}$.

Output: Return $\widehat{\lambda} = \underset{\lambda}{\arg\min}\ \mathrm{Loss}(\lambda)$.

---

main text can still identify the potential low-rank structure by using Algorithm 3 to select $\lambda$. Furthermore, $\widehat{\mathbf{B}}^{\text{Our}}$ from Algorithm 1 in the main text is no worse than $\widehat{\mathbf{B}}^{\text{Avg}}$ from the averaging method based on the estimation error under the Frobenius norm, suggesting that $\mathbf{A}$ possibly plays a less important role compared to $\widehat{\mathbf{Z}}$ when estimating $\mathbf{B}^*$.

| | | Our Method | | Avg. Method | |
|---|---|---|---|---|---|
| | | $n = 10^3$ | $n = 10^4$ | $n = 10^3$ | $n = 10^4$ |
| Setting 1 with $d = 2$ | Estimation Error | $0.0473(2.3 \times 10^{-3})$ | $0.0130(8.0 \times 10^{-4})$ | $0.0473(2.3 \times 10^{-3})$ | $0.0130(7.0 \times 10^{-4})$ |
| | Estimated Dimensionality | $2(0)$ | $2(0)$ | $2(0)$ | $2(0)$ |
| Setting 2 with $d = 1$ | Estimation Error | $0.0432(2.5 \times 10^{-3})$ | $0.0120(7.0 \times 10^{-4})$ | $0.0446(2.4 \times 10^{-3})$ | $0.0127(7.0 \times 10^{-4})$ |
| | Estimated Dimensionality | $1.29(0.05)$ | $1.24(0.04)$ | $2(0)$ | $2(0)$ |

Table 6: The average (standard error) estimation errors and dimensionality for $L = 1$, $\rho = \log n/n$, and $\mathbf{B}^*$ per Section S3.1.

## S3.2 MonoSBM with perfectly estimated node memberships

**Dimensionality $d = K$ and $d < K$:** Consider SBMs with $K = 2$, $n \in \{10^3, 10^4\}$, $\rho \in \{1/n, \log n/n\}$, $L \in \{20, 100\}$, and $(n_1, n_2) = (0.25n, 0.75n)$ as follows:

1. The full-rank SBM with $d = 2$ and probability matrix

$$\mathbf{B}^* = \begin{bmatrix} 0.5 & 0.3 \\ 0.3 & 0.4 \end{bmatrix}.$$

2. The low-rank SBM with $d = 1$ and probability matrix

$$\mathbf{B}^* = \boldsymbol{v}\boldsymbol{v}^{\text{T}}, \quad \boldsymbol{v} = (0.7, 0.7^2)^{\text{T}}.$$

We first investigate the estimation errors of $\widehat{\mathbf{B}}^{\text{Avg}}$ and $\widehat{\mathbf{B}}^{\text{Our}}$ under the Frobenius norm. Table 7 summarizes the average estimation errors with their standard errors across 100 independent replications. The performance of our method is similar to the averaging method when $d = K$ but produces a slightly more accurate estimate when $d < K$. One possible explanation is that the additional variability introduced by the additional eigenvectors

4

of $\widehat{\mathbf{B}}^{\text{Avg}}$ does not have a dramatic influence on estimation. As $L$, $\rho$, or $n$ increase, both methods yield better estimates of $\mathbf{B}^*$ due to larger effective sample size.

|  |  | Our Method $n = 10^3$ | | Avg. Method $n = 10^3$ | | Our Method $n = 10^4$ | | Avg. Method $n = 10^4$ | |
|---|---|---|---|---|---|---|---|---|---|
|  |  | $L = 20$ | $L = 100$ | $L = 20$ | $L = 100$ | $L = 20$ | $L = 100$ | $L = 20$ | $L = 100$ |
| Setting 1 with $d = 2$ | $\rho = \frac{1}{n}$ | $0.0297(1.5 \times 10^{-3})$ | $0.0124(7.0 \times 10^{-4})$ | $0.0298(1.5 \times 10^{-3})$ | $0.0123(7.0 \times 10^{-4})$ | $0.0092(5.0 \times 10^{-4})$ | $0.0038(2.0 \times 10^{-4})$ | $0.0092(5.0 \times 10^{-4})$ | $0.0038(2.0 \times 10^{-4})$ |
|  | $\rho = \frac{\log n}{n}$ | $0.0097(6.0 \times 10^{-4})$ | $0.0052(3.0 \times 10^{-4})$ | $0.0097(6.0 \times 10^{-4})$ | $0.0052(3.0 \times 10^{-4})$ | $0.0028(1.0 \times 10^{-4})$ | $0.0012(1.0 \times 10^{-4})$ | $0.0028(1.0 \times 10^{-4})$ | $0.0012(1.0 \times 10^{-4})$ |
| Setting 2 with $d = 1$ | $\rho = \frac{1}{n}$ | $0.0255(1.4 \times 10^{-3})$ | $0.0109(7.0 \times 10^{-4})$ | $0.0284(1.4 \times 10^{-4})$ | $0.0124(7.0 \times 10^{-4})$ | $0.0079(5.0 \times 10^{-4})$ | $0.0036(2.0 \times 10^{-4})$ | $0.0088(5.0 \times 10^{-4})$ | $0.0041(2.0 \times 10^{-4})$ |
|  | $\rho = \frac{\log n}{n}$ | $0.0098(6.0 \times 10^{-4})$ | $0.0039(2.0 \times 10^{-4})$ | $0.0103(5.0 \times 10^{-4})$ | $0.0045(2.0 \times 10^{-4})$ | $0.0026(2.0 \times 10^{-4})$ | $0.0012(1.0 \times 10^{-4})$ | $0.0029(2.0 \times 10^{-4})$ | $0.0014(1.0 \times 10^{-4})$ |

Table 7: The average (standard error) estimation errors in Frobenius norm. See Section S3.2.

To better summarize the results, write

$$\mathbf{\Delta}^{(i)} := \|\widehat{\mathbf{B}}^{\text{Our}(i)} - \mathbf{B}^*\|_{\text{F}} - \|\widehat{\mathbf{B}}^{\text{Avg}(i)} - \mathbf{B}^*\|_{\text{F}}, \tag{S3.1}$$

where $i \in [\![100]\!]$ indicates the $i$-th simulation iteration. For each setting, we plot the histogram of $\mathbf{\Delta}^{(i)}$. Here, we see that our method is consistently better than the averaging method when $d < K$ but without asserting that it necessarily gives accurate estimates of $\mathbf{B}^*$ (so we do not use the relative measure to visualize the results). In addition, if the estimation errors from our method and the averaging method are both large, then the relative measure will fail to reflect the discrepancy. We summarize the results in Figure 6. To make the histograms comparable in each row of Figure 6, two outlier values ($-0.001$ and $0.002$) are not shown (row 1, column 2).

In the first and third rows of Figure 6, when $d = K$, all histograms are approximately symmetric around zero. When $L$ or $n$ increases, the histograms are more concentrated around zero, which further indicates that our method and the averaging method are roughly comparable in the full rank case. When $d < K$, all histograms are left-skewed, which also indicates that $\widehat{\mathbf{B}}^{\text{Our}}$ performs better than $\widehat{\mathbf{B}}^{\text{Avg}}$ on the basis of estimation error.

Next, we investigate how well our method estimates the embedding dimensionality of $\mathbf{B}^*$. When $\widehat{\mathbf{Z}} = \mathbf{Z}$, the community structure is not affected, so we expect the estimated dimensionality to be equal to the true rank of $\mathbf{B}^*$. For each setting, we compute the

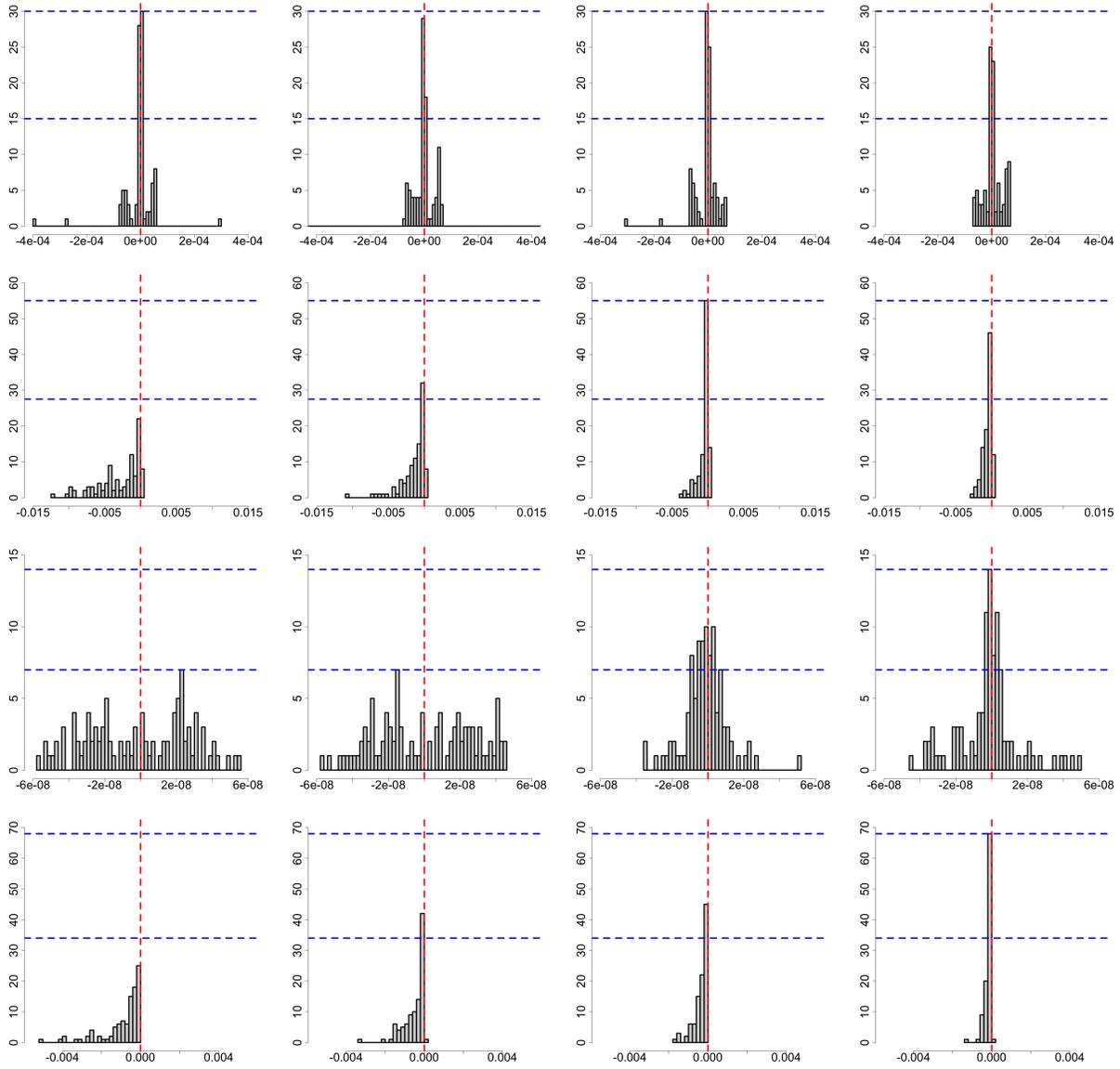

Figure 6: Histograms of $\boldsymbol{\Delta}^{(i)}$ per Eq. (S3.1). The dashed red line corresponds to $\boldsymbol{\Delta}^{(i)} = 0$. The upper (bottom) blue line corresponds to the maximum (half maximum) frequency across all histograms in each row. The first two rows are for $n = 10^3$ with $d = 2$ and $d = 1$, while the last two rows are for $n = 10^4$ with $d = 2$ and $d = 1$. The first two columns are for $\rho = 1/n$ and $L \in \{20, 100\}$, while the last two columns are for $\rho = \log n/n$ and $L \in \{20, 100\}$. See Section S3.2.

6

|  |  | Our Method | | Avg. Method | | Our Method | | Avg. Method | |
|  |  | $n = 10^3$ | | $n = 10^3$ | | $n = 10^4$ | | $n = 10^4$ | |
|  |  | $L = 20$ | $L = 100$ | $L = 20$ | $L = 100$ | $L = 20$ | $L = 100$ | $L = 20$ | $L = 100$ |
| Setting 1 with $d = 2$ | $\rho = \frac{1}{n}$ | 2(0) | 2(0) | 2(0) | 2(0) | 2(0) | 2(0) | 2(0) | 2(0) |
|  | $\rho = \frac{\log n}{n}$ | 2(0) | 2(0) | 2(0) | 2(0) | 2(0) | 2(0) | 2(0) | 2(0) |
| Setting 2 with $d = 1$ | $\rho = \frac{1}{n}$ | 1.02(0.01) | 1(0) | 2(0) | 2(0) | 1(0) | 1(0) | 2(0) | 2(0) |
|  | $\rho = \frac{\log n}{n}$ | 1.32(0.05) | 1.09(0.03) | 2(0) | 2(0) | 1.09(0.03) | 1(0) | 2(0) | 2(0) |

Table 8: The average (standard error) estimated dimensionality. See Section S3.2.

rank $\widehat{d}^{\mathrm{Our}(i)}$ and $\widehat{d}^{\mathrm{Avg}(i)}$ of $\widehat{\mathbf{B}}^{\mathrm{Our}(i)}$ and $\widehat{\mathbf{B}}^{\mathrm{Avg}(i)}$ across independent trials. We then calculate the average estimated dimensionality and corresponding standard errors across these 100 replications for both methods, summarized in Table 8. We find that the averaging method gives full-rank estimates regardless of whether the SBM is full rank or low rank. In contrast, our method provides more accurate estimates. When $d = K$, our method still estimates the true rank. When $d < K$, the average estimated rank from our method approaches the true rank as $L$ or $n$ increases.

These simulation examples show that our method can be useful when $d$ is unknown.

## S3.3 MonoSBM with imperfectly estimated node memberships

**Dimensionality $d = K$ and $d < K$:** Throughout this subsection, $K = 2$, $d \in \{1, 2\}$, $n = 10^3$, $(n_1, n_2) = (0.25n, 0.75n)$, $L \in \{40, 100\}$, $\rho \in \{\sqrt{\log n}/n, \log n/n\}$, and we construct $\mathbf{B}^*$ so that for each layer $\rho = \sqrt{\log n}/n$, $L = 40$, and the adjusted Rand index (ARI) between $\widehat{\boldsymbol{g}}$ and $\boldsymbol{g}$ is at least 0.8. We consider the following settings:

1. The full-rank SBM with $d = 2$ and probability matrix

$$\mathbf{B}^* = \begin{bmatrix} 0.5 & 0.2 \\ 0.2 & 0.5 \end{bmatrix}.$$

2. The low-rank SBM with $d = 1$ and probability matrix

$$\mathbf{B}^* = \boldsymbol{v}\boldsymbol{v}^{\mathrm{T}}, \quad \boldsymbol{v} = (0.9, 0.4)^{\mathrm{T}}.$$

Table 9 summarizes the average estimation errors, and Table 10 summarizes the average estimated dimensionality values. We find that the results are consistent with the previous section, even when some nodes are not correctly clustered. The rank estimation from our method is still very close to the true rank, especially when $L$ is large. Collectively, these findings suggest that our method is somewhat robust.

We also compare the results obtained from the spectral embedding method and the low-rank approximation method in Tang et al. (2022). The spectral embedding method to estimate $\mathbf{B}^*$ proceeds as follows:

1. Let $\overline{\mathbf{A}}_{\text{trunc}} := \widehat{\mathbf{U}}\widehat{\mathbf{\Lambda}}\widehat{\mathbf{U}}^{\text{T}}$, where $\widehat{\mathbf{\Lambda}}$ is the diagonal matrix with the $\widehat{d}$ largest eigenvalues of $\overline{\mathbf{A}}$ in modulus and the columns of $\widehat{\mathbf{U}} \in \mathbb{R}^{n \times \widehat{d}}$ are corresponding eigenvectors of $\overline{\mathbf{A}}$.

2. For each pair $(k, \ell)$, define $\widehat{\mathbf{B}}_{k\ell}^{\text{Trunc}} := \frac{1}{\rho \widehat{n}_k \widehat{n}_\ell} \widehat{\mathbf{Z}}_{\bullet k}^{\text{T}} \overline{\mathbf{A}}_{\text{trunc}} \widehat{\mathbf{Z}}_{\bullet \ell}$.

Notably, the spectral embedding method always estimates the rank of $\mathbf{B}^*$ to be $\widehat{d}$ since $\text{rk}(\overline{\mathbf{A}}_{\text{trunc}}) = \widehat{d}$, so we do not include this result in Table 10. Tang et al. (2022) mentions estimating $d$ in not-too-sparse networks via an eigenvalue threshold test; however, in the sparse regime, estimating $d$ becomes a more difficult problem. Therefore, in what follows, we consider truncating $\overline{\mathbf{A}}$ with dimension $K$ and $d$. Table 9 indicates that the spectral embedding method may yield smaller estimation errors than the averaging method in certain cases, although the improvement is not substantial. However, it can also produce noticeably larger estimation errors, especially when $d = 2$ and $\rho = \sqrt{\log n}/n$, which further illustrates why we do not consider the spectral embedding of $\mathbf{A}$. We also find that when $d = 1$, the estimation errors from the spectral embedding method with dimensions $K$ and $d$ are similar. Therefore, if $d$ is unknown, the correct dimension might not be selected, resulting in a less accurate dimensionality estimate of $\mathbf{B}^*$.

The low-rank approximation method for $\widehat{\mathbf{B}}^{\text{Avg}}$ is given by

$$\widehat{\mathbf{B}}^{\text{AvgLR}} = \underset{\mathbf{M}:\text{rk}(\mathbf{M})=d}{\arg\min} \|\mathbf{M} - \widehat{\mathbf{B}}^{\text{Avg}}\|_{\text{F}}. \tag{S3.2}$$

The low-rank matrix $\widehat{\mathbf{B}}^{\text{AvgLR}}$ is obtained by the spectral embedding of $\widehat{\mathbf{B}}^{\text{Avg}}$ and shares similarities with our method since it applies hard-thresholding on the singular values of $\widehat{\mathbf{B}}^{\text{Avg}}$. On the other hand, the low-rank approximation method requires us to know $d$ or determine $\widehat{d}$, whereas our method can choose $\widehat{d}$ by tuning $\lambda$. Due to the difficulty of estimating $d$, we do not assume $d$ is known in Eq. (S3.2) but rather use five-fold cross-validation and then compare the average rank estimation with our method. The corresponding estimation errors are consistently smaller than or equal to those for the averaging method. Still, our ADMM-based method can estimate the rank more accurately than the low-rank approximation method. We also find that even when $L$ increases, the rank estimation from low-rank approximation will not dramatically improve. Our method estimates $d$ by tuning a wide range of $\lambda$ values and using soft-thresholding, enabling greater flexibility.

|  |  | Our Method | | Avg. Method | | AvgLR Method | | Spectral Embedding with $K$ | | Spectral Embedding with $d$ | |
|---|---|---|---|---|---|---|---|---|---|---|---|
|  |  | $L=40$ | $L=100$ | $L=40$ | $L=100$ | $L=40$ | $L=100$ | $L=40$ | $L=100$ | $L=40$ | $L=100$ |
| Setting 1 with $d=2$ | $\rho=\frac{\sqrt{\log n}}{n}$ | 0.0160(9.0×10⁻⁴) | 0.0074(4.0×10⁻⁴) | 0.0160(9.0×10⁻⁴) | 0.0074(4.0×10⁻⁴) | 0.0160(9.0×10⁻⁴) | 0.0074(4.0×10⁻⁴) | 0.1010(4.2×10⁻³) | 0.0106(6.0×10⁻⁴) | 0.1010(4.2×10⁻³) | 0.0106(6.0×10⁻⁴) |
|  | $\rho=\frac{\log n}{n}$ | 0.0072(4.0×10⁻⁴) | 0.0047(3.0×10⁻⁴) | 0.0072(4.0×10⁻⁴) | 0.0047(3.0×10⁻⁴) | 0.0072(4.0×10⁻⁴) | 0.0047(3.0×10⁻⁴) | 0.0097(6.0×10⁻⁴) | 0.0048(3.0×10⁻⁴) | 0.0097(6.0×10⁻⁴) | 0.0048(3.0×10⁻⁴) |
| Setting 2 with $d=1$ | $\rho=\frac{\sqrt{\log n}}{n}$ | 0.0290(1.7×10⁻³) | 0.0089(6.0×10⁻⁴) | 0.0300(1.7×10⁻³) | 0.0094(6.0×10⁻⁴) | 0.0299(1.7×10⁻³) | 0.0093(6.0×10⁻⁴) | 0.0277(1.6×10⁻³) | 0.0096(6.0×10⁻⁴) | 0.0277(1.6×10⁻³) | 0.0096(6.0×10⁻⁴) |
|  | $\rho=\frac{\log n}{n}$ | 0.0084(6.0×10⁻⁴) | 0.0049(4.0×10⁻⁴) | 0.0088(6.0×10⁻⁴) | 0.0053(4.0×10⁻⁴) | 0.0086(6.0×10⁻⁴) | 0.0052(4.0×10⁻⁴) | 0.0089(5.0×10⁻⁴) | 0.0047(4.0×10⁻⁴) | 0.0089(5.0×10⁻⁴) | 0.0048(4.0×10⁻⁴) |

Table 9: The average (standard error) estimation errors under the Frobenius norm with $n=10^3$. For Setting 1, the smallest (average) ARI with $L \in \{40, 100\}$, $\rho = \sqrt{\log n}/n$ is 0.8908(0.9436) and 0.9957(0.9999). For Setting 2, the smallest (average) ARI with $L \in \{40, 100\}$, $\rho = \sqrt{\log n}/n$ is 0.8353(0.9121) and 0.9915(0.9983). See Section S3.3.

|  |  | Our Method | | Avg. Method | | AvgLR Method | |
|---|---|---|---|---|---|---|---|
|  |  | $L=40$ | $L=100$ | $L=40$ | $L=100$ | $L=40$ | $L=100$ |
| Setting 1 with $d=2$ | $\rho=\frac{\sqrt{\log n}}{n}$ | 2(0) | 2(0) | 2(0) | 2(0) | 2(0) | 2(0) |
|  | $\rho=\frac{\log n}{n}$ | 2(0) | 2(0) | 2(0) | 2(0) | 2(0) | 2(0) |
| Setting 2 with $d=1$ | $\rho=\frac{\sqrt{\log n}}{n}$ | 1.42(0.05) | 1.02(0.01) | 2(0) | 2(0) | 1.78(0.04) | 1.39(0.05) |
|  | $\rho=\frac{\log n}{n}$ | 1.31(0.05) | 1.11(0.03) | 2(0) | 2(0) | 1.47(0.05) | 1.46(0.05) |

Table 10: The average (standard error) of estimated dimensionality. See Section S3.3.

All methods considered in this section may not have significant estimation error dif-

9

ferences when $d$ is not much smaller than $K$, (except for some cases with the spectral embedding method) since none of the methods can significantly reduce the error from $\overline{\mathbf{A}}$ and the error from $\widehat{\mathbf{Z}}$ in the extremely sparse regime. However, the main benefit of our method here is for dimensionality estimation, since other methods are not able to estimate the dimensionality accurately when $d$ is unknown and the graph is very sparse.

**Dimensionality** $d \ll K$: In Section 5.2 in the main text, we illustrate that when $\widehat{\mathbf{Z}} \neq \mathbf{Z}$, even if the true rank is used, the low-rank approximation method with $d$ may produce a worse estimate of $\mathbf{B}^*$ compared to the averaging method. One possible reason is that if $\widehat{\mathbf{Z}} \neq \mathbf{Z}$, then the network block structure from the SBM is incorrectly aggregated. For illustration, we choose one replicate (from one hundred) as an example and summarize the results from this replicate in Figure 7. The scree plot of $\widehat{\mathbf{B}}^{\mathrm{Avg}}$ in Figure 7 indicates the dimensionality might be two instead of one. The heatmaps of $|\widehat{B}_{k\ell}^{\mathrm{Avg}} - B_{k\ell}^*|$ (middle) and $|\widehat{B}_{k\ell}^{\mathrm{AvgLR}} - B_{k\ell}^*|$ (right) in Figure 7 suggest that the connectivity related to community 10 is destroyed due to $\widehat{\mathbf{Z}}$.

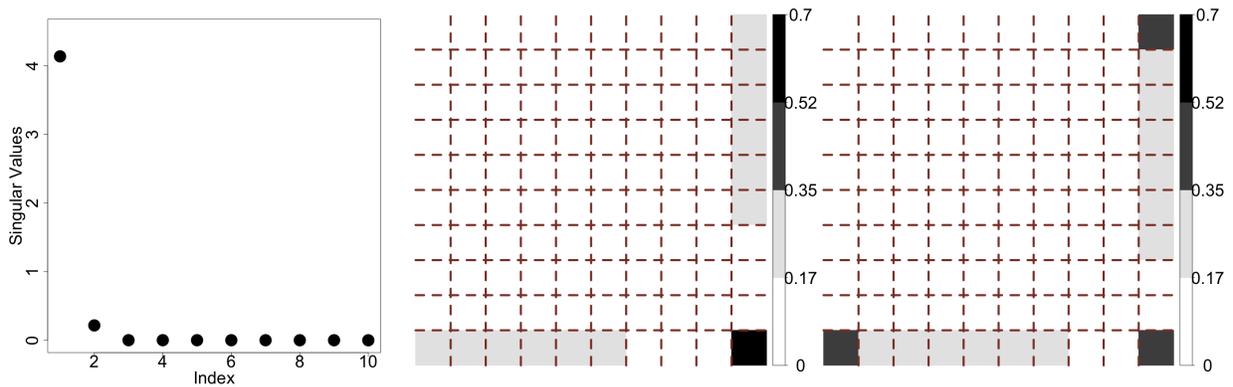

Figure 7: Left: scree plot of $\widehat{\mathbf{B}}^{\mathrm{Avg}}$. Middle: heatmap of $|\widehat{B}_{k\ell}^{\mathrm{Avg}} - B_{k\ell}^*|$. Right: heatmap of $|\widehat{B}_{k\ell}^{\mathrm{AvgLR}} - B_{k\ell}^*|$. Brown dashed lines distinguish the $K = 10$ communities. See Section S3.3.

## S3.4 Summary of numerical studies and limitations

The numerical studies provided in this supplement and in the main text illustrate the performance of our proposed method in a diverse collection of settings. We have shown that the proposed method can produce more accurate connectivity estimates than several other methods when the membership matrix can be perfectly recovered. If the membership matrix $\mathbf{Z}$ is imperfectly recovered, then our method can still estimate the dimensionality more accurately than the averaging method and the low-rank approximation method. Throughout, our method applies when obtaining $\widehat{\mathbf{Z}}$ using existing clustering methods.

Unsurprisingly, when $d \leq K$ but not $d \ll K$, the method proposed in this paper does not significantly improve the estimation accuracy even when $\widehat{\mathbf{Z}} = \mathbf{Z}$. Still, motivated by the refitting procedure from Mazumder et al. (2010) and employed in MacDonald et al. (2022), one could re-estimate the eigenvalues using the estimated rank and eigenvectors, which might result in improved estimation. We leave this direction for future work.

To be clear, our method outperforms the averaging-based method when $d \ll K$ and $\widehat{\mathbf{Z}} = \mathbf{Z}$, both in terms of estimating connectivity and dimensionality, though the improvement in connectivity estimation is reduced when $\widehat{\mathbf{Z}} \neq \mathbf{Z}$. If $\widehat{d} = d$, then the performance of our method is similar to the low-rank approximation method but more computationally expensive. That being said, the selection of $\widehat{d}$ for the low-rank approximation method is itself a challenging task.

## S4 Additional results for real data analysis

Figure 8 shows the scree plot of the matrix consisting of the upper triangular elements of matrices $\widehat{\mathbf{B}}^{(\ell)}$. In addition, the method in Zhu and Ghodsi (2006) also suggests that $\widetilde{L} = 3$.

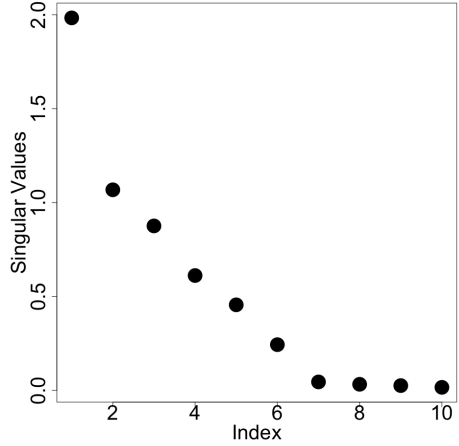

Figure 8: The scree plot of the matrix consisting of the upper triangular elements of $\widehat{\mathbf{B}}^{(\ell)}$.

# S5 Proofs of theoretical guarantees

Recall from the main text that the model of interest involves the expressions

$$\overline{\mathbf{A}} = \mathbb{E}(\mathbf{A}) + \boldsymbol{\Delta}_1, \quad \rho\widehat{\mathbf{Z}}\mathbf{B}^*\widehat{\mathbf{Z}}^{\mathrm{T}} = \mathbb{E}(\mathbf{A}) + \boldsymbol{\Delta}_2,$$

so,

$$\overline{\mathbf{A}} = \rho\widehat{\mathbf{Z}}\mathbf{B}^*\widehat{\mathbf{Z}}^{\mathrm{T}} + \boldsymbol{\Delta}_1 - \boldsymbol{\Delta}_2,$$

where for notational convenience we shall write $\boldsymbol{\Delta}_3 \coloneqq \boldsymbol{\Delta}_1 - \boldsymbol{\Delta}_2$.

We seek to establish upper bounds for $\|\widehat{\mathbf{B}}-\mathbf{B}^*\|_{\mathrm{F}}$ and $\|\widehat{\mathbf{B}}-\mathbf{B}^*\|_*$. Our proof of Theorem 1 is based on Lemma 1, Lemma 2, the following Lemma 3, and Eqs. (B.2) to (B.5) in Hamdi and Bayati (2022), which itself builds on the developments in Klopp (2014).

## S5.1 Eqs. (B.2) to (B.5) in Hamdi and Bayati (2022)

Given a collection of vectors $S$, let $\mathbf{P}_S$ denote the orthogonal projection onto the linear subspace spanned by the vectors in $S$. For any matrix $\mathbf{B} \in \mathbb{R}^{K \times K}$, let $S_r(\mathbf{B})$ and $S_c(\mathbf{B})$ denote the linear subspace spanned by the left and right orthonormal singular vectors of

**B**, respectively. For any choice $\mathbf{C} \in \mathbb{R}^{K \times K}$, define

$$\mathbf{P_B}(\mathbf{C}) := \mathbf{C} - \mathbf{P}_\mathbf{B}^\perp(\mathbf{C}), \quad \mathbf{P}_\mathbf{B}^\perp(\mathbf{C}) := \mathbf{P}_{S_r^\perp(\mathbf{B})} \mathbf{C} \mathbf{P}_{S_c^\perp(\mathbf{B})}.$$

Alternatively,

$$\begin{aligned} \mathbf{P_B}(\mathbf{C}) &= \mathbf{C} - \mathbf{P}_\mathbf{B}^\perp(\mathbf{C}) \\ &= \mathbf{P}_{S_r(\mathbf{B})} \mathbf{C} + \mathbf{P}_{S_r^\perp(\mathbf{B})} \mathbf{C} - \mathbf{P}_\mathbf{B}^\perp(\mathbf{C}) \\ &= \mathbf{P}_{S_r(\mathbf{B})} \mathbf{C} + \mathbf{P}_{S_r^\perp(\mathbf{B})} \left[ \mathbf{C} - \mathbf{C} \mathbf{P}_{S_c^\perp(\mathbf{B})} \right] \\ &= \mathbf{P}_{S_r(\mathbf{B})} \mathbf{C} + \mathbf{P}_{S_r^\perp(\mathbf{B})} \mathbf{C} \mathbf{P}_{S_c(\mathbf{B})}. \end{aligned} \quad (\text{S5.1})$$

Both $S_r(\mathbf{B})$ and $S_c(\mathbf{B})$ have dimension equal to $\mathrm{rk}(\mathbf{B})$, hence from Eq. (S5.1)

$$\mathrm{rk}(\mathbf{P_B}(\mathbf{C})) \leq 2\,\mathrm{rk}(\mathbf{B}). \quad (\text{S5.2})$$

By the definition of $\mathbf{P}_\mathbf{B}^\perp$, the left and right singular vectors of $\mathbf{P}_\mathbf{B}^\perp(\mathbf{C})$ are orthogonal to those of $\mathbf{B}$, hence

$$\|\mathbf{B} + \mathbf{P}_\mathbf{B}^\perp(\mathbf{C})\|_* = \|\mathbf{B}\|_* + \|\mathbf{P}_\mathbf{B}^\perp(\mathbf{C})\|_*.$$

An application of the triangle inequality and some algebra yields

$$\begin{aligned} \|\mathbf{B} + \mathbf{C}\|_* &= \|\mathbf{B} + \mathbf{P}_\mathbf{B}^\perp(\mathbf{C}) + \mathbf{P_B}(\mathbf{C})\|_* \\ &\geq \|\mathbf{B} + \mathbf{P}_\mathbf{B}^\perp(\mathbf{C})\|_* - \|\mathbf{P_B}(\mathbf{C})\|_* \\ &= \|\mathbf{B}\|_* + \|\mathbf{P}_\mathbf{B}^\perp(\mathbf{C})\|_* - \|\mathbf{P_B}(\mathbf{C})\|_*. \end{aligned} \quad (\text{S5.3})$$

## S5.2 Proof of Lemma 2

*Proof of Lemma 2.* The objective is to obtain a high-probability upper bound for the expression $n^{-1}\|\widehat{\mathbf{Z}}^{\mathrm{T}}\boldsymbol{\Delta}_3\widehat{\mathbf{Z}}\|_{\mathrm{op}}$. To begin, by applying the triangle inequality, we have

$$\frac{1}{n}\left\|\widehat{\mathbf{Z}}^{\mathrm{T}}\boldsymbol{\Delta}_3\widehat{\mathbf{Z}}\right\|_{\mathrm{op}}$$
$$= \frac{1}{n}\left\|(\mathbf{Z}+\widehat{\mathbf{Z}}-\mathbf{Z})^{\mathrm{T}}\boldsymbol{\Delta}_3(\mathbf{Z}+\widehat{\mathbf{Z}}-\mathbf{Z})\right\|_{\mathrm{op}}$$
$$\leq \underbrace{\frac{1}{n}\left\|\mathbf{Z}^{\mathrm{T}}\boldsymbol{\Delta}_3\mathbf{Z}\right\|_{\mathrm{op}}}_{\mathrm{I}_1}$$
$$+ \underbrace{\frac{1}{n}\left\|(\widehat{\mathbf{Z}}-\mathbf{Z})^{\mathrm{T}}\boldsymbol{\Delta}_3(\widehat{\mathbf{Z}}-\mathbf{Z})\right\|_{\mathrm{op}} + \frac{2}{n}\left\|(\widehat{\mathbf{Z}}-\mathbf{Z})^{\mathrm{T}}\boldsymbol{\Delta}_3\mathbf{Z}\right\|_{\mathrm{op}}}_{\mathrm{I}_2}.$$

**Upper bound for $\mathrm{I}_1$:** Again by the triangle inequality,

$$\mathrm{I}_1 = \frac{1}{n}\left\|\mathbf{Z}^{\mathrm{T}}\left(\overline{\mathbf{A}}-\rho\mathbf{Z}\mathbf{B}^*\mathbf{Z}^{\mathrm{T}}+\rho\mathbf{Z}\mathbf{B}^*\mathbf{Z}^{\mathrm{T}}-\rho\widehat{\mathbf{Z}}\mathbf{B}^*\widehat{\mathbf{Z}}^{\mathrm{T}}\right)\mathbf{Z}\right\|_{\mathrm{op}}$$
$$\leq \underbrace{\frac{1}{n}\left\|\mathbf{Z}^{\mathrm{T}}\left(\overline{\mathbf{A}}-\rho\mathbf{Z}\mathbf{B}^*\mathbf{Z}^{\mathrm{T}}\right)\mathbf{Z}\right\|_{\mathrm{op}}}_{\widetilde{\mathrm{I}}_1} + \underbrace{\frac{1}{n}\left\|\mathbf{Z}^{\mathrm{T}}\left(\rho\mathbf{Z}\mathbf{B}^*\mathbf{Z}^{\mathrm{T}}-\rho\widehat{\mathbf{Z}}^{\mathrm{T}}\mathbf{B}^*\widehat{\mathbf{Z}}^{\mathrm{T}}\right)\mathbf{Z}\right\|_{\mathrm{op}}}_{\widetilde{\mathrm{I}}_2}.$$

**Upper bound for $\widetilde{\mathrm{I}}_1$:** Write $\widetilde{\mathbf{B}} \coloneqq \mathbf{Z}^{\mathrm{T}}\left(\overline{\mathbf{A}}-\rho\mathbf{Z}\mathbf{B}^*\mathbf{Z}^{\mathrm{T}}\right)\mathbf{Z}$ and $\mathbf{P} \coloneqq \mathbb{E}(\mathbf{A}) = \rho\mathbf{Z}\mathbf{B}^*\mathbf{Z}^{\mathrm{T}}$. Thus, $\widetilde{B}_{qk} = \sum_{1\leq i,j\leq n} Z_{iq}Z_{jk}\sum_{1\leq \ell\leq L}(A_{ij}^{(\ell)}-P_{ij})/L$. We shall decompose $\widetilde{\mathbf{B}}$ into the upper-triangular part $\widetilde{\mathbf{B}}^+$ (including the main diagonal elements) and the lower-triangular part $\widetilde{\mathbf{B}}^-$ (excluding the main diagonal elements). Namely, $\widetilde{\mathbf{B}} = \widetilde{\mathbf{B}}^+ + \widetilde{\mathbf{B}}^-$, where the elements within $\widetilde{\mathbf{B}}^+$ and $\widetilde{\mathbf{B}}^-$ are independent, respectively. We shall use the conventional $\epsilon$-net argument (Vershynin, 2018) to bound $\|\widetilde{\mathbf{B}}^+\|_{\mathrm{op}}$ and $\|\widetilde{\mathbf{B}}^-\|_{\mathrm{op}}$.

**Upper bound for $\|\widetilde{\mathbf{B}}^+\|_{\mathrm{op}}$:** Set $\epsilon = 1/4$. By standard covering and packing arguments, there exist $\epsilon$-nets $\mathcal{N}$ and $\mathcal{M}$ each of the unit sphere in $\mathbb{R}^K$ satisfying $|\mathcal{N}|,|\mathcal{M}|\leq 9^K$ where

$$\|\widetilde{\mathbf{B}}^+\|_{\mathrm{op}} \leq 2\max_{\boldsymbol{x}\in\mathcal{N},\boldsymbol{y}\in\mathcal{M}}\boldsymbol{x}^{\mathrm{T}}\widetilde{\mathbf{B}}^+\boldsymbol{y}. \tag{S5.4}$$

Observe that for any choice of deterministic unit vectors $\boldsymbol{x}$ and $\boldsymbol{y}$,

$$\left|\boldsymbol{x}^{\mathrm{T}}\widetilde{\mathbf{B}}^+\boldsymbol{y}\right| = \left|\sum_{q=1}^{K}\sum_{k=1}^{K} \widetilde{B}_{qk}^+ x_q y_k\right|$$

$$= \left|\sum_{1\leq q\leq k\leq K} x_q y_k \sum_{1\leq i,j\leq n} Z_{iq}Z_{jk} \sum_{1\leq \ell \leq L}(A_{ij}^{(\ell)} - P_{ij})/L\right|.$$

Here, $\mathrm{Var}(x_q y_k A_{ij}^{(\ell)}) \leq x_q^2 y_k^2 \rho$ and $|x_q y_k (A_{ij}^{(\ell)} - P_{ij})| \leq 1$ holds uniformly, so for all values $|\lambda| < 1$ (not a tuning parameter) it holds that

$$\mathbb{E}\left(\exp\left\{\lambda\left(\sum_{1\leq i,j\leq n} Z_{iq}Z_{jk} \sum_{1\leq \ell \leq L} x_q y_k \left(A_{ij}^{(\ell)} - P_{ij}\right)\right)\right\}\right)$$

$$= \prod_{i,j,\ell} \mathbb{E}\left(\exp\left\{\lambda\left(Z_{iq}Z_{jk}x_q y_k \left(A_{ij}^{(\ell)} - P_{ij}\right)\right)\right\}\right)$$

$$\leq \left(\exp\left\{\frac{\lambda^2 x_q^2 y_k^2 \rho}{1 - |\lambda|}\right\}\right)^{n_q n_k L}$$

$$\leq \exp\left\{\frac{\lambda^2 n_{\max}^2 L \rho x_q^2 y_k^2}{1 - |\lambda|}\right\}.$$

Therefore,

$$\mathbb{E}\left(\exp\left\{\lambda \sum_{1\leq q\leq k\leq K}\left(\sum_{1\leq i,j\leq n} Z_{iq}Z_{jk} \sum_{1\leq \ell \leq L} x_q y_k \left(A_{ij}^{(\ell)} - P_{ij}\right)\right)\right\}\right)$$

$$\leq \exp\left\{\frac{\lambda^2 n_{\max}^2 L\rho \sum_{1\leq q\leq k\leq K} x_q^2 y_k^2}{1 - |\lambda|}\right\}$$

$$\leq \exp\left\{\frac{\lambda^2 n_{\max}^2 L\rho}{1 - |\lambda|}\right\}.$$

By applying Bernstein's inequality, for all $t > 0$,

$$\mathbb{P}\left(\left|\sum_{1\leq q\leq k\leq K}\left(\sum_{1\leq i,j\leq n} Z_{iq}Z_{jk} \sum_{1\leq \ell \leq L} x_q y_k \left(A_{ij}^{(\ell)} - P_{ij}\right)\right)\right| > Lt\right)$$

$$\leq 2\exp\left\{-\frac{L^2 t^2}{2n_{\max}^2 L\rho + 2Lt}\right\}$$

$$\leq 2\exp\left\{-\frac{L^2 K^2 t^2}{2c_1^2 n^2 L\rho + 2LK^2 t}\right\},$$

where the second inequality holds by an application of Assumption 1.

Now, choose $t = 4\sqrt{2}c_1 \frac{n}{K}\sqrt{\frac{\rho}{L}}(\sqrt{K} + \sqrt{\log n})$ and assume the condition $c_1 n\sqrt{L\rho} \geq$

15

$4\sqrt{2}K(\sqrt{K} + \sqrt{\log n})$ holds. By taking a union bound, we obtain

$$\mathbb{P}\left(\max_{\boldsymbol{x}\in\mathcal{N}, \boldsymbol{y}\in\mathcal{M}} \left|\boldsymbol{x}^{\mathrm{T}}\widetilde{\mathbf{B}}^{+}\boldsymbol{y}\right| > t\right)$$
$$\leq 2 \times 81^K \exp\left\{-\frac{L^2 K^2 t^2}{2c_1^2 n^2 L\rho + 2LK^2 t}\right\} \quad \text{(S5.5)}$$
$$\leq 2n^{-8}.$$

By combining Eqs. (S5.4) and (S5.5), the bound

$$\|\widetilde{\mathbf{B}}^{+}\|_{\mathrm{op}} \leq 8\sqrt{2}c_1 \frac{n}{K}\sqrt{\frac{\rho}{L}}\left(\sqrt{K} + \sqrt{\log n}\right)$$

holds with probability at least $1 - O(n^{-8})$.

Similarly, under the same conditions as above, the bound

$$\|\widetilde{\mathbf{B}}^{-}\|_{\mathrm{op}} \leq 8\sqrt{2}c_1 \frac{n}{K}\sqrt{\frac{\rho}{L}}\left(\sqrt{K} + \sqrt{\log n}\right)$$

holds with probability at least $1 - O(n^{-8})$.

Consequently, under the condition $n\sqrt{L\rho} \gtrsim K(\sqrt{K} + \sqrt{\log n})$, we have

$$\frac{1}{n}\left\|\mathbf{Z}^{\mathrm{T}}\left(\overline{\mathbf{A}} - \rho\mathbf{Z}\mathbf{B}^{*}\mathbf{Z}\right)\mathbf{Z}\right\|_{\mathrm{op}} \leq \frac{1}{n}\left(\|\widetilde{\mathbf{B}}^{+}\|_{\mathrm{op}} + \|\widetilde{\mathbf{B}}^{-}\|_{\mathrm{op}}\right)$$
$$\leq 16\sqrt{2}c_1 \frac{1}{K}\sqrt{\frac{\rho}{L}}\left(\sqrt{K} + \sqrt{\log n}\right) \quad \text{(S5.6)}$$

with probability at least $1 - O(n^{-8})$.

**Upper bound for $\mathbb{E}(\frac{1}{n}\|\mathbf{Z}^{\mathrm{T}}(\overline{\mathbf{A}} - \rho\mathbf{Z}\mathbf{B}^{*}\mathbf{Z}^{\mathrm{T}})\mathbf{Z}\|_{\mathrm{op}})$:** From Eq. (S5.5), we have

$$\mathbb{P}\left(\max_{\boldsymbol{x}\in\mathcal{N}, \boldsymbol{y}\in\mathcal{M}} \left|\boldsymbol{x}^{\mathrm{T}}\widetilde{\mathbf{B}}^{+}\boldsymbol{y}\right| > t\right) \leq 2 \times 81^K \exp\left\{-\frac{L^2 K^2}{4c_1^2 n^2 L\rho}t^2\right\} \quad \text{for } t \leq t^*, \quad \text{(S5.7)}$$

and

$$\mathbb{P}\left(\max_{\boldsymbol{x}\in\mathcal{N}, \boldsymbol{y}\in\mathcal{M}} \left|\boldsymbol{x}^{\mathrm{T}}\widetilde{\mathbf{B}}^{+}\boldsymbol{y}\right| > t\right) \leq 2 \times 81^K \exp\left\{-\frac{L}{4}t\right\} \quad \text{for } t \geq t^*, \quad \text{(S5.8)}$$

where $t^* = c_1^2 n^2 \rho / K^2$.

Let $v_1 := L^2 K^2/(4c_1^2 n^2 L\rho)$ and $v_2 := L/4$. By Hölder's inequality,

$$\mathbb{E}\left(\max_{\bm{x}\in\mathcal{N},\bm{y}\in\mathcal{M}}\left|\bm{x}^{\mathrm{T}}\widetilde{\mathbf{B}}^+\bm{y}\right|\right) \leq \left(\mathbb{E}\left(\max_{\bm{x}\in\mathcal{N},\bm{y}\in\mathcal{M}}\left|\bm{x}^{\mathrm{T}}\widetilde{\mathbf{B}}^+\bm{y}\right|\right)^{2\log 81K}\right)^{1/(2\log 81K)}$$

$$= \left(\int_0^{+\infty}\mathbb{P}\left(\max_{\bm{x}\in\mathcal{N},\bm{y}\in\mathcal{M}}\left|\bm{x}^{\mathrm{T}}\widetilde{\mathbf{B}}^+\bm{y}\right|>t^{1/(2\log 81K)}\right)\mathrm{d}t\right)^{1/(2\log 81K)}$$

$$\leq \left(2\times 81^K\int_0^{+\infty}\exp\{-v_1 t^{1/(\log 81K)}\mathrm{d}t\}\right.$$

$$\left.+2\times 81^K\int_0^{+\infty}\exp\{-v_2 t^{1/(2\log 81K)}\}\right)^{1/(2\log 81K)}$$

$$\leq 2^{1/(2\log 81)}\sqrt{e}\left(\log 81K v_1^{-\log 81K}\Gamma(\log 81K)\right.$$

$$\left.+2\log 81K v_2^{-2\log 81K}\Gamma(2\log 81K)\right)^{1/(2\log 81K)}$$

$$\leq 2^{1/(2\log 81)}\sqrt{e}\left((\log 81K)^{\log 81K}v_1^{-\log 81K}2^{1-\log 81K}\right.$$

$$\left.+2(\log 81K)^{2\log 81K}v_2^{-2\log 81K}\right)^{1/(2\log 81K)}$$

$$\leq 6^{1/(2\log 81)}\sqrt{\log 81e}\sqrt{\frac{K}{v_1}}$$

$$= 6^{1/(2\log 81)}\sqrt{\log 81e}\times 2c_1\frac{n}{\sqrt{K}}\sqrt{\frac{\rho}{L}},$$

where the fourth inequality is due to the fact that the Gamma function satisfies

$$\Gamma(x) \leq \left(\frac{x}{2}\right)^{x-1}, \qquad \text{provided } x \geq 2.$$

The last inequality holds by the assumption $c_1 n\sqrt{L\rho} \geq 2\sqrt{\log 81}K^{3/2}$.

Similarly, under the same condition, we obtain

$$\mathbb{E}\left(\|\widetilde{\mathbf{B}}^-\|_{\mathrm{op}}\right) \leq 6^{1/(2\log 81)}\sqrt{\log 81e}\times 4c_1\frac{n}{\sqrt{K}}\sqrt{\frac{\rho}{L}}.$$

Therefore, under the condition $n\sqrt{L\rho} \gtrsim K^{3/2}$, we have

$$\mathbb{E}\left(\frac{1}{n}\|\mathbf{Z}^{\mathrm{T}}(\overline{\mathbf{A}}-\rho\mathbf{Z}\mathbf{B}^*\mathbf{Z}^{\mathrm{T}})\mathbf{Z}\|_{\mathrm{op}}\right) \leq 6^{1/(2\log 81)}\sqrt{\log 81e}\times 8c_1\frac{1}{\sqrt{K}}\sqrt{\frac{\rho}{L}}. \tag{S5.9}$$

Hence, if $|\Omega^c|=0$, then

$$\mathbb{E}\left(\|\widehat{\mathbf{B}}-\mathbf{B}^*\|_{\mathrm{F}}\right) \lesssim K^{3/2}\sqrt{d}\frac{1}{\sqrt{n}}\frac{1}{\sqrt{Ln\rho}}.$$

**Upper bound for $\widetilde{I}_2$:** As in the main text, define the index set $\Omega$ of correctly clustered nodes as

$$\Omega := \{i : \widehat{\mathbf{Z}}_{i\bullet} = \mathbf{Z}_{i\bullet}\}.$$

We decompose $\widehat{\mathbf{Z}}$ and $\mathbf{Z}$ into two parts by setting the entries of rows not in $\Omega$ to be zero in the first part and setting the entries of rows in $\Omega$ to be zero in the second part, namely

$$\widehat{\mathbf{Z}} = \widehat{\mathbf{Z}}_0 + \widehat{\mathbf{Z}}_1,$$

$$\mathbf{Z} = \mathbf{Z}_0 + \mathbf{Z}_1,$$

where,

$$\widehat{\mathbf{Z}}_{0,\Omega\bullet} = \mathbf{Z}_{0,\Omega\bullet} = \mathbf{Z}_{\Omega\bullet}, \quad \widehat{\mathbf{Z}}_{0,\Omega^c\bullet} = \mathbf{Z}_{0,\Omega^c\bullet} = \mathbf{0},$$

$$\widehat{\mathbf{Z}}_{1,\Omega\bullet,} = \mathbf{Z}_{1,\Omega\bullet} = \mathbf{0}, \quad \widehat{\mathbf{Z}}_{1,\Omega^c\bullet} = \widehat{\mathbf{Z}}_{\Omega^c\bullet}, \quad \mathbf{Z}_{1,\Omega^c\bullet} = \mathbf{Z}_{\Omega^c\bullet}.$$

This allows us to write

$$\widehat{\mathbf{Z}}\mathbf{B}^*\widehat{\mathbf{Z}}^{\mathrm{T}} - \mathbf{Z}\mathbf{B}^*\mathbf{Z}^{\mathrm{T}}$$

$$= \widehat{\mathbf{Z}}_0\mathbf{B}^*\widehat{\mathbf{Z}}_0^{\mathrm{T}} + \widehat{\mathbf{Z}}_0\mathbf{B}^*\widehat{\mathbf{Z}}_1^{\mathrm{T}} + \widehat{\mathbf{Z}}_1\mathbf{B}^*\widehat{\mathbf{Z}}_0^{\mathrm{T}} + \widehat{\mathbf{Z}}_1\mathbf{B}^*\widehat{\mathbf{Z}}_1^{\mathrm{T}}$$

$$- \mathbf{Z}_0\mathbf{B}^*\mathbf{Z}_0^{\mathrm{T}} - \mathbf{Z}_0\mathbf{B}^*\mathbf{Z}_1^{\mathrm{T}} - \mathbf{Z}_1\mathbf{B}^*\mathbf{Z}_0^{\mathrm{T}} - \mathbf{Z}_1\mathbf{B}^*\mathbf{Z}_1^{\mathrm{T}}$$

$$= \mathbf{Z}_0\mathbf{B}^*(\widehat{\mathbf{Z}}_1^{\mathrm{T}} - \mathbf{Z}_1^{\mathrm{T}}) + (\widehat{\mathbf{Z}}_1 - \mathbf{Z}_1)\mathbf{B}^*\mathbf{Z}_0^{\mathrm{T}} + \widehat{\mathbf{Z}}_1\mathbf{B}^*\widehat{\mathbf{Z}}_1^{\mathrm{T}} - \mathbf{Z}_1\mathbf{B}^*\mathbf{Z}_1^{\mathrm{T}}.$$

Consequently,

$$\left\|\rho\mathbf{Z}\mathbf{B}^*\mathbf{Z}^{\mathrm{T}} - \rho\widehat{\mathbf{Z}}\mathbf{B}^*\widehat{\mathbf{Z}}^{\mathrm{T}}\right\|_{\mathrm{op}}$$

$$= \rho\left\|\mathbf{Z}_0\mathbf{B}^*(\widehat{\mathbf{Z}}_1^{\mathrm{T}} - \mathbf{Z}_1^{\mathrm{T}}) + (\widehat{\mathbf{Z}}_1 - \mathbf{Z}_1)\mathbf{B}^*\mathbf{Z}_0^{\mathrm{T}} + \widehat{\mathbf{Z}}_1\mathbf{B}^*\widehat{\mathbf{Z}}_1^{\mathrm{T}} - \mathbf{Z}_1\mathbf{B}^*\mathbf{Z}_1^{\mathrm{T}}\right\|_{\mathrm{op}}$$

$$\leq 2\rho\underbrace{\left\|\mathbf{Z}_0\mathbf{B}^*(\widehat{\mathbf{Z}}_1 - \mathbf{Z}_1)^{\mathrm{T}}\right\|_{\mathrm{op}}}_{\widetilde{I}_{21}} + \rho\underbrace{\left\|\widehat{\mathbf{Z}}_1\mathbf{B}^*\widehat{\mathbf{Z}}_1^{\mathrm{T}} - \mathbf{Z}_1\mathbf{B}^*\mathbf{Z}_1^{\mathrm{T}}\right\|_{\mathrm{op}}}_{\widetilde{I}_{22}}.$$

We now proceed to bound each of these terms.

**Upper bound for $\widetilde{I}_{21}$:** By standard properties of matrix norms,

$$\begin{aligned}\widetilde{I}_{21} &\leq 2\rho\|\mathbf{Z}_0\|_{\text{op}}\|\mathbf{B}^*\|_{\text{op}}\left\|\widehat{\mathbf{Z}}_1 - \mathbf{Z}_1\right\|_{\text{op}} \\ &\leq 2\rho\|\mathbf{Z}_0\|_{\text{op}}\|\mathbf{B}^*\|_{\text{op}}\left\|\widehat{\mathbf{Z}}_1 - \mathbf{Z}_1\right\|_{\text{F}} \\ &\leq 2\rho\sqrt{n_{\max}}\|\mathbf{B}^*\|_{\text{op}}\left\|\widehat{\mathbf{Z}}_1 - \mathbf{Z}_1\right\|_{\text{F}} \\ &\leq 2\rho\sqrt{c_1 n/K}\|\mathbf{B}^*\|_{\text{op}}\sqrt{2|\Omega^c|},\end{aligned} \quad (\text{S5.10})$$

since $(\widehat{\mathbf{Z}}_1 - \mathbf{Z}_1)_{ij} \in \{-1, 0, 1\}$, and only $2|\Omega^c|$ elements are nonzero.

**Upper bound for $\widetilde{I}_{22}$:** For the matrix difference $\widehat{\mathbf{Z}}_1\mathbf{B}^*\widehat{\mathbf{Z}}_1^{\text{T}} - \mathbf{Z}_1\mathbf{B}^*\mathbf{Z}_1^{\text{T}}$, there are at most $|\Omega^c|^2$ nonzero elements, each with absolute value smaller than or equal to $\|\mathbf{B}^*\|_{\max}$. So,

$$\widetilde{I}_{22} \leq \rho|\Omega^c|\|\mathbf{B}^*\|_{\max}. \quad (\text{S5.11})$$

Combining Eqs. (S5.10) and (S5.11) yields

$$\begin{aligned}\widetilde{I}_2 &\leq \frac{1}{n}\|\mathbf{Z}\|_{\text{op}}^2\left(\widetilde{I}_{21} + \widetilde{I}_{22}\right) \\ &\leq \frac{c_1}{K}\left(2\rho\sqrt{c_1 n/K}\sqrt{2|\Omega^c|}\|\mathbf{B}^*\|_{\text{op}} + \rho|\Omega^c|\|\mathbf{B}^*\|_{\max}\right).\end{aligned} \quad (\text{S5.12})$$

Hence, combining Eqs. (S5.6) and (S5.12) and provided $n\sqrt{L\rho} \gtrsim K(\sqrt{K} + \sqrt{\log n})$, we obtain

$$\begin{aligned}I_1 = \frac{1}{n}\left\|\mathbf{Z}^{\text{T}}\boldsymbol{\Delta}_3\mathbf{Z}\right\|_{\text{op}} &\leq 16\sqrt{2}c_1\frac{1}{K}\sqrt{\frac{\rho}{L}}\left(\sqrt{K} + \sqrt{\log n}\right) \\ &+ \frac{c_1}{K}\left(2\rho\sqrt{c_1 n/K}\sqrt{2|\Omega^c|}\|\mathbf{B}^*\|_{\text{op}} + \rho|\Omega^c|\|\mathbf{B}^*\|_{\max}\right)\end{aligned} \quad (\text{S5.13})$$

which holds with probability at least $1 - O(n^{-8})$.

Next, we bound $\|\boldsymbol{\Delta}_3\|_{\text{op}}$ before we bound $I_2$ and $I_3$. We have

$$\begin{aligned}\|\boldsymbol{\Delta}_3\|_{\text{op}} &= \left\|\overline{\mathbf{A}} - \rho\mathbf{Z}\mathbf{B}^*\mathbf{Z}^{\text{T}} + \rho\mathbf{Z}\mathbf{B}^*\mathbf{Z}^{\text{T}} - \rho\widehat{\mathbf{Z}}\mathbf{B}^*\widehat{\mathbf{Z}}^{\text{T}}\right\|_{\text{op}} \\ &\leq \underbrace{\left\|\overline{\mathbf{A}} - \rho\mathbf{Z}\mathbf{B}^*\mathbf{Z}^{\text{T}}\right\|_{\text{op}}}_{I_4} + \underbrace{\left\|\rho\mathbf{Z}\mathbf{B}^*\mathbf{Z}^{\text{T}} - \rho\widehat{\mathbf{Z}}\mathbf{B}^*\widehat{\mathbf{Z}}^{\text{T}}\right\|_{\text{op}}}_{I_5}.\end{aligned}$$

**Upper bound for $I_4$:** Since $\mathbb{E}(\mathbf{A}) = \rho\mathbf{Z}\mathbf{B}^*\mathbf{Z}^{\text{T}}$, applying (Lei and Lin, 2023, Theorem 3)

19

for the choice $t = 12\sqrt{2}\sqrt{n\rho/L}\sqrt{\log n}$ yields

$$\mathbb{P}\left(\left\|\sum_{\ell=1}^{L}(\mathbf{A}^{(\ell)} - \mathbb{E}(\mathbf{A}))\right\|_{\mathrm{op}} > Lt\right) \leq 8n\exp\left\{-\frac{L^2 t^2/8}{vLn + RLt}\right\}$$

$$\leq 8n\exp\left\{-\frac{Lt^2}{16n\rho + 8t}\right\}$$

$$= 8n\exp\left\{-\frac{288\sqrt{L}n\rho\log n}{16\sqrt{L}n\rho + 96\sqrt{2}\sqrt{n\rho\log n}}\right\}$$

$$\leq 8n\exp\left\{-(\log n)^9\right\}$$

$$= 8n^{-8}.$$

The above derivation uses the fact that $A_{ij}^{(\ell)} - \mathbb{E}(A_{ij})$ satisfies the $(v, R)$-Bernstein tail condition in (Lei and Lin, 2023, Definition 1) with $v = 2\rho$ and $R = 1$. The final inequality holds under the condition $Ln\rho \gtrsim \log n$ (i.e., $Ln\rho \geq 72\log n$). In summary,

$$\left\|\overline{\mathbf{A}} - \rho\mathbf{Z}\mathbf{B}^*\mathbf{Z}^\mathrm{T}\right\|_{\mathrm{op}} \leq 12\sqrt{2}\sqrt{\frac{n\rho\log n}{L}} \tag{S5.14}$$

holds with probability at least $1 - O(n^{-8})$.

**Upper bound for** $\mathrm{I}_5$: We have

$$\begin{aligned}\mathrm{I}_5 &\leq \widetilde{\mathrm{I}}_{21} + \widetilde{\mathrm{I}}_{22} \\ &\leq 2\rho\sqrt{c_1 n/K}\sqrt{2|\Omega^c|}\|\mathbf{B}^*\|_{\mathrm{op}} + \rho|\Omega^c|\|\mathbf{B}^*\|_{\max}.\end{aligned} \tag{S5.15}$$

Hence, by combining Eqs. (S5.14) and (S5.15) and provided that $Ln\rho \gtrsim \log n$ (i.e., $Ln\rho \geq 72\log n$), we obtain the bound

$$\|\boldsymbol{\Delta}_3\|_{\mathrm{op}} \leq 12\sqrt{2}\sqrt{\frac{n\rho\log n}{L}} + 2\rho\sqrt{c_1 n/K}\sqrt{2|\Omega^c|}\|\mathbf{B}^*\|_{\mathrm{op}} + \rho|\Omega^c|\|\mathbf{B}^*\|_{\max} \tag{S5.16}$$

which holds with probability at least $1 - O(n^{-8})$.

For the special case $L = 1$, a sharper bound holds for $\mathrm{I}_4$. Namely, if $n\rho \gtrsim \log n$, then from (Lei and Rinaldo, 2015, Theorem 5.2) and (Chen et al., 2021, Theorem 3.4), there exists a constant $C > 0$ such that the inequality

$$\left\|\mathbf{A} - \rho\mathbf{Z}\mathbf{B}^*\mathbf{Z}^\mathrm{T}\right\|_{\mathrm{op}} \leq C\sqrt{n\rho}$$

holds with probability at least $1 - O(n^{-8})$. This implies that

$$\|\boldsymbol{\Delta}_3\|_{\text{op}} \leq C\sqrt{n\rho} + 2\rho\sqrt{c_1 n/K}\sqrt{2|\Omega^c|}\|\mathbf{B}^*\|_{\text{op}} + \rho|\Omega^c|\|\mathbf{B}^*\|_{\max} \quad (S5.17)$$

holds with probability at least $1 - O(n^{-8})$.

**Upper bound for** $\text{I}_2$: We have

$$\begin{aligned}
\text{I}_2 &= \frac{1}{n}\left\|(\widehat{\mathbf{Z}} - \mathbf{Z})^{\text{T}}\boldsymbol{\Delta}_3(\widehat{\mathbf{Z}} - \mathbf{Z})\right\|_{\text{op}} + \frac{2}{n}\left\|(\widehat{\mathbf{Z}} - \mathbf{Z})^{\text{T}}\boldsymbol{\Delta}_3\mathbf{Z}\right\|_{\text{op}} \\
&\leq \frac{1}{n}\left\|\widehat{\mathbf{Z}} - \mathbf{Z}\right\|_{\text{op}}^2 \|\boldsymbol{\Delta}_3\|_{\text{op}} + \frac{2}{n}\left\|\widehat{\mathbf{Z}} - \mathbf{Z}\right\|_{\text{op}} \|\mathbf{Z}\|_{\text{op}} \|\boldsymbol{\Delta}_3\|_{\text{op}} \\
&\leq \frac{1}{n}\left\|\widehat{\mathbf{Z}}_1 - \mathbf{Z}_1\right\|_{\text{F}}^2 \|\boldsymbol{\Delta}_3\|_{\text{op}} + \frac{2}{n}\left\|\widehat{\mathbf{Z}}_1 - \mathbf{Z}_1\right\|_{\text{F}} \|\mathbf{Z}\|_{\text{op}} \|\boldsymbol{\Delta}_3\|_{\text{op}} \\
&\leq \left(2\frac{|\Omega^c|}{n} + 2\sqrt{2c_1}\sqrt{\frac{1}{K}}\sqrt{\frac{|\Omega^c|}{n}}\right)\|\boldsymbol{\Delta}_3\|_{\text{op}}.
\end{aligned} \quad (S5.18)$$

By combining Eqs. (S5.13) and (S5.16) to (S5.18), if $n\sqrt{L}\rho \gtrsim K(\sqrt{K} + \sqrt{\log n})$ and $Ln\rho \gtrsim \log n$, then

$$\begin{aligned}
&\frac{1}{n}\left\|\widehat{\mathbf{Z}}^{\text{T}}\boldsymbol{\Delta}_3\widehat{\mathbf{Z}}\right\|_{\text{op}} \\
&\leq 16\sqrt{2}c_1 \frac{1}{K}\sqrt{\frac{\rho}{L}}\left(\sqrt{K} + \sqrt{\log n}\right) \\
&\quad + \frac{c_1}{K}\left(2\rho\sqrt{c_1 n/K}\sqrt{2|\Omega^c|}\|\mathbf{B}^*\|_{\text{op}} + \rho|\Omega^c|\|\mathbf{B}^*\|_{\max}\right) \\
&\quad + \left(2\frac{|\Omega^c|}{n} + 2\sqrt{2c_1}\sqrt{\frac{1}{K}}\sqrt{\frac{|\Omega^c|}{n}}\right) 12\sqrt{2}\sqrt{\frac{n\rho\log n}{L}} \\
&\quad + \left(2\frac{|\Omega^c|}{n} + 2\sqrt{2c_1}\sqrt{\frac{1}{K}}\sqrt{\frac{|\Omega^c|}{n}}\right)\left(2\rho\sqrt{c_1 n/K}\sqrt{2|\Omega^c|}\|\mathbf{B}^*\|_{\text{op}} + \rho|\Omega^c|\|\mathbf{B}^*\|_{\max}\right)
\end{aligned}$$

holds with probability at least $1 - O(n^{-8})$. If $L = 1$, then the alternative bound

$$\begin{aligned}
&\frac{1}{n}\left\|\widehat{\mathbf{Z}}^{\text{T}}\boldsymbol{\Delta}_3\widehat{\mathbf{Z}}\right\|_{\text{op}} \\
&\leq 16\sqrt{2}c_1 \frac{1}{K}\sqrt{\rho}\left(\sqrt{K} + \sqrt{\log n}\right) \\
&\quad + \frac{c_1}{K}\left(2\rho\sqrt{c_1 n/K}\sqrt{2|\Omega^c|}\|\mathbf{B}^*\|_{\text{op}} + \rho|\Omega^c|\|\mathbf{B}^*\|_{\max}\right) \\
&\quad + \left(2\frac{|\Omega^c|}{n} + 2\sqrt{2c_1}\sqrt{\frac{1}{K}}\sqrt{\frac{|\Omega^c|}{n}}\right) C\sqrt{n\rho} \\
&\quad + \left(2\frac{|\Omega^c|}{n} + 2\sqrt{2c_1}\sqrt{\frac{1}{K}}\sqrt{\frac{|\Omega^c|}{n}}\right)\left(2\rho\sqrt{c_1 n/K}\sqrt{2|\Omega^c|}\|\mathbf{B}^*\|_{\text{op}} + \rho|\Omega^c|\|\mathbf{B}^*\|_{\max}\right)
\end{aligned}$$

holds with probability at least $1 - O(n^{-8})$. This completes the proof of Lemma 2. $\square$

## S5.3 Proof of Lemma 3

Lemma 3 facilitates establishing a nuclear norm error bound. The proof of Lemma 3 follows the proof of (Hamdi and Bayati, 2022, Lemma B.1) which adapts earlier work in Klopp (2014).

**Lemma 3** (Adapted from Hamdi and Bayati (2022))**.** Write $\mathbf{B}_\rho \coloneqq \rho\mathbf{B}$. If it holds that $\lambda \geq 3n^{-1}\|\widehat{\mathbf{Z}}^{\mathrm{T}}\boldsymbol{\Delta}_3\widehat{\mathbf{Z}}\|_{\mathrm{op}}$, then

$$\|\mathbf{P}_{\mathbf{B}_\rho^*}^{\perp}(\widehat{\mathbf{B}}_\rho - \mathbf{B}_\rho^*)\|_* \leq 5\|\mathbf{P}_{\mathbf{B}_\rho^*}(\widehat{\mathbf{B}}_\rho - \mathbf{B}_\rho^*)\|_*.$$

*Proof of Lemma 3.* Conditional on $\widehat{\mathbf{Z}}$ and $\overline{\mathbf{A}}$, define $\mathcal{L}(\mathbf{B}_\rho) \coloneqq n^{-1}\|\overline{\mathbf{A}} - \widehat{\mathbf{Z}}\mathbf{B}_\rho\widehat{\mathbf{Z}}^{\mathrm{T}}\|_{\mathrm{F}}^2$ which is convex in the matrix variable $\mathbf{B}_\rho$. As a result,

$$\frac{1}{n}\|\overline{\mathbf{A}} - \widehat{\mathbf{Z}}\widehat{\mathbf{B}}_\rho\widehat{\mathbf{Z}}^{\mathrm{T}}\|_{\mathrm{F}}^2 - \frac{1}{n}\|\overline{\mathbf{A}} - \widehat{\mathbf{Z}}\mathbf{B}_\rho^*\widehat{\mathbf{Z}}^{\mathrm{T}}\|_{\mathrm{F}}^2 \geq -\frac{2}{n}\left\langle\widehat{\mathbf{Z}}^{\mathrm{T}}\boldsymbol{\Delta}_3\widehat{\mathbf{Z}}, \widehat{\mathbf{B}}_\rho - \mathbf{B}_\rho^*\right\rangle$$

$$\geq -\frac{2}{n}\|\widehat{\mathbf{Z}}^{\mathrm{T}}\boldsymbol{\Delta}_3\widehat{\mathbf{Z}}\|_{\mathrm{op}}\|\widehat{\mathbf{B}}_\rho - \mathbf{B}_\rho^*\|_*$$

$$\geq -\frac{2}{3}\lambda\|\widehat{\mathbf{B}}_\rho - \mathbf{B}_\rho^*\|_*.$$

Next, replacing $\mathbf{B}$ and $\mathbf{C}$ in Eq. (S5.3) with $\mathbf{B}_\rho^*$ and $\widehat{\mathbf{B}}_\rho - \mathbf{B}_\rho^*$, respectively, yields

$$\frac{2}{3}\lambda\|\widehat{\mathbf{B}}_\rho - \mathbf{B}_\rho^*\|_* \geq \frac{1}{n}\|\overline{\mathbf{A}} - \widehat{\mathbf{Z}}\mathbf{B}_\rho^*\widehat{\mathbf{Z}}^{\mathrm{T}}\|_{\mathrm{F}}^2 - \frac{1}{n}\|\overline{\mathbf{A}} - \widehat{\mathbf{Z}}\widehat{\mathbf{B}}_\rho\widehat{\mathbf{Z}}^{\mathrm{T}}\|_{\mathrm{F}}^2$$

$$\geq \lambda\|\widehat{\mathbf{B}}_\rho\|_* - \lambda\|\mathbf{B}_\rho^*\|_*$$

$$\geq \lambda\|\mathbf{P}_{\mathbf{B}_\rho^*}^{\perp}(\widehat{\mathbf{B}}_\rho - \mathbf{B}_\rho^*)\|_* - \lambda\|\mathbf{P}_{\mathbf{B}_\rho^*}(\widehat{\mathbf{B}}_\rho - \mathbf{B}_\rho^*)\|_*.$$

Thus, by the triangle inequality,

$$\|\mathbf{P}_{\mathbf{B}_\rho^*}^{\perp}(\widehat{\mathbf{B}}_\rho - \mathbf{B}_\rho^*)\|_* \leq 5\|\mathbf{P}_{\mathbf{B}_\rho^*}(\widehat{\mathbf{B}}_\rho - \mathbf{B}_\rho^*)\|_*.$$

This completes the proof of Lemma 3. □

## S5.4 Proof of Theorem 1, Frobenius norm bound

*Proof of Theorem 1.* Let $\mathcal{L}(\mathbf{B}_\rho) \coloneqq n^{-1}\|\overline{\mathbf{A}} - \widehat{\mathbf{Z}}\mathbf{B}_\rho\widehat{\mathbf{Z}}^{\mathrm{T}}\|_{\mathrm{F}}^2$, where $\mathbf{B}_\rho \coloneqq \rho\mathbf{B}$. Since $\widehat{\mathbf{B}}_\rho$ is the solution to our optimization problem, we have

$$\frac{1}{n}\|\overline{\mathbf{A}} - \widehat{\mathbf{Z}}\widehat{\mathbf{B}}_\rho\widehat{\mathbf{Z}}^{\mathrm{T}}\|_{\mathrm{F}}^2 + \lambda\|\widehat{\mathbf{B}}_\rho\|_* \leq \frac{1}{n}\|\overline{\mathbf{A}} - \widehat{\mathbf{Z}}\mathbf{B}_\rho^*\widehat{\mathbf{Z}}^{\mathrm{T}}\|_{\mathrm{F}}^2 + \lambda\|\mathbf{B}_\rho^*\|_*.$$

Therefore, by expanding $\mathcal{L}(\mathbf{B}_\rho)$ and after some algebra, we have

$$\frac{1}{n}\|\widehat{\mathbf{Z}}(\mathbf{B}_\rho^* - \widehat{\mathbf{B}}_\rho)\widehat{\mathbf{Z}}^{\mathrm{T}}\|_{\mathrm{F}}^2 + \frac{2}{n}\left\langle \widehat{\mathbf{Z}}^{\mathrm{T}}\boldsymbol{\Delta}_3\widehat{\mathbf{Z}}, \mathbf{B}_\rho^* - \widehat{\mathbf{B}}_\rho \right\rangle + \lambda\|\widehat{\mathbf{B}}_\rho\|_* \leq \lambda\|\mathbf{B}_\rho^*\|_*,$$

where as before $\boldsymbol{\Delta}_3 = \boldsymbol{\Delta}_1 - \boldsymbol{\Delta}_2$. Due to the trace duality property, we obtain

$$\frac{1}{n}\|\widehat{\mathbf{Z}}(\mathbf{B}_\rho^* - \widehat{\mathbf{B}}_\rho)\widehat{\mathbf{Z}}^{\mathrm{T}}\|_{\mathrm{F}}^2 + \lambda\|\widehat{\mathbf{B}}_\rho\|_* \leq \frac{2}{n}\|\widehat{\mathbf{Z}}^{\mathrm{T}}\boldsymbol{\Delta}_3\widehat{\mathbf{Z}}\|_{\mathrm{op}}\|\mathbf{B}_\rho^* - \widehat{\mathbf{B}}_\rho\|_* + \lambda\|\mathbf{B}_\rho^*\|_*. \tag{S5.19}$$

By replacing $\mathbf{B}$ and $\mathbf{C}$ in Eq. (S5.19) with $\mathbf{B}_\rho^*$ and $\widehat{\mathbf{B}}_\rho - \mathbf{B}_\rho^*$, respectively, and subsequently applying Lemma 2, we obtain

$$\begin{aligned}
\frac{1}{n}&\|\widehat{\mathbf{Z}}(\mathbf{B}_\rho^* - \widehat{\mathbf{B}}_\rho)\widehat{\mathbf{Z}}^{\mathrm{T}}\|_{\mathrm{F}}^2 \\
&\leq \frac{2}{n}\|\widehat{\mathbf{Z}}^{\mathrm{T}}\boldsymbol{\Delta}_3\widehat{\mathbf{Z}}\|_{\mathrm{op}}\|\mathbf{B}_\rho^* - \widehat{\mathbf{B}}_\rho\|_* + \lambda\|\mathbf{P}_{\mathbf{B}_\rho^*}(\widehat{\mathbf{B}}_\rho - \mathbf{B}_\rho^*)\|_* \\
&\quad - \lambda\|\mathbf{P}_{\mathbf{B}_\rho^*}^{\perp}(\widehat{\mathbf{B}}_\rho - \mathbf{B}_\rho^*)\|_* \\
&\leq \left(\frac{2}{n}\|\widehat{\mathbf{Z}}^{\mathrm{T}}\boldsymbol{\Delta}_3\widehat{\mathbf{Z}}\|_{\mathrm{op}} + \lambda\right)\|\mathbf{P}_{\mathbf{B}_\rho^*}(\widehat{\mathbf{B}}_\rho - \mathbf{B}_\rho^*)\|_* \\
&\quad + \left(\frac{2}{n}\|\widehat{\mathbf{Z}}^{\mathrm{T}}\boldsymbol{\Delta}_3\widehat{\mathbf{Z}}\|_{\mathrm{op}} - \lambda\right)\|\mathbf{P}_{\mathbf{B}_\rho^*}^{\perp}(\widehat{\mathbf{B}}_\rho - \mathbf{B}_\rho^*)\|_* \\
&\leq \frac{5}{3}\lambda\|\mathbf{P}_{\mathbf{B}_\rho^*}(\widehat{\mathbf{B}}_\rho - \mathbf{B}_\rho^*)\|_*.
\end{aligned} \tag{S5.20}$$

Using this observation and the fact in Eq. (S5.2) that $\mathrm{rk}(\mathbf{P}_{\mathbf{B}_\rho^*}(\widehat{\mathbf{B}}_\rho - \mathbf{B}_\rho^*)) \leq 2\,\mathrm{rk}(\mathbf{B}_\rho^*)$, we can apply the Cauchy–Schwarz inequality to the singular values of $\mathbf{P}_{\mathbf{B}_\rho^*}(\widehat{\mathbf{B}}_\rho - \mathbf{B}_\rho^*)$ to obtain

$$\begin{aligned}
\|\mathbf{P}_{\mathbf{B}_\rho^*}&(\widehat{\mathbf{B}}_\rho - \mathbf{B}_\rho^*)\|_* \\
&= \sum_{i=1}^{\mathrm{rk}(\mathbf{P}_{\mathbf{B}_\rho^*}(\widehat{\mathbf{B}}_\rho - \mathbf{B}_\rho^*))} \sigma_i(\mathbf{P}_{\mathbf{B}_\rho^*}(\widehat{\mathbf{B}}_\rho - \mathbf{B}_\rho^*)) \\
&\leq \sqrt{\mathrm{rk}(\mathbf{P}_{\mathbf{B}_\rho^*}(\widehat{\mathbf{B}}_\rho - \mathbf{B}_\rho^*))}\left(\sum_{i=1}^{\mathrm{rk}(\mathbf{P}_{\mathbf{B}_\rho^*}(\widehat{\mathbf{B}}_\rho - \mathbf{B}_\rho^*))} \sigma_i^2(\mathbf{P}_{\mathbf{B}_\rho^*}(\widehat{\mathbf{B}}_\rho - \mathbf{B}_\rho^*))\right)^{1/2} \\
&= \sqrt{\mathrm{rk}(\mathbf{P}_{\mathbf{B}_\rho^*}(\widehat{\mathbf{B}}_\rho - \mathbf{B}_\rho^*))}\|\mathbf{P}_{\mathbf{B}_\rho^*}(\widehat{\mathbf{B}}_\rho - \mathbf{B}_\rho^*)\|_{\mathrm{F}} \\
&\leq \sqrt{2\,\mathrm{rk}(\mathbf{B}_\rho^*)}\|\mathbf{P}_{\mathbf{B}_\rho^*}(\widehat{\mathbf{B}}_\rho - \mathbf{B}_\rho^*)\|_{\mathrm{F}}.
\end{aligned} \tag{S5.21}$$

By using Eqs. (S5.20) and (S5.21), we have

$$\begin{aligned}
\frac{1}{n}\|\widehat{\mathbf{Z}}(\mathbf{B}_\rho^* - \widehat{\mathbf{B}}_\rho)\widehat{\mathbf{Z}}^{\mathrm{T}}\|_{\mathrm{F}}^2 &\leq \frac{5}{3}\lambda\sqrt{2\,\mathrm{rk}(\mathbf{B}_\rho^*)}\|\mathbf{P}_{\mathbf{B}_\rho^*}(\widehat{\mathbf{B}}_\rho - \mathbf{B}_\rho^*)\|_{\mathrm{F}} \\
&\leq \frac{5}{3}\lambda\sqrt{2\,\mathrm{rk}(\mathbf{B}_\rho^*)}\|\widehat{\mathbf{B}}_\rho - \mathbf{B}_\rho^*\|_{\mathrm{F}}.
\end{aligned}$$

Then, by Lemma 1, we have

$$\frac{n}{(c_2 K)^2}\|\widehat{\mathbf{B}}_\rho - \mathbf{B}^*_\rho\|_{\mathrm{F}}^2 \leq \frac{1}{n}\|\widehat{\mathbf{Z}}(\widehat{\mathbf{B}}_\rho - \mathbf{B}^*_\rho)\widehat{\mathbf{Z}}^{\mathrm{T}}\|_{\mathrm{F}}^2 \qquad (S5.22)$$
$$\leq \frac{5}{3}\lambda\sqrt{2\,\mathrm{rk}(\mathbf{B}^*_\rho)}\|\widehat{\mathbf{B}}_\rho - \mathbf{B}^*_\rho\|_{\mathrm{F}}.$$

Eq. (S5.22) is equivalent to

$$\|\widehat{\mathbf{B}} - \mathbf{B}^*\|_{\mathrm{F}} \leq \frac{5}{3}c_2^2 K^2\sqrt{2\,\mathrm{rk}(\mathbf{B}^*)}\frac{\lambda}{n\rho}.$$

Thus, from Lemma 2, if $n\sqrt{L\rho} \gtrsim K(\sqrt{K} + \sqrt{\log n})$ and $Ln\rho \gtrsim \log n$, then

$$\|\widehat{\mathbf{B}} - \mathbf{B}^*\|_{\mathrm{F}}$$
$$\leq 5c_2^2 K^2\sqrt{2\,\mathrm{rk}(\mathbf{B}^*)}\Bigg[16\sqrt{2}\frac{c_1}{K}\sqrt{\frac{1}{Ln^2\rho}}\left(\sqrt{K} + \sqrt{\log n}\right)$$
$$+ \left(\frac{c_1}{K} + 2\sqrt{2}\sqrt{\frac{c_1}{K}}\sqrt{\frac{|\Omega^c|}{n}} + 2\frac{|\Omega^c|}{n}\right) \qquad (S5.23)$$
$$\times \left(2\sqrt{2}\sqrt{\frac{c_1}{K}}\|\mathbf{B}^*\|_{\mathrm{op}}\sqrt{\frac{|\Omega^c|}{n}} + \|\mathbf{B}^*\|_{\max}\frac{|\Omega^c|}{n}\right)$$
$$+ \left(2\sqrt{2}\sqrt{\frac{c_1}{K}}\sqrt{\frac{|\Omega^c|}{n}} + 2\frac{|\Omega^c|}{n}\right)12\sqrt{2}\sqrt{\frac{\log n}{Ln\rho}}\Bigg],$$

while if $L = 1$, under the same condition as above,

$$\|\widehat{\mathbf{B}} - \mathbf{B}^*\|_{\mathrm{F}} \leq 5c_2^2 K^2\sqrt{2\,\mathrm{rk}(\mathbf{B}^*)}\Bigg[16\sqrt{2}\frac{c_1}{K}\sqrt{\frac{1}{n^2\rho}}\left(\sqrt{K} + \sqrt{\log n}\right)$$
$$+ \left(\frac{c_1}{K} + 2\sqrt{2}\sqrt{\frac{c_1}{K}}\sqrt{\frac{|\Omega^c|}{n}} + 2\frac{|\Omega^c|}{n}\right) \qquad (S5.24)$$
$$\times \left(2\sqrt{2}\sqrt{\frac{c_1}{K}}\|\mathbf{B}^*\|_{\mathrm{op}}\sqrt{\frac{|\Omega^c|}{n}} + \|\mathbf{B}^*\|_{\max}\frac{|\Omega^c|}{n}\right)$$
$$+ \left(2\sqrt{2}\sqrt{\frac{c_1}{K}}\sqrt{\frac{|\Omega^c|}{n}} + 2\frac{|\Omega^c|}{n}\right)C\sqrt{\frac{1}{n\rho}}\Bigg],$$

each with probability at least $1 - O(n^{-8})$. This completes the proof of the Frobenius norm error bound in Theorem 1. □

## S5.5 Proof of Theorem 1, nuclear norm bound

*Proof of Theorem 1.* By the triangle inequality, Lemma 3 and Eqs. (S5.2) and (S5.21),

$$\|\widehat{\mathbf{B}} - \mathbf{B}^*\|_* = \|\mathbf{P}_{\mathbf{B}^*}(\widehat{\mathbf{B}} - \mathbf{B}^*) + \mathbf{P}_{\mathbf{B}^*}^\perp(\widehat{\mathbf{B}} - \mathbf{B}^*)\|_*$$

$$\leq \|\mathbf{P}_{\mathbf{B}^*}(\widehat{\mathbf{B}} - \mathbf{B}^*)\|_* + \|\mathbf{P}_{\mathbf{B}^*}^\perp(\widehat{\mathbf{B}} - \mathbf{B}^*)\|_*$$

$$\leq 6\|\mathbf{P}_{\mathbf{B}^*}(\widehat{\mathbf{B}} - \mathbf{B}^*)\|_*$$

$$\leq \sqrt{72\,\mathrm{rk}(\mathbf{B}^*)}\|\mathbf{P}_{\mathbf{B}^*}(\widehat{\mathbf{B}} - \mathbf{B}^*)\|_\mathrm{F}$$

$$\leq \sqrt{72\,\mathrm{rk}(\mathbf{B}^*)}\|\widehat{\mathbf{B}} - \mathbf{B}^*\|_\mathrm{F}.$$

Finally, applying the Frobenius norm error bound derived in the previous proof yields the stated result. This completes the proof of the nuclear norm error bound in Theorem 1. □

 % << arXiv

\end{document}